\documentclass[acmsmall, authorversion]{acmart}

\AtEndPreamble{
  \hypersetup{           % Now apply the settings after the preamble
    colorlinks=true,     % Enable colored links
    linkcolor=ACMPurple,      % Internal links in blue
    citecolor=ACMPurple,     % Citations in green
    urlcolor=ACMDarkBlue         % URLs in red
  }
}

\usepackage{stmaryrd}
\usepackage{mathrsfs}
\usepackage{bbold}
\usepackage{algorithm}
\usepackage{algpseudocode}
\usepackage{enumitem}
\usepackage{subcaption}  % Add this in the preamble
\usepackage{tikz}
\usepackage{graphicx}
\usepackage[absolute,overlay]{textpos}
\usepackage{xcolor}
\DeclareCaptionLabelFormat{custom}{\MakeUppercase{#1}#2:}
\AtBeginDocument{%
  }

 \setcopyright{acmlicensed}
 \copyrightyear{2025}
 \acmYear{2025}
 \acmDOI{10.1145/3728312}

 \acmJournal{PACMCGIT}
 \acmVolume{8}
 \acmNumber{1}
 \acmArticle{7}
 \acmMonth{5}

\begin{document}

%copy right : https://www.weather.gov/apx/severe-2017Jun11
%     \centering
%     \begin{minipage}[b]{0.24\textwidth}
%         \includegraphics[width=2.67cm,height=1.5cm]{illstrate/Teaser/multicell.jpeg}
%     \end{minipage}%
%     \hspace{0.1cm}
%     \begin{minipage}[b]{0.24\textwidth}
%         \includegraphics[width=2.67cm,height=1.5cm]{illstrate/Teaser/supercell.jpg}
%     \end{minipage}%
%     \hspace{0.1cm}
%     \begin{minipage}[b]{0.24\textwidth}
%         \includegraphics[width=2.67cm,height=1.5cm]{illstrate/6-3/japan_thunder.jpg}
%     \end{minipage}%
%     \hspace{0.1cm}
%     \begin{minipage}[b]{0.24\textwidth}
%         \includegraphics[width=2.67cm,height=1.5cm]{illstrate/6-3/california_thunder.jpg}
%     \end{minipage}%
%     \caption{Different types of thunderstorms from left to right: Single cell, multicell, squall line, and supercell.}
%     \Description{Illustrations of four distinct types of thunderstorms: single cell, multicell, squall line, and supercell. Each image represents a unique thunderstorm system observed in meteorological studies.}

\begin{teaserfigure}
    \centering
    \includegraphics[width=0.24\linewidth, height=2cm]{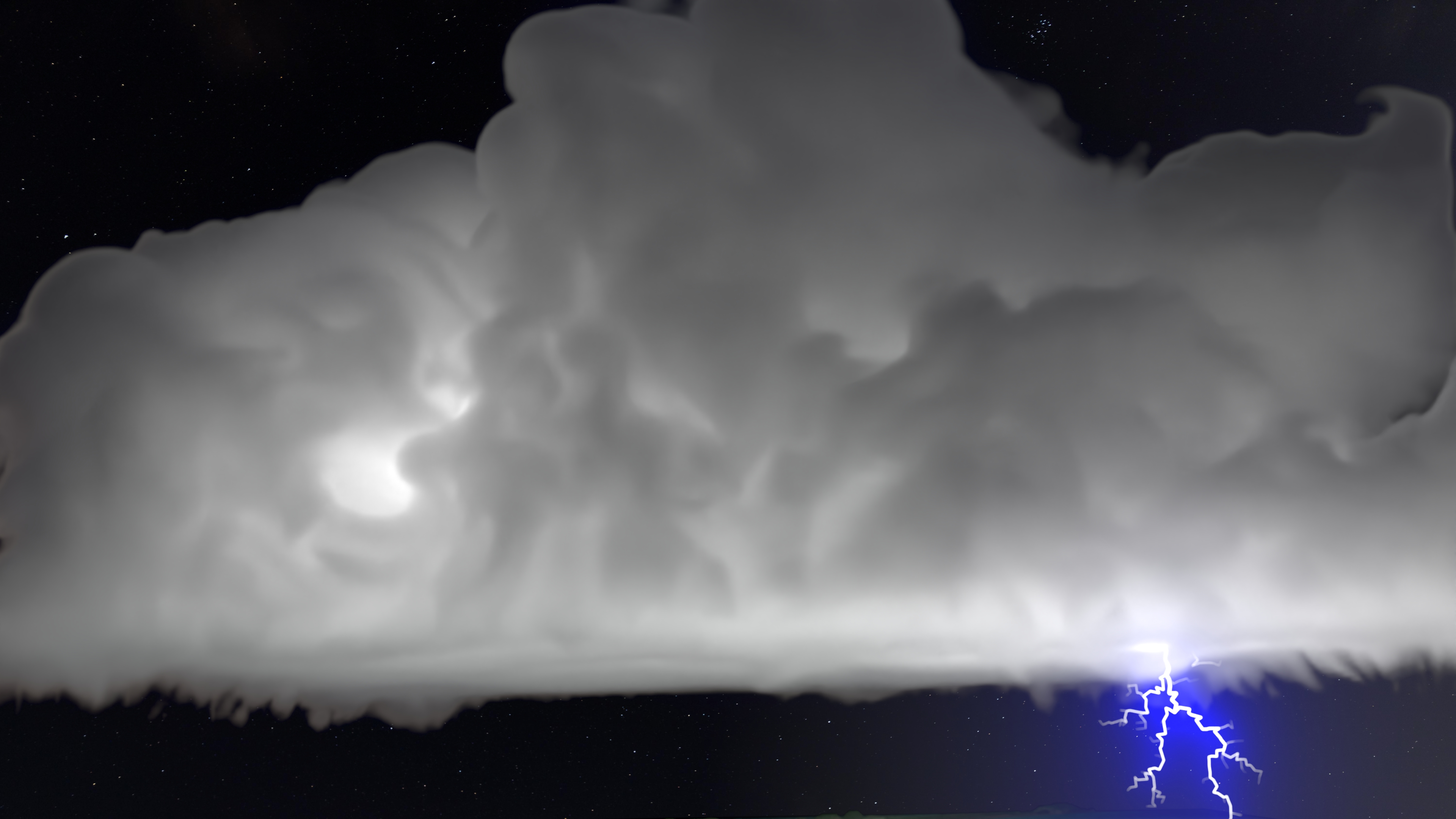}%
    \hspace{0.1cm}
    \includegraphics[width=0.24\linewidth, height=2cm]{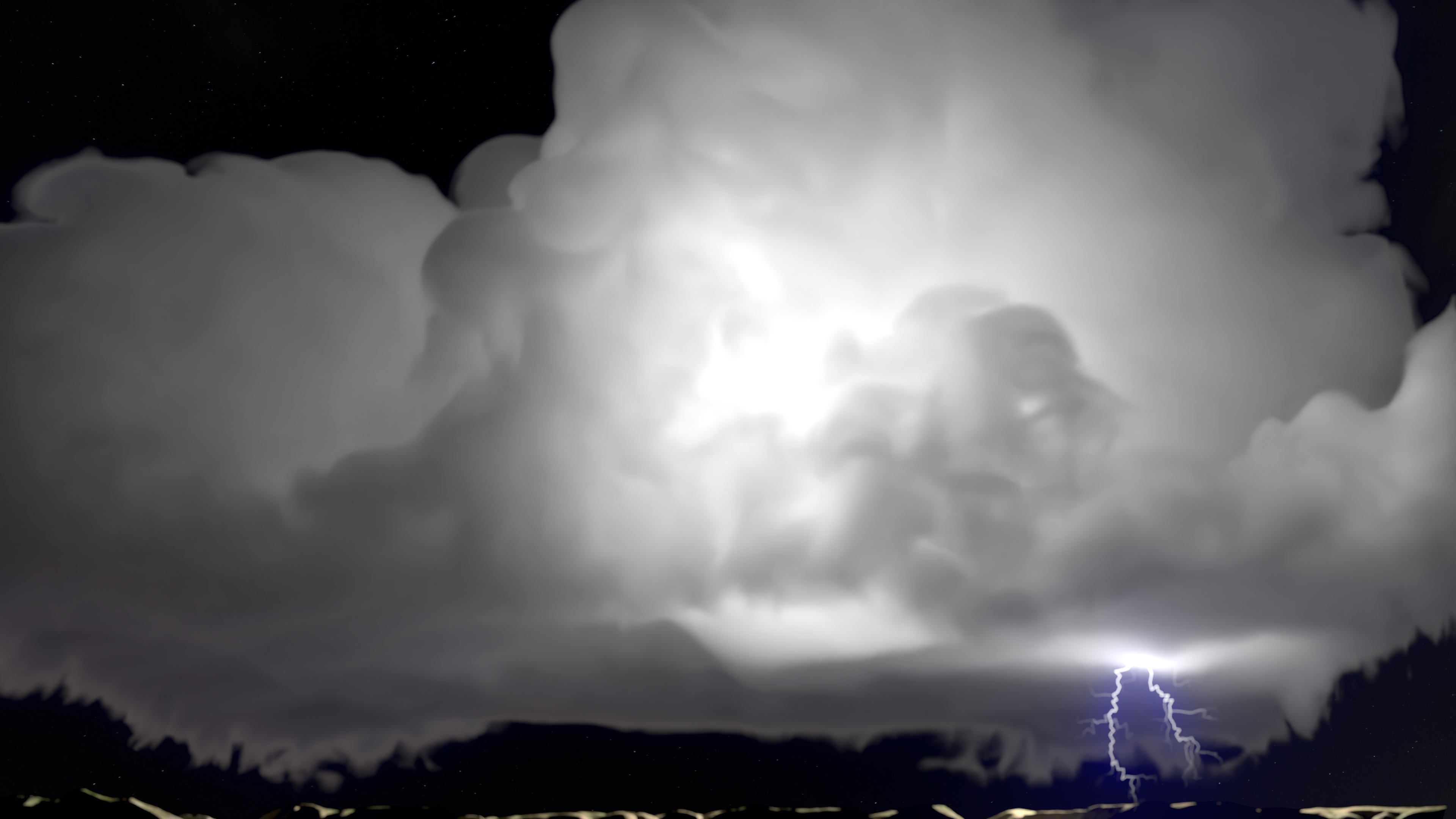}%
    \hspace{0.1cm}
    \includegraphics[width=0.24\linewidth, height=2cm]{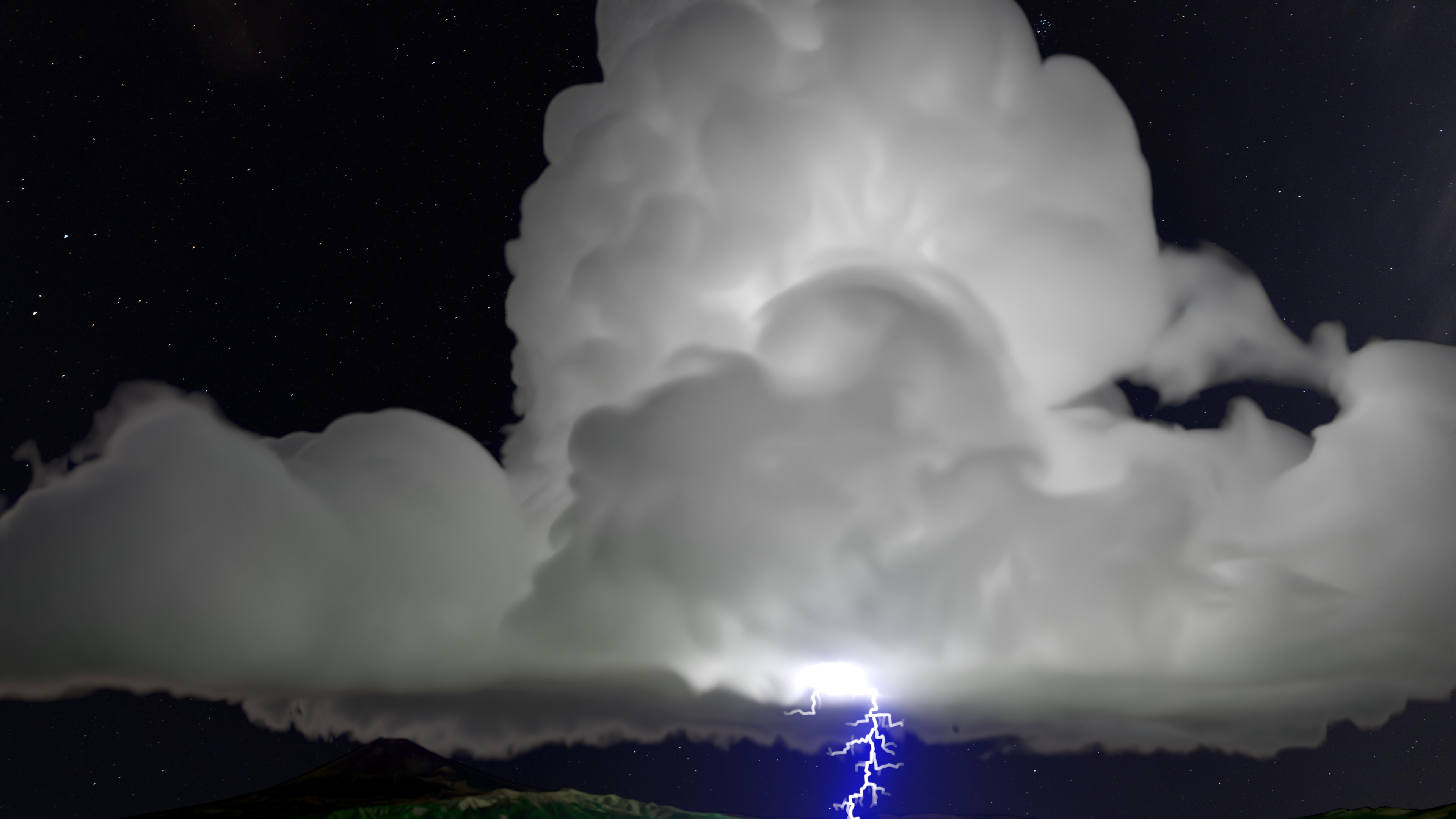}%
    \hspace{0.1cm}
    \includegraphics[width=0.24\linewidth, height=2cm]{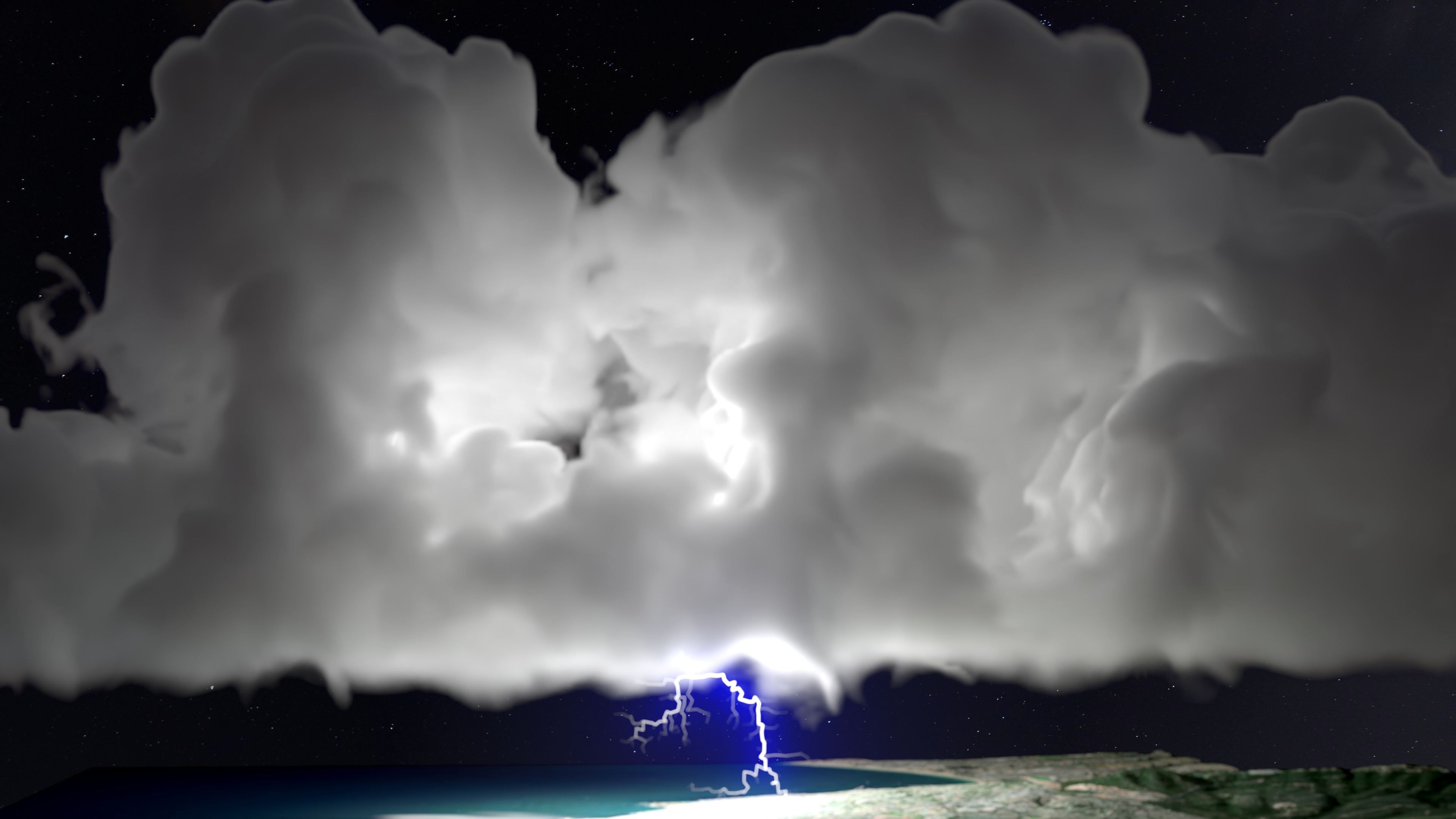}%
    \caption{MCSs from different locations and seasons, from left to right: tropical thunderstorms in Florida and New Mexico, a winter thunderstorm in Japan, and an atmospheric river in California. Each image represents a typical thunderstorm system observed in meteorological studies.}
    \Description{Illustrations of four distinct types of thunderstorms: multi-cell, supercell, winter thunderstorm, atmospheric river. Each image represents a unique thunderstorm system observed in meteorological studies.}
\end{teaserfigure}

%\title{Thunderscapes: interactive mesoscale simulation of thunderstorm}
% \title{Thunderscapes: interactive simulation of mesoscale convective system}
% \title{Thunderscapes: realistic simulation of mesoscale convective system}
% \title{Thunderscapes: fast simulation of mesoscale convective system}
% \title{Thunderscapes: efficient simulation of mesoscale convective system}
\title{Thunderscapes: Simulating the Dynamics of Mesoscale Convective System}

\author{TIANCHEN HAO}
\email{2023223040024@stu.scu.edu.cn}
\orcid{0009-0007-2499-0079}
\affiliation{%
	\institution{Sichuan University}
	\country{China}
}

\author{JINXIAN PAN}
\email{lemonxorp@gmail.com}
\orcid{0009-0000-7564-5039}
\affiliation{%
	\institution{Sichuan University}
	\country{China}
}

\author{YANGCHENG XIANG}
\email{yangcheng.xiang.k0@elms.hokudai.ac.jp}
\orcid{0009-0007-6163-6028}
\affiliation{%
	\institution{Hokkaido University}
	\country{Japan}
}

\author{XIANGDA SHEN}
\email{2022223045139@stu.scu.edu.cn}
\orcid{0000-0003-1449-8826}
\affiliation{%
	\institution{Sichuan University}
	\country{China}
}

\author{XINSHENG LI}
\affiliation{%
	\institution{Sichuan University}
	% \city{Hekla}
	\country{China}}
\email{lixinsheng@scu.edu.cn}
\orcid{0009-0009-5894-5465}

\author{YANCI ZHANG}
\affiliation{%
	\institution{Sichuan University}
	% \city{New York}
	\country{China}}
\email{yczhang@scu.edu.cn}
\orcid{0000-0001-7045-185X}


% \renewcommand{\shortauthors}{Hao et al.}

% \begin{abstract}
% A Mesoscale Convective System (MCS) is a collection of thunderstorms that function as a system, representing a widely discussed phenomenon in both the natural sciences and visual effects industries, and embodying the untamed forces of nature.In this paper, we present the first efficient, physically based mesoscale thunderstorms simulation model that integrates Grabowski-style cloud microphysics with hydrometeor electrification processes. Our model simulates thunderclouds development and lightning flashes within a unified meteorological framework, providing a realistic and efficient approach for graphical applications. By incorporating key physical principles, it effectively links cloud formation, electrification, and lightning generation. The simulation also encompasses various thunderstorm types and their corresponding lightning activities.
% \end{abstract}

\begin{abstract}

A Mesoscale Convective System (MCS) is a collection of thunderstorms operating as a unified system, showcasing nature's untamed power. They represent a phenomenon widely referenced in both the natural sciences and the visual effects (VFX) industries.However, in computer graphics, visually accurate simulation of MCS dynamics remains a significant challenge due to the inherent complexity of atmospheric microphysical processes.To achieve a high level of visual quality while ensuring practical performance, we introduce \textit{Thunderscapes}, the first physically based simulation framework for visually realistic MCS tailored to graphical applications.Our model integrates mesoscale cloud microphysics with hydrometeor electrification processes to simulate thunderstorm development and lightning flashes. By capturing various thunderstorm types and their associated lightning activities, \textit{Thunderscapes} demonstrates the versatility and physical accuracy of the proposed approach.

\end{abstract}


% \ccsdesc[500]{Computing methodologies~Physical simulation;Procedural animation}



\maketitle

\section{INTRODUCTION}

Thunderstorms represent the wild power of nature and are a common atmospheric element in the visual effects (VFX) industry. Notable works, such as \textit{Horizon Forbidden West: Burning Shores} and \textit{Ghost of Tsushima},  demonstrate the importance of realistic atmospheric effects. Modern applications demand tools for efficient and realistic simulation of thunderstorms.

However, current general-purpose VFX software, such as Houdini and Maya, relies on traditional Eulerian fluid dynamics (e.g., stable fluids \cite{stam2023stable}) to simulate phenomena like smoke or explosions. They fall short in capturing the inherent complexity of atmospheric microphysical processes, which are essential for accurately simulating the dynamics of complex thunderstorm systems, such as MCS, and for producing consistent, high-quality results.

This paper introduces \textit{Thunderscapes}, a physically based simulation framework for visually realistic MCS. By incorporating a Grabowski-style cloud microphysics model~\cite{grabowski1998toward} alongside hydrometeor electrification processes, our approach delivers high visual fidelity while maintaining practical performance levels, capturing the core dynamics of thunderstorm evolution and lightning activities within an MCS. Furthermore, the framework integrates seamlessly with Houdini, a widely used tool in the VFX industry, through intuitive and lightweight parameters, allowing artists to utilize its advanced simulation features without requiring expertise in meteorology or knowledge of supercomputing.

The key contributions of this work are as follows:
\begin{enumerate}[label=(\arabic*), itemsep=0pt, topsep=5pt, partopsep=0pt]
    \item We develop a realistic, physically based simulation framework for MCS, integrating mesoscale cloud microphysics with hydrometeor electrification processes to simulate diverse thunderstorm phenomena.
    \item We present a lightweight set of parameters that allows artists to intuitively create thunderstorm animations within general-purpose VFX software.
    \item We validate the framework using diverse meteorological datasets, demonstrating its capability to generate visually realistic and physically consistent atmospheric effects.
\end{enumerate}
\section{RELATED WORK}
\subsection{Simulating Thunderstorm in Computer Graphics}

The visual simulation of atmospheric phenomena such as convective clouds has been extensively explored through a variety of computational methods. Webanck et al.~\cite{webanck2018procedural} proposed a procedural approach for generating cloudscapes, while Miyazaki et al.~\cite{miyazaki2002simulation} simulated cumulus clouds by coupling computational fluid dynamics (CFD) with fundamental water transport equations. Ferreira et al.~\cite{ferreira2015adaptive} and Zhang et al.~\cite{zhang2020target} utilized position-based fluids (PBF) for adaptive cloud simulations. Smoothed Particle Hydrodynamics (SPH) techniques, as demonstrated by Goswami and Neyret~\cite{goswami2017real}, focus on real-time simulations of convective clouds. Additionally, Vimont et al.~\cite{vimont2020interactive} proposed a hybrid, 2D-layered atmospheric model to simulate mesoscale skyscapes.
However, these approaches lack a detailed focus on cloud microphysics, limiting their ability to produce realistic and complex cloudscapes.

% Some studies have specifically focused on the simulation of volcanic cloud dynamics. Lastic et al.~\cite{lastic2022interactive} employed Lagrangian dynamics to simulate volcanic plumes and pyroclastic flows, while Pretorius et al.~\cite{pretorius2024volcanic} integrated volcanic eruptions with atmospheric simulations to produce coherent skyscapes. 
% In their framework, lightning effects were incorporated, but the electrification process, which is essential for lightning formation, was not considered, resulting in a physically inaccurate simulation..

In the domain of lightning simulation, Reed and Wyvill~\cite{reed1994visual}, Kim and Lin~\cite{kim2007fast}, and Yun et al.~\cite{yun2017physically} developed methods for the efficient development of lightning branches, contributing to a more realistic representation of the method of electrical discharges.

Recently, more sophisticated microphysical schemes from atmospheric science have been incorporated into computer graphics research. Garcia-Dorado et al.~\cite{garcia2017fast} and Hädrich et al.~\cite{hadrich2020stormscapes} adopted the classic Kessler warm cloud microphysics scheme~\cite{kessler1969distribution} in cloud simulations. Herrera et al.~\cite{herrera2021weatherscapes} extended cloud simulations to include multiphase cloud dynamics. Amador Herrera et al.~\cite{amador2024cyclogenesis} developed a framework to simulate hurricane and tornado dynamics.However, these methods overlook the hydrometeor electrification process, which limits their ability to simulate consistent and realistic thunderstorms with  associated lightning phenomena.
% incorporating turbulent microphysics to model these extreme atmospheric events with greater accuracy.

\subsection{Thunderstorm models in atmospheric sciences}

Thunderstorm microphysics and electrification have been widely discussed topics in the field of atmospheric science.Kessler~\cite{kessler1969distribution} proposed a fundamental framework for the distribution and continuity of water substance in atmospheric circulations, which remains influential in the parameterization of the microphysics of warm clouds.

One notable development in cloud microphysics is the work by Grabowski~\cite{grabowski1998toward}, who introduced an extended warm cloud microphysics scheme for large-scale tropical circulations. His method divides the parameterization of warm and cold clouds using a temperature interpolation scheme, a significant inspiration for our approach.

In terms of thunderstorm electrification, Solomon et al.~\cite{solomon2005explicit} introduced a 1.5-dimensional explicit microphysics thunderstorm model that incorporates a lightning parameterization, addressing key aspects of thunderstorm electrification. Furthermore, Mansell et al.~\cite{mansell2002simulated} simulated three-dimensional branched lightning in a numerical thunderstorm model, providing insights on the complex activities of lightning formation. Barthe and Pinty~\cite{barthe2007simulation} further advanced the field by simulating a supercell storm using a three-dimensional mesoscale model with an explicit lightning flash scheme, capturing the lightning activities specific to supercell thunderstorms.Mansell et al.~\cite{mansell2010simulated} also examined the electrification of small thunderstorms using a two-moment bulk microphysics scheme, extending the understanding of lightning activity in small multicell thunderstorms.

% Recent studies have focused on exploring the relationship between thundercloud structures and electrification processes. Formenton et al.~\cite{formenton2013using} used a cloud electrification model to study the relationship between lightning activity and cloud microphysical structure, providing valuable insights into the interaction between cloud formation and lightning generation. More recently, Wu et al.~\cite{wu2023thundercloud} employed coherent Doppler wind lidar to detect and analyze thundercloud structures, offering a novel method for observing thunderstorm activities.

% These contributions highlight the ongoing efforts to simulate and understand thunderstorm activities, including cloud microphysics, lightning activity, and the complex interactions between these phenomena.

% Additionally, our simulation includes common phenomena generated by thundercloud electrification, such as cloud-to-ground (CG) and intra-cloud (IC) lightning\footnote{\url{https://www.nssl.noaa.gov/education/svrwx101/lightning/types/}}.

\section{OVERVIEW}

The primary motivation for our approach is to  visually capture the realistic development of MCS. We propose a physically based simulation framework that is intuitive for artists, enabling the creation of visually realistic thunderstorm animations while managing physical complexity and ensuring practical performance.

Simulating MCS is challenging due to the complex microphysical and electrostatic dynamics and the interactions among hydrometeors (ice, snow, and rain) that induce charge imbalances. Traditional meteorological models capture these phenomena with high accuracy but rely on numerous variables and detailed phase transitions\cite{barthe2007simulation}, making them computationally expensive and difficult for non-experts to manage.For example,the conventional warm-cloud model typically involves three or four phase transitions and works well for ordinary cumulus clouds, whereas modeling complex thunderstorms generally requires a combined warm-cold cloud scheme with more than ten phase transitions (e.g., melting and riming \cite{rutledge1984mesoscale}). Although such detailed modeling improves numerical weather prediction, its prohibitively long computation times necessitate the use of supercomputers.

In contrast,our Grabowski-style framework extends the basic warm-cloud model with a temperature interpolation scheme to incorporate cold-cloud processes, capturing the essential material transformations during thunderstorm development with fewer phase transitions. This simplification sacrifices some microscopic detail yet preserves a physically accurate macroscopic evolution, reduces the number of user-controlled variables, and ensures practical performance.

\begin{figure}[h]
    \centering
    \includegraphics[width=\linewidth]{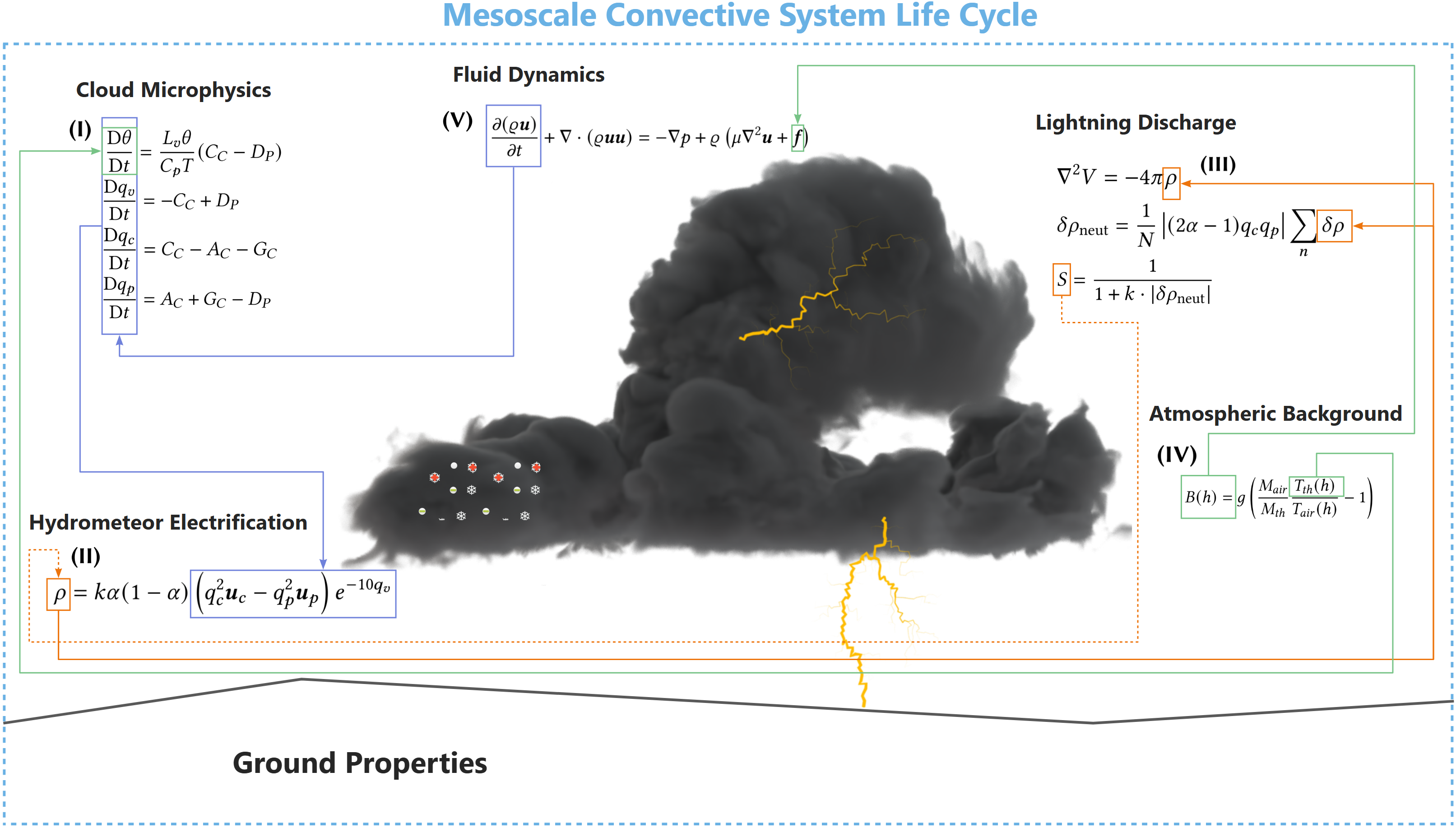}
    \caption{Schematic illustration of our MCS life cycle, encompassing the interplay and feedback among key modules. Detailed explanations are provided in Section \hyperref[sec:Method]{4}.\textbf{(I)} Cloud Microphysics: Describes the phase transitions of hydrometeors during MCS development.\textbf{(II)} Hydrometeor Electrification: Explains the accumulation of charge density resulting from the collision and coalescence of hydrometeors.\textbf{(III)} Lightning Discharge: Highlights the process of lightning channel formation and propagation, triggered when static charge exceeds a predefined electric field threshold, leading to the neutralization of hydrometeors along the lightning channel.\textbf{(IV)} Atmospheric Background: Introduces vertical buoyancy forces to the system driven by temperature variations caused by hydrometeor phase transitions.\textbf{(V)} Fluid Dynamics: Drives the system's temporal evolution and introduces forces influencing the dynamics of the MCS life cycle.}
  \Description{A diagram illustrating the stages of hydrometeor phase transitions and electrification during thunderstorm development.}
  \label{fig:Thunderstorm_scheme_0}
\end{figure}

Building on these advantages, we develop a comprehensive framework to describe the essential microphysical processes of hydrometeor phase transitions and electrification.Our model employs six key parameters to characterize the atmospheric state: the vapor field $q_v$, cloud field $q_c$, precipitation field $q_p$, temperature field $\theta$, charge density field $\rho$, and velocity field $\mathbf{u}$.Our simulation framework, as illustrated in Figure~\ref{fig:Thunderstorm_scheme_0}, consists of the following components:\textbf{(I)} Cloud microphysics, describing hydrometeor phase transitions such as condensation, evaporation, and precipitation; \textbf{(II)} Hydrometeor electrification, modeling charge accumulation during hydrometeor interactions; \textbf{(III)} Lightning discharge, computing charge neutralization as lightning channels propagate; \textbf{(IV)} Atmospheric background, accounting for buoyancy forces arising from temperature gradients;  \textbf{(V)} Fluid dynamics, driving field evolution with an Eulerian solver for MCS development.

% \begin{figure}[h]
%   \centering
%   \includegraphics[width=0.7\linewidth]{illstrate/Fig2.png}
%   \caption{Illustration of our MCS microphysics scheme, which integrates the Grabowski-style cloud microphysics with the hydrometeor electrification process.}
%   \Description{A schematic representation of the thunderstorm microphysics scheme, coupling the extended warm cloud microphysics with the electrification process of hydrometeors.}
%   \label{fig:Thunderstorm_scheme}
% \end{figure}

\section{METHODOLOGY}
\label{sec:Method}

Our microphysics model, depicted in Figure~\ref{fig:Thunderstorm_scheme_0}, illustrates the interactions among hydrometeor phase transitions and electrification mechanisms. Key processes include the condensation of water vapor into droplets and ice crystals, which subsequently form precipitation through autoconversion and accretion. Evaporation recycles hydrometeors back into the vapor phase, as detailed in Section \hyperref[sec:cloud_microphysics]{4.1}. Collisions and coalescence facilitate charge separation, leading to lightning when the electric field strength exceeds a critical threshold, as described in  Section \hyperref[sec:hydrometeor_electrification]{4.2}. The buoyancy forces generated by atmospheric temperature variations are discussed in Section \hyperref[sec:atmospheric_background]{4.3}, while the system's dynamics updates are outlined in Section \hyperref[sec:fluid_dynamics]{4.4}. This feedback loop provides a comprehensive framework for understanding thunderstorm development within an MCS.

\subsection{Cloud Microphysics}
\label{sec:cloud_microphysics}

The fundamental warm-cloud microphysics equations govern the evolution of potential temperature, water vapor, cloud condensate, and precipitation. These equations are formulated using the material derivative \cite{kundu2012fluid}:\begin{equation}
\frac{\mathrm{D} \phi}{\mathrm{D} t} = \frac{\partial \phi}{\partial t} + \boldsymbol{u} \cdot \nabla \phi ,
\end{equation}which provides the foundation for expressing the microphysics equations:\begin{align}
& \frac{\mathrm{D} \theta}{\mathrm{D} t} = \frac{L_v \theta}{C_p T} (C_C - D_P) ,\\
& \frac{\mathrm{D} q_v}{\mathrm{D} t} = -C_C + D_P , \\
& \frac{\mathrm{D} q_c}{\mathrm{D} t} = C_C - A_C - G_C, \\
& \frac{\mathrm{D} q_p}{\mathrm{D} t} = A_C + G_C - D_P, 
\end{align}where \(\theta\) represents the potential temperature, \(L_V\) is the latent heat of condensation or evaporation, \(T\) is the absolute temperature, and \(C_p\) is the specific heat capacity at constant pressure. Additionally, \(q_v\), \(q_c\), and \(q_p\) denote the mixing ratios of water vapor, cloud condensate, and precipitation, respectively. The terms \(C_C\), \(D_P\), \(A_C\), and \(G_C\) correspond to the condensation rate, diffusional growth rate of precipitation, autoconversion rate, and accretion rate.

We adopt the equilibrium approach proposed by Grabowski \cite{grabowski1998toward}, incorporating a temperature-dependent factor \(\alpha\) to differentiate warm and cold cloud microphysics. The parameter \(\alpha\) varies linearly with temperature: warm clouds dominate above \(0^\circ \mathrm{C}\), cold clouds dominate below \(-20^\circ \mathrm{C}\). The unified relationship for vapor saturation, cloud, and precipitation components is expressed as:
\begin{equation}
q_{mc} = \alpha q_{wc} + (1 - \alpha) q_{cc}, 
\end{equation}
where \(q_{mc}\) represents the total quantity of a variable \(q_{vs}, q_c, q_p\), combining its warm-phase component \(q_{wc}\) and cold-phase component \(q_{cc}\). Specifically, for vapor saturation, \(q_{wc} = q_{vw}\) (saturation over water) and \(q_{cc} = q_{vi}\) (saturation over ice); for cloud condensate, \(q_{wc} = q_w\) (cloud water) and \(q_{cc} = q_i\) (cloud ice); and for precipitation, \(q_{wc} = q_r\) (rain) and \(q_{cc} = q_s\) (snow).

Building on this framework, we parameterize diverse microphysical processes, unified by the temperature-dependent factor \(\alpha\).   
The saturation vapor mixing ratio \(q_{vs}\) combines contributions from vapor saturation over water and ice\cite{yau1996short}:
\begin{equation}
q_{vs} = \frac{380.16}{p} \left[ \alpha \exp\left(\frac{17.67 \cdot T}{T + 243.50}\right) + (1 - \alpha) \exp\left(\frac{24.46 \cdot T}{T + 272.62}\right) \right], 
\end{equation}
where \(p\) represents the pressure, and \(T\) denotes the temperature. The saturation vapor is modeled as an exponential distribution for both liquid water and ice.
The condensation rate \(C_C\) governs cloud formation from water vapor, integrating water and ice contributions \cite{dudhia1989numerical}, where \(\beta_w, \beta_i\) are phase-specific coefficients:
\begin{equation}
C_C = \beta_w \cdot \left( \alpha \cdot q_{vs} - q_v \right)^+ + \beta_i \cdot \left( (1 - \alpha) \cdot q_{vs} - q_v \right)^+. 
\end{equation}
The autoconversion rate \(A_C\) describes cloud condensate aggregation into precipitation \cite{kessler1995continuity,lin1983bulk}:
\begin{equation}
A_C = \beta_r \cdot (\alpha \cdot q_c - 10^{-3})^+ + \beta_s \cdot e^{0.025  T} \cdot ((1 - \alpha) \cdot q_c - 10^{-3})^+ .
\end{equation}
The accretion rate \(G_C\) models the collection of cloud condensate by precipitation particles \cite{morrison2015parameterization}:
\begin{equation}
G_C = q_c \cdot q_p \cdot \left( \beta_r \cdot \alpha^2 + \beta_s \cdot (1 - \alpha)^2 \right).
\end{equation}
The diffusional growth rate \(D_P\) describes the conversion between precipitation and vapor through deposition and evaporation processes \cite{klemp1978simulation}, and is modeled as follows:
\begin{equation}
D_P = \alpha (1 - \alpha) \frac{q_{v s} - q_{v}}{\hat{\rho} q_{v s}} \frac{\left( \hat{\rho} q_{p} \right)^{0.525}}{5.4 \cdot 10^{5} + 2.55 \cdot 10^{8} \left( p q_{v s} \right)^{-1}},
\end{equation}
where \(\hat{\rho}\) is the density of humid air,  calculated in the same way as in \cite{hadrich2020stormscapes}, and \(p\) represents the atmospheric pressure, which will be described in Section \hyperref[sec:atmospheric_background]{4.3}.
The terminal velocity \(\boldsymbol{u}_p\)  of precipitation  during deposition is modeled as a weighted combination of rain \(\boldsymbol{u}_r\) and snow \(\boldsymbol{u}_s\)  velocities.Since precipitation particles quickly reach a uniform terminal velocity, we use typical values of \(\boldsymbol{u}_r = -10 \, \mathrm{m/s}\) for rain and \(\boldsymbol{u}_s = -2 \, \mathrm{m/s}\) for snow:
\begin{align}
& \boldsymbol{u}_p = \alpha \boldsymbol{u}_r + (1-\alpha) \boldsymbol{u}_s,
\end{align}

%TODO不要用XX mixture很low
%%%%%%%%%%%%%%%%%%%%%%%%%% part three
\subsection{Hydrometeor Electrification}
\label{sec:hydrometeor_electrification}

% Based on the Reynolds-Brook theory of thunderstorm electrification \cite{latham1965role}, the downward-pointing fair-weather electric field (\(E\)) induces equal amounts of negative and positive charge on the tops and bottoms of precipitation particles of different sizes. Due to their varying directions and velocities, these particles collide, leading to partial neutralization of their charges during contact. After the collision, each particle retains a net charge. 

% Collisions between two water droplets often result in coalescence, particularly during glancing interactions, where charge separation is minimal because charges induced near the equators of the droplets are not neutralized. Similarly, collisions between ice particles result in limited charge transfer due to their short contact time and the lower conductivity of ice compared to water. Interactions involving water-ice or ice-riming particles have been considered, but the inductive process, starting from a fair-weather electric field, is generally insufficient to generate the intense electric fields characteristic of thunderstorms.

% The electrification process can be further modeled using the following equations. The charge density (\(\rho\)) generated through microphysical interactions is expressed as:

Building on the Reynolds-Brook theory of thunderstorm electrification \cite{latham1965role}, the fair weather electric field induces opposing charges on precipitation particles. As these particles collide and coalesce, their velocities and directions influence partial charge neutralization, leaving a residual net charge.  

% Water droplet collisions generally coalesce without significant charge separation, while ice particle interactions are limited by brief contact times and low conductivity. 
% Though water-ice and ice-riming collisions contribute, the inductive process alone is insufficient for thunderstorm-scale electric fields.
This theory underpins the following mathematical model for electrification, with charge density \(\rho\) defined as:
\begin{equation}
\rho = k \alpha (1 - \alpha) \left( q_c^2 \boldsymbol{u}_c - q_p^2 \boldsymbol{u}_p \right) e^{-10 q_v} ,
\end{equation}
where \(k\) is a user-defined constant, \(\boldsymbol{u}_c\) and \(\boldsymbol{u}_p\) are their respective velocities of cloud and precipitation.
The threshold electric field \(E_{tr}\) required for lightning initiation depends on the altitude \(\rho_{A}(h)\) as defined by \cite{marshall1995electric}:
\begin{equation}
E_{tr} = \pm 167 \, \rho_{A}(h) \quad \text{where} \quad \rho_{A}(h) = 1.208 \exp \left( \frac{-h}{8.4} \right). 
\end{equation}
To analyze the electrodynamics in the atmosphere, we solve the Poisson equation, a fundamental equation derived from Gauss's law in electrostatics. This equation describes the simplified spatial variation of the potential \(V\) due to charge density \(\rho\)\cite{kim2007fast}:  
\begin{equation}
\nabla^2 V = -4 \pi \rho .
\end{equation}  
Charge neutralization processes, occurring after a lightning discharge\cite{barthe2007simulation}, are governed by the following relationship:
\begin{equation}
\delta \rho = 
\begin{cases}
\pm\left(|\rho| - \rho_{\text{excess}}\right), & \text{if } |\rho| > \rho_{\text{excess}}, \\
0, & \text{if } |\rho| \leq \rho_{\text{excess}},
\end{cases}
\end{equation}
where \(\delta \rho\) represents the net charge change, and \(\rho_{\text{excess}}\) denotes the threshold for excess charge density. The total neutralized charge density, accounting for the collective contributions of lightning growth points, is expressed as:
\begin{equation}
\delta \rho_{\text{neut}} = \frac{1}{N} \left| (2\alpha - 1) q_c q_p \right| \sum_{n} \delta \rho .
\end{equation}
Finally, a suppression factor \(S\) is introduced to modulate lightning activity as thunderstorms dissipate:
\begin{equation}
S = \frac{1}{1 + k \cdot |\delta \rho_{\text{neut}}|}. 
\end{equation}

\subsection{Atmospheric Background}
\label{sec:atmospheric_background}

Our atmospheric background is based on the theory proposed by Hädrich et al. \cite{hadrich2020stormscapes}. Specifically, we assume that the atmosphere is initially electroneutral, with the charge density, denoted as $\rho$, being zero.

The isentropic exponent \(\gamma_{\text{th}}\) for the air-water mixture \cite{anderson1990modern} is calculated as a weighted average of the vapor-specific exponent \(\gamma_{\text{vapor}}\) and the air-specific exponent \(\gamma_{\text{air}}\), as shown below:
\begin{equation}
\gamma_{\text{th}} = Y_{\text{vapor}} \gamma_{\text{vapor}} + \left(1 - Y_{\text{vapor}}\right) \gamma_{\text{air}}, 
\end{equation}
where \(Y_{\text{vapor}}\) represents the mass fraction of water vapor in the air. The values of \(\gamma_{\text{air}}\) and \(\gamma_{\text{vapor}}\) are taken as 1.4 and 1.33, respectively, based on standard thermodynamic properties of dry air and water vapor.

The atmospheric temperature profile \cite{atmosphere1975international}, \(T(h)\), is modeled by a piecewise function that accounts for the lapse rate, including the effect of the inversion layer at a height \(h_1\). The temperature at a given altitude \(h\) is expressed as:
\begin{equation}
T(h) = 
\begin{cases}
T_0 + \Gamma_0 h, & 0 \leq h \leq h_1, \\
T_0 + \Gamma_0 h_1 + \Gamma_1 (h - h_1), & h_1 \leq h,
\end{cases}
\end{equation}
where \(T_0\) is the base temperature at sea level, \(\Gamma_0\) and \(\Gamma_1\) represent the lapse rates in the lower and upper layers.

The atmospheric pressure profile, \(p(h)\), is derived from the hydrostatic equation \cite{houze2014cloud}, considering the effect of gravity and the ideal gas law. It is given by:
\begin{equation}
p(h) = p_0 \left( 1 - \frac{\Gamma h}{T_0} \right)^{\frac{g}{R T_0}},
\end{equation}
where \(p_0\) is the pressure at sea level, \(g\) is the acceleration due to gravity, and \(R\) is the specific gas constant.

To model the thermodynamic properties of humid air, we calculate the average molar mass of the air-water mixture \(\mathit{M}_{\mathit{th}}\). This is given by:
\begin{equation}
\mathit{M}_{\mathit{th}} = X_{\text{vapor}} \mathit{M}_{\mathit{water}} + \left( 1 - X_{\text{vapor}} \right) \mathit{M}_{\mathit{air}}, 
\end{equation}
where \(X_{\text{vapor}}\) is the mole fraction of water vapor, and \(\mathit{M}_{\mathit{water}}\) and \(\mathit{M}_{\mathit{air}}\) are the molar masses of water (18.02 g/mol) and dry air (28.96 g/mol), respectively.

The mass fraction of water vapor in the humid air, \(Y_{\text{vapor}}\), is related to the mole fraction \(X_{\text{vapor}}\) by:
\begin{equation}
Y_{\text{vapor}} = X_{\text{vapor}} \frac{\mathit{M}_{\mathit{water}}}{\mathit{M}_{\mathit{th}}} .
\end{equation}
The temperature of the air in the atmosphere can also be related to pressure changes through the isentropic relation, which governs the temperature \(T_{\mathrm{th}}(h)\) at height \(h\) in terms of the pressure profile. This is given by:
\begin{equation}
T_{\mathrm{th}}(h) = T_0 \left( \frac{p(h)}{p_0} \right)^{\frac{\gamma_{\mathrm{th}} - 1}{\gamma_{\mathrm{th}}}}.
\end{equation}
Finally, the buoyancy force, which drives the upward movement of thundercloud, is calculated based on Archimedes' principle and Newton's second law. The buoyancy force \(B(h)\) at height \(h\) is given by:
\begin{equation}
B(h) = g \left( \frac{\mathit{M}_{\mathit{air}}}{\mathit{M}_{\mathit{th}}} \frac{T_{\mathit{th}}(h)}{T_{\mathit{air}}(h)} - 1 \right).
\end{equation}

\subsection{Fluid Dynamics}
\label{sec:fluid_dynamics}

The motion of atmospheric fluids is governed by the incompressible Navier-Stokes equations, which represent the principles of momentum and mass conservation in a fluid medium \cite{wendt2008computational}. These equations account for the effects of inertial forces, pressure gradients, viscosity, and external forces:
\begin{equation}
\frac{\partial (\varrho \boldsymbol{u})}{\partial t} + \nabla \cdot (\varrho \boldsymbol{u} \boldsymbol{u}) = - \nabla p + \varrho \left( \mu \nabla^2 \boldsymbol{u} + \boldsymbol{f} \right), \quad \nabla \cdot \boldsymbol{u} = 0,
\end{equation}
where \(\varrho\) represents the fluid density, \(\boldsymbol{u}\) is the velocity, \(p\) is the pressure, \(\mu\) denotes the dynamic viscosity, and \(\mathbf{f}\) encompasses buoyancy and other external forces. The equation \(\nabla \cdot \boldsymbol{u} = 0\) represents the continuity equation, ensuring the conservation of mass in incompressible flows.The pressure term, \(p\), is determined via the Helmholtz–Hodge decomposition by solving the Poisson equation:

\begin{equation}
\nabla^2 p = \nabla \cdot \boldsymbol{u}.
\end{equation}

\section{ALGORITHMICS}

The theoretical framework discussed in the previous section is translated into a numerical procedure, as outlined in Algorithm~\ref{alg:Thunderstorm_scheme}. Figure~\ref{fig:Thunderstorm_scheme_0} visualizes the key interrelationships governing MCS development. We first present our algorithm for simulating the life cycle of a MCS, followed by an explanation of the implementation details of our procedure.
%这里叫: NumericalIntegration

%\subsection{Thunderstorms Motion}
\subsection{Mesoscale Convective System Cycle}

The evolution of the MCS system follows Algorithm~\ref{alg:Thunderstorm_scheme}, iteratively updating state variables, including potential temperature \(\theta\), pressure \(p\), charge density \(\rho\), velocity field \(\boldsymbol{u}\), and hydrometeor quantities \(q_v\), \(q_c\), and \(q_p\). The process begins with updating atmospheric background conditions \(\theta, p, \rho\) following Eqs.~(20--21). The velocity field \(\boldsymbol{u}\) is then advected and diffused, followed by the computation of thermal buoyancy \(b\) based on Eq.~(25). Buoyancy, wind, and vorticity confinement forces are subsequently integrated into velocity field \(\boldsymbol{u}\) to reflect key atmospheric dynamics.

%TODO 公式编号已经更新需要改变：
%还是不分开了,直接一起说更方便
\begin{algorithm}
\caption{Thunderscapes Algorithm}
\begin{algorithmic}[1]
\State \textbf{Input:} Current \textbf{MCS} state $(\theta, p, \rho, \boldsymbol{u}, q_v, q_c, q_p)$. 
\State \textbf{Output:} Updated \textbf{MCS} state.
\Procedure{}{}
    \State $\theta, p, \rho \gets$ Update atmospheric background conditions \hfill Eqs.(20--21)
    \State $\boldsymbol{u} \gets$ Advect and diffuse velocity field
    \State $\boldsymbol{b} \gets$ Compute thermal buoyancy \hfill Eq.(25)
    \State $\boldsymbol{u} \gets \boldsymbol{u} + \boldsymbol{b} + \boldsymbol{f}_w + \boldsymbol{f}_v$ \hfill \Comment{Apply buoyancy, wind, vorticity confinement}
    \State $q_v, q_c, q_p \gets$ Advect hydrometeor quantities \hfill 
    \State $\boldsymbol{u} \gets$ Pressure projection \hfill Eq.(27)
    \State $q_v, q_c, q_p, \theta \gets$ Update cloud microphysics and temperature \hfill Eqs.(2--5)
    \State $\rho \gets$ Hydrometeor electrification \hfill Eqs.(13)   
    \If{$S \cdot \rho > E_{tr}$}   \hfill Eq.(14)
        \State $V, \rho_{\text{neut}} \gets$ lightning discharge \hfill Eqs.(15--17)
        \State $S \gets$ Apply lightning neutralization \hfill Eq.(18)
    \EndIf
\EndProcedure
\end{algorithmic}
\label{alg:Thunderstorm_scheme}
\end{algorithm}

Hydrometeor quantities \(q_v, q_c, q_p\) are transported through advection in accordance with the updated velocity field  \(\boldsymbol{u}\). To ensure mass conservation, the velocity field  \(\boldsymbol{u}\) is corrected via pressure projection following Eq.~(27). Next, cloud microphysics are updated by solving parameterized equations based on Eqs.~(2--5) to account for condensation, evaporation, and precipitation.

Hydrometeor electrification is modeled using parameterized equations following Eq.~(13). When the charge density, modulated by the suppress factor \(S \cdot \rho\), exceeds the electric field threshold \(E_{tr}\) based on Eq.~(14), lightning discharge is triggered. During the discharge, lightning channel growth is stochastically modeled using principles from the Dielectric Breakdown Model (DBM) \cite{kim2007fast}. Finally, neutralized charge density \(\rho_{\text{neut}}\) is updated following Eqs.~(15--17), and the suppress factor is recalibrated to reduce the likelihood of subsequent lightning events as thunderstorms dissipate based on Eq.~(18).

%\subsection{Lightning Generation}
%删减一些：大论文可以唬人，但小论文言多必失
%可以藏的更深,这节整个删掉,上一节,直接说利用Kim的DBM模型,就类似把原来的electrodynamics和electrification合并
% \subsection{Lightning Channel Growth}
\subsection{Implementation}
Our tool is implemented as a Houdini Digital Asset (HDA) using Houdini 20.0.625's microsolver framework, with OpenCL for GPU acceleration. The hardware setup comprises an NVIDIA® GeForce® RTX A6000 GPU, a 13th Gen Intel® Core™ i9-13900 processor, and 128GB of RAM.

The framework is designed to align with the traditional workflow of digital artists, utilizing custom volume fields as input sources, including temperature field, vapor field, and height field.These fields, representing ground properties, serve as the foundation for driving the MCS simulation.
Our framework operates on a uniform 3D grid, utilizing a semi-Lagrangian scheme to advect system quantities. To simulate horizontal thunderstorm movement, we apply a uniform horizontal wind field. Combined with vertical buoyancy forces, this approach produces a more realistic depiction of thunderstorm surges. 

To mitigate numerical dissipation, which can cause small-scale vortices to vanish prematurely, we integrate vorticity confinement techniques \cite{miyazaki2002simulation}. This enhancement ensures that fine-scale details, such as swirling motions within the storm, are preserved throughout the simulation.

Lightning channel growth during hydrometeor discharge events is modeled using the Dielectric Breakdown Model (DBM) proposed by Kim et al. \cite{kim2007fast}. The resulting lightning channels are converted from grid points to geometric points. To achieve a more natural branching structure, a random dithering technique is applied to these points, which are subsequently connected to form the final geometry of the lightning branches.

Both the sparse Poisson problem for fluid pressure projection and the DBM processes are efficiently handled using the compact Poisson filter \cite{rabbani2022compact}, a GPU-friendly method specifically optimized for solving large sparse linear systems.

\section{VISUALIZATION}
%\Implementation: as a DCC Plug in, based on houdini 20.0.625 microsolver opencl for gpu acceleration NVIDIA®GeForce®RTX 3060 Ti + 13th Gen Intel(R) Core(TM) i5-13600KF ;32GB RAM
%we would show the simulation result made by this tool,include multipe types of thunderstorms and lightning and couple with some complex scenario based on real weather events,the render is use houdini karma renderer

% The implementation is open-sourced and available at \href{https://github.com/logic-three-body/Thunderscapes}{https://github.com/logic-three-body/Thunderscapes}.
%这句话慎重，可能会有版权纠纷，就说built-in:
% Houdini's Karma renderer for high-quality visual outputs
%如果后续中稿，需要考虑是否换成blender cycles渲染
%若中稿需要考虑：
%https://www.sidefx.com/buy/#houdini-education
The simulation results demonstrate various thunderstorm phenomena, integrated with scenarios inspired by real-world weather events. As shown in Table~\ref{tab:lightweight_params}, we summarize the spatial scale used for the scenes presented in this section. All renders are produced using Houdini's native volumetric rendering engine to ensure high-quality visual outputs.For dynamic details, please refer to our supplementary video.

\begin{table}[h]
    \caption{The table provides an overview of the spatial scale used in the scenes presented in this paper. Moreover, resolution ($R$) and computation time ($T_{\text{mcs}}$) measured in seconds per frame are listed. A constant time step size of $\Delta t = 1 \, \mathrm{min}$ is used. The background parameters defining the atmosphere are set to $\Gamma = -8.5 \, \mathrm{K} / \mathrm{km}$ and $h_{1} = 8 \, \mathrm{km}$.}

    \label{tab:lightweight_params}

\centering
  \begin{tabular}{cccccc}
    \toprule
    Fig. & Scene              & $D$ (km)         & $R$               & $T_{mcs}$ (s) \\
    \midrule
    ~\ref{fig:thundercloud_comparison} & Single Cell        & $18 \times 12 \times 18$ & $450 \times 301 \times 450$ & 0.48  \\
    ~\ref{fig:thundercloud_comparison} & Multicell          & $20 \times 15 \times 20$ & $500 \times 376 \times 500$ & 0.81  \\
    ~\ref{fig:thundercloud_comparison} & Squall Line        & $30 \times 15 \times 30$ & $500 \times 251 \times 500$ & 0.54  \\
    ~\ref{fig:thundercloud_comparison} & Super Cell         & $30 \times 15 \times 30 $& $500 \times 251 \times 500$ & 0.68  \\
    ~\ref{fig:weather_simulation} & Florida            & $11 \times 12 \times 11$ & $287 \times 309 \times 287$ & 0.24  \\
    ~\ref{fig:weather_simulation} & New Mexico         & $11 \times 13 \times 11$ & $383 \times 438 \times 383$ & 0.59  \\
    ~\ref{fig:weather_simulation} & Japan              & $11 \times 13 \times 11$ & $383 \times 444 \times 383$ & 0.64  \\
    ~\ref{fig:weather_simulation} & California         & $10 \times 7 \times 10$ & $340 \times 247 \times 340$ & 0.28  \\
    \bottomrule
  \end{tabular}
\end{table}

\subsection{Thunderstorm Variation}

%由于我们以效果优先，所以dbm的substep的事情可以放心说了
%Explain :we simulate four common type of thunderstorms in a MCS: 
%single cell: ...
%multicell : ...
%squall line : a collection of storms that form a line hundreds of miles long with shelf cloud feature
%supercell :very intensive,capable of producing tornadoes
% as shown ~\ref{fig:thundercloud_comparison}

We simulate four common types of thunderstorms that occur within a MCS according to the National Severe Storms Laboratory (NSSL)\footnote{\url{https://www.nssl.noaa.gov/education/svrwx101/thunderstorms/types/}}, capturing their distinctive characteristics and dynamic behaviors:

% \begin{itemize}
%     \item \textbf{Single Cell:} This is the simplest form of thunderstorm, typically short-lived and consisting of a single updraft and downdraft cycle. It often forms in isolated conditions and usually dissipates within an hour.
%     \item \textbf{Multicell:} Multicell thunderstorms are composed of multiple convective cells at various stages of their life cycle. These storms exhibit greater longevity and intensity compared to single cells, as new cells continuously form along the gust front of the system.
%     \item \textbf{Squall Line:} A squall line is a collection of thunderstorms that align into a long, narrow band, often stretching hundreds of miles. This type of storm is typically associated with intense winds, heavy rainfall, and the characteristic "shelf cloud" feature that marks the leading edge of the gust front.
%     \item \textbf{Supercell:} The supercell is the most intense and organized type of thunderstorm, characterized by a rotating updraft known as a mesocyclone. These storms are capable of producing severe weather phenomena, including large hail, damaging winds, and tornadoes.
% \end{itemize}

\begin{itemize} \item \textbf{Single Cell:} Single cell thunderstorms, often referred to as “popcorn” convection, are small, isolated storms characterized by their brief lifespan, typically lasting less than an hour. These storms are visually compact with a single updraft and downdraft cycle, forming as isolated towering cumulus clouds. 
\item \textbf{Multicell:} Multicell thunderstorms consist of clusters of individual convective cells, each at varying stages of development. They appear as a dynamic structure where new cells continuously form along the gust front, sustaining the system for several hours. 

\item \textbf{Squall Line:} A squall line is a linear arrangement of thunderstorms, visually identifiable by a long and narrow band of cumulonimbus clouds. These storms can extend hundreds of miles in length but are typically narrow, often around 10-20 miles wide.

\item \textbf{Supercell:} The supercell is the most visually striking and organized thunderstorm type, dominated by a large, rotating updraft known as a mesocyclone. These storms feature massive, tilted, and rotating cloud structures, often rising up to 50,000 feet. The mesocyclone can span up to 10 miles in diameter and persists for hours.   
\end{itemize}

%参考：https://www.andreaskj.com/thunder-clouds-in-houdini-20/#simulation

%说明一下图片版权来源national weather services：https://www.weather.gov/fsd/20230713_hail_sesdswmnnwia

\begin{figure}[htbp]
    \centering
    % 第一行
    \begin{minipage}{\textwidth}
        \centering
        \begin{subfigure}[t]{0.24\textwidth}
            \centering
            \begin{tikzpicture}
                \node[anchor=north west, inner sep=0] (image) at (-0.5,0.5) 
                {\includegraphics[width=3.5cm,height=2cm]{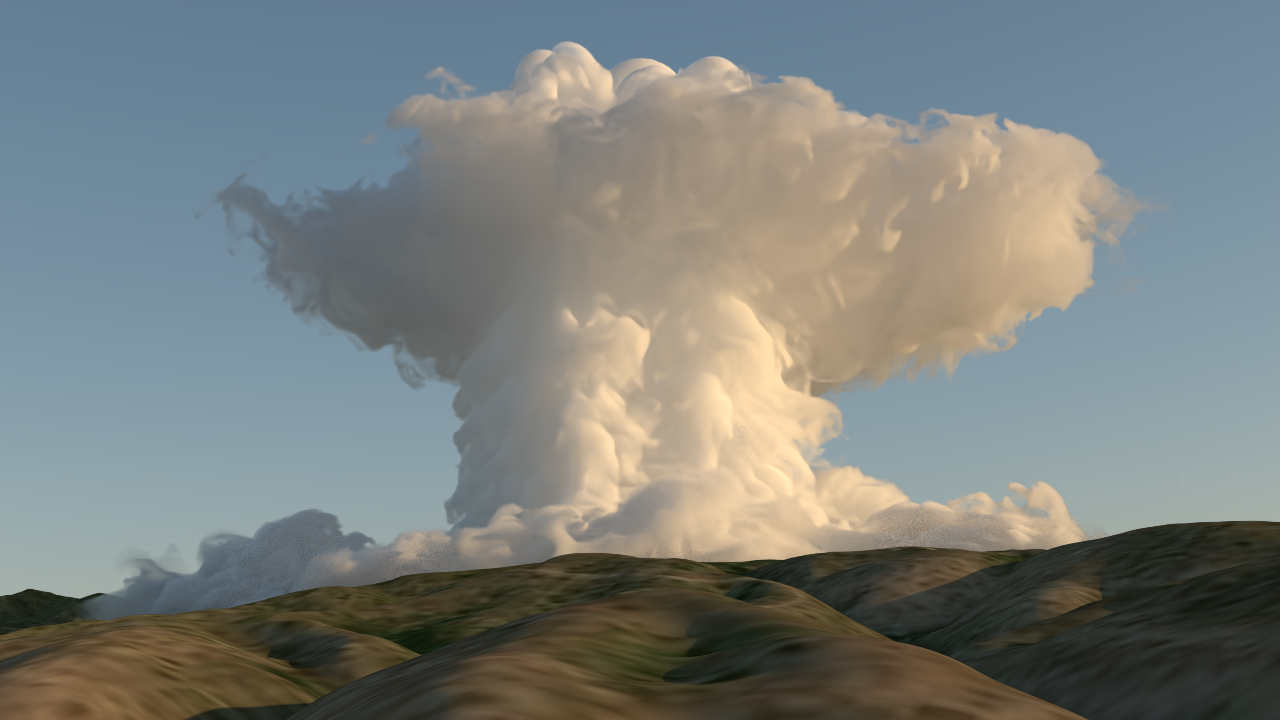}};
                \node[white] at (0, 0) {\textbf{\large (a)}};
            \end{tikzpicture}
        \end{subfigure}
        \hfill
        \begin{subfigure}[t]{0.24\textwidth}
            \centering
            \begin{tikzpicture}
                \node[anchor=north west, inner sep=0] (image) at (-0.5,0.5) 
                {\includegraphics[width=3.5cm,height=2cm]{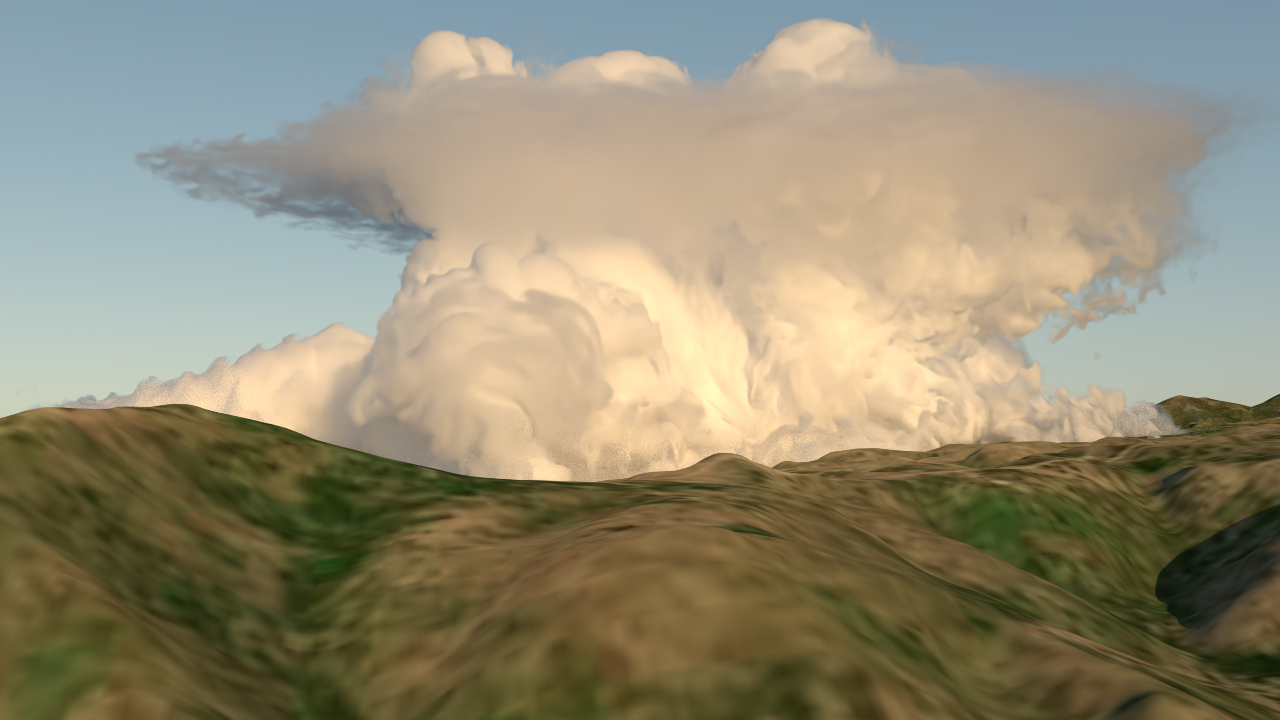}};
                \node[white] at (0, 0) {\textbf{\large (b)}};
            \end{tikzpicture}
        \end{subfigure}
        \hfill
        \begin{subfigure}[t]{0.24\textwidth}
            \centering
            \begin{tikzpicture}
                \node[anchor=north west, inner sep=0] (image) at (-0.5,0.5)  
                {\includegraphics[width=3.5cm,height=2cm]{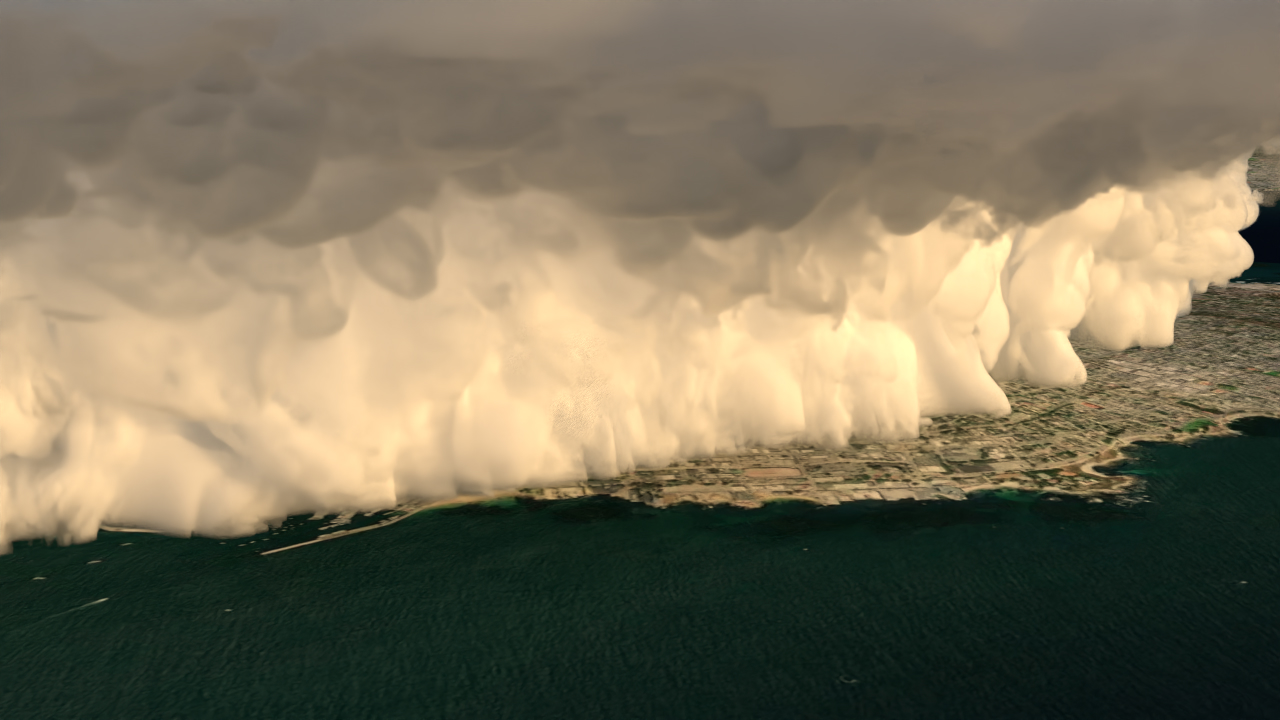}};
                \node[white] at (0, 0) {\textbf{\large (c)}};
            \end{tikzpicture}
        \end{subfigure}
        \hfill
        \begin{subfigure}[t]{0.24\textwidth}
            \centering
            \begin{tikzpicture}
                \node[anchor=north west, inner sep=0] (image) at (-0.5,0.5) 
                {\includegraphics[width=3.5cm,height=2cm]{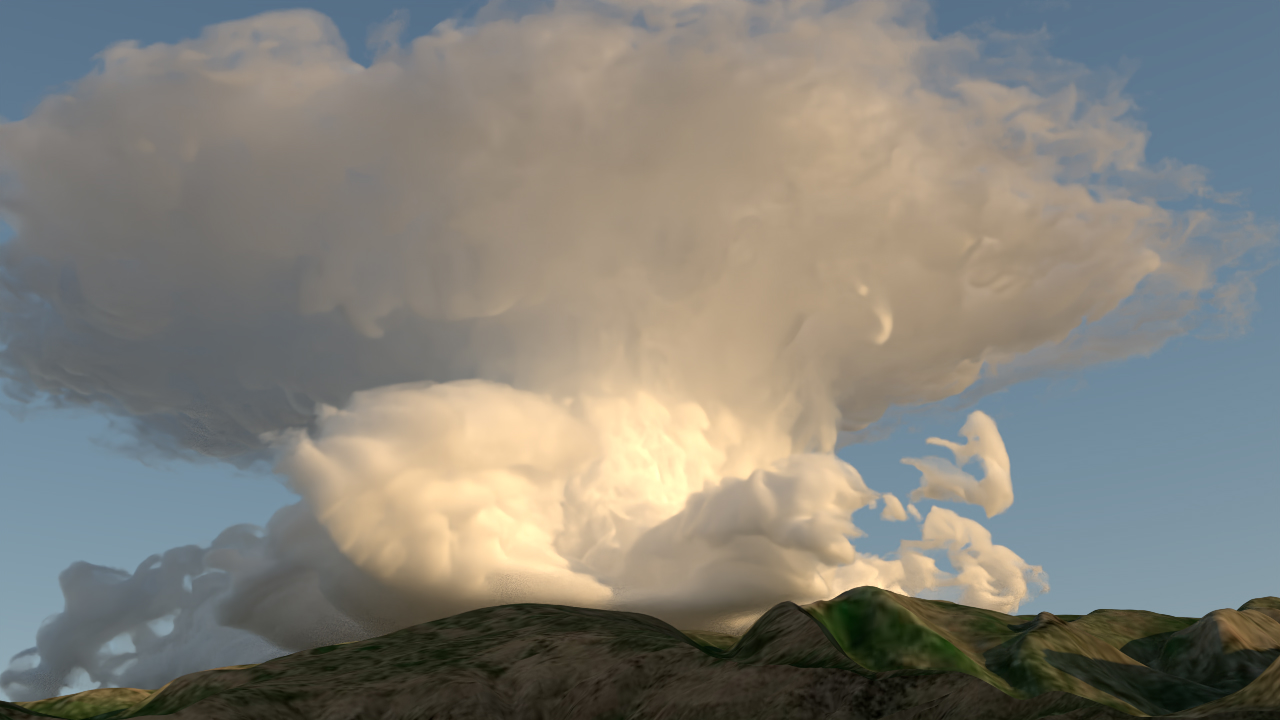}};
                \node[white] at (0, 0) {\textbf{\large (d)}};
            \end{tikzpicture}
        \end{subfigure}
    \end{minipage}
    \\[0.5em] % 行间距
    % 第二行
    \begin{minipage}{\textwidth}
        \centering
        \begin{subfigure}[t]{0.24\textwidth}
            \centering
            \begin{tikzpicture}
                \node[anchor=north west, inner sep=0] (image) at (-0.5,0.5) 
                {\includegraphics[width=3.5cm,height=2cm]{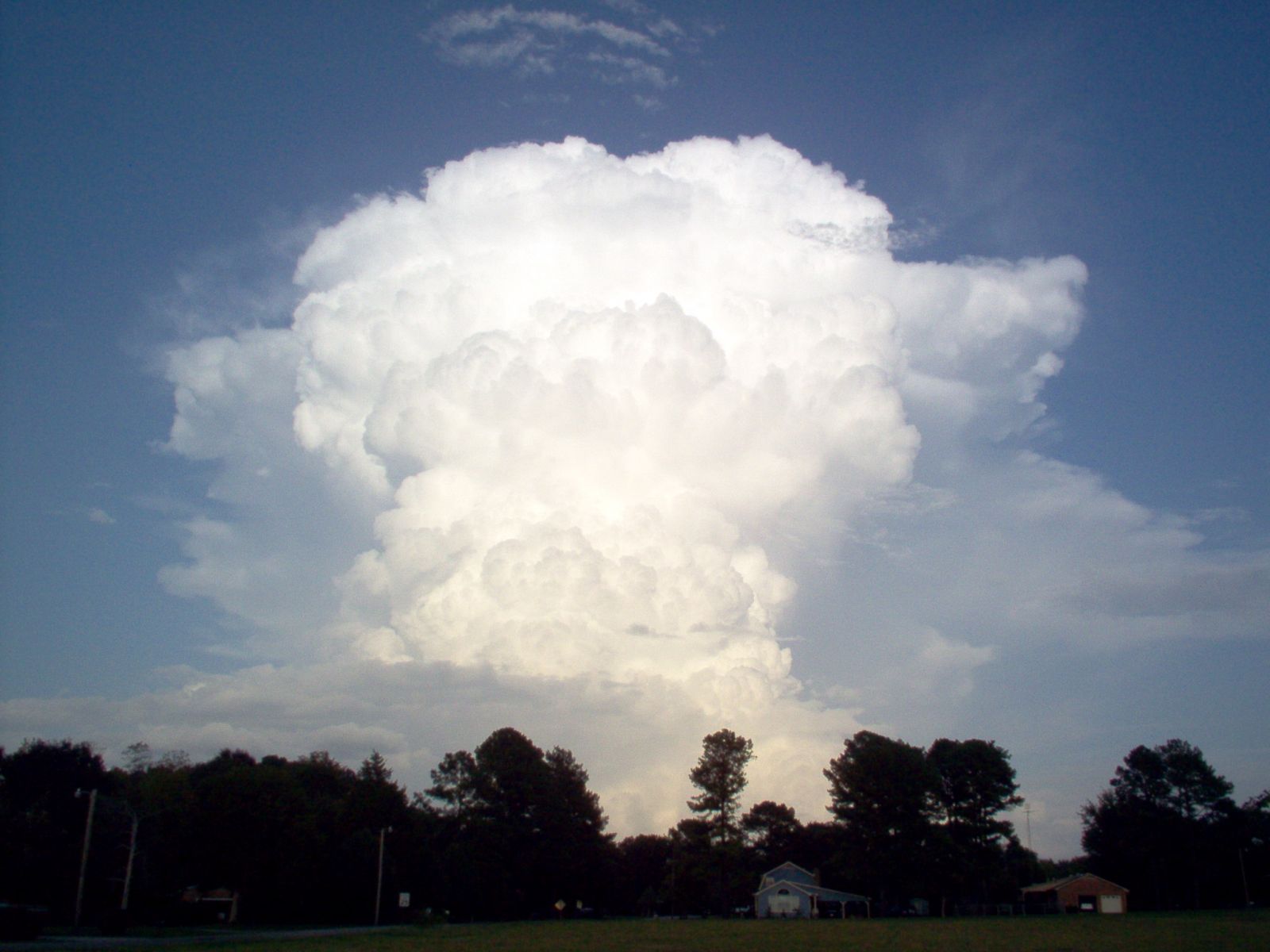}};
                \node[white] at (0, 0) {\textbf{\large (e)}};
            \end{tikzpicture}
        \end{subfigure}
        \hfill
        \begin{subfigure}[t]{0.24\textwidth}
            \centering
            \begin{tikzpicture}
                \node[anchor=north west, inner sep=0] (image) at (-0.5,0.5) 
                {\includegraphics[width=3.5cm,height=2cm]{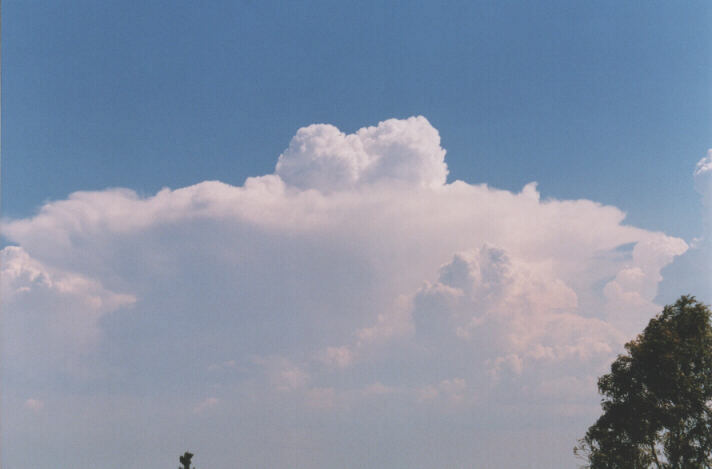}};
                \node[white] at (0, 0) {\textbf{\large (f)}};
            \end{tikzpicture}
        \end{subfigure}
        \hfill
        \begin{subfigure}[t]{0.24\textwidth}
            \centering
            \begin{tikzpicture}
                \node[anchor=north west, inner sep=0] (image) at (-0.5,0.5) 
                {\includegraphics[width=3.5cm,height=2cm]{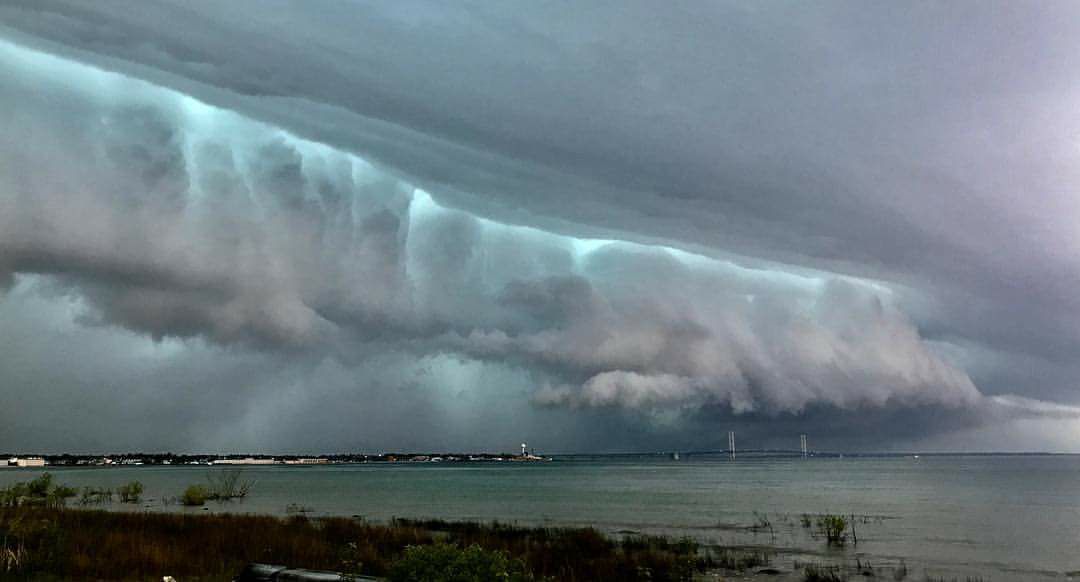}};
                \node[white] at (0, 0) {\textbf{\large (g)}};
            \end{tikzpicture}
        \end{subfigure}
        \hfill
        \begin{subfigure}[t]{0.24\textwidth}
            \centering
            \begin{tikzpicture}
                \node[anchor=north west, inner sep=0] (image) at (-0.5,0.5)  
                {\includegraphics[width=3.5cm,height=2cm]{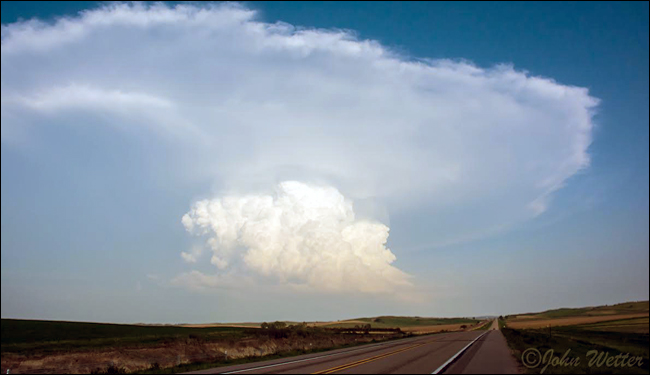}};
                \node[white] at (0, 0) {\textbf{\large (h)}};
            \end{tikzpicture}
        \end{subfigure}
    \end{minipage}
        % 第三行
    \begin{minipage}{\textwidth}
        \centering
        \begin{subfigure}[t]{0.24\textwidth}
            \centering
            \begin{tikzpicture}
                \node[anchor=north west, inner sep=0] (image) at (-0.5,0.5) 
                {\includegraphics[width=3.5cm,height=3.5cm]{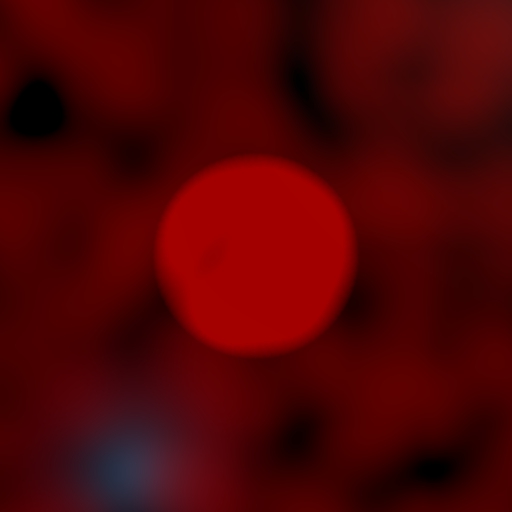}};
                \node[white] at (0, 0) {\textbf{\large (i)}};
            \end{tikzpicture}
        \end{subfigure}
        \hfill
        \begin{subfigure}[t]{0.24\textwidth}
            \centering
            \begin{tikzpicture}
                \node[anchor=north west, inner sep=0] (image) at (-0.5,0.5) 
                {\includegraphics[width=3.5cm,height=3.5cm]{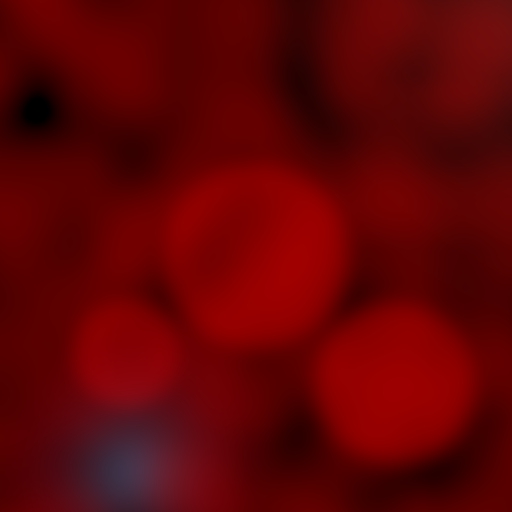}};
                \node[white] at (0, 0) {\textbf{\large (j)}};
            \end{tikzpicture}
        \end{subfigure}
        \hfill
        \begin{subfigure}[t]{0.24\textwidth}
            \centering
            \begin{tikzpicture}
                \node[anchor=north west, inner sep=0] (image) at (-0.5,0.5) 
                {\includegraphics[width=3.5cm,height=3.5cm]{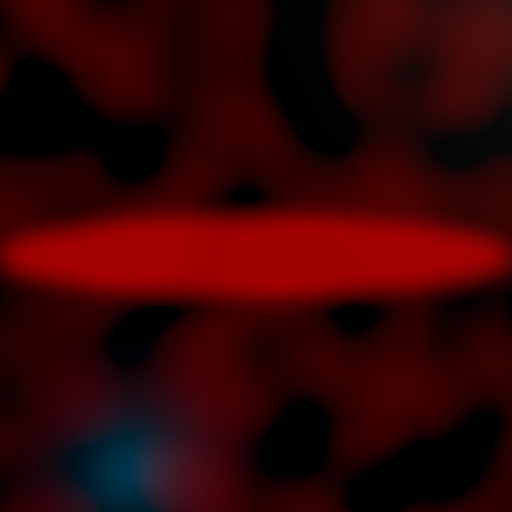}};
                \node[white] at (0, 0) {\textbf{\large (k)}};
            \end{tikzpicture}
        \end{subfigure}
        \hfill
        \begin{subfigure}[t]{0.24\textwidth}
            \centering
            \begin{tikzpicture}
                \node[anchor=north west, inner sep=0] (image) at (-0.5,0.5)  
                {\includegraphics[width=3.5cm,height=3.5cm]{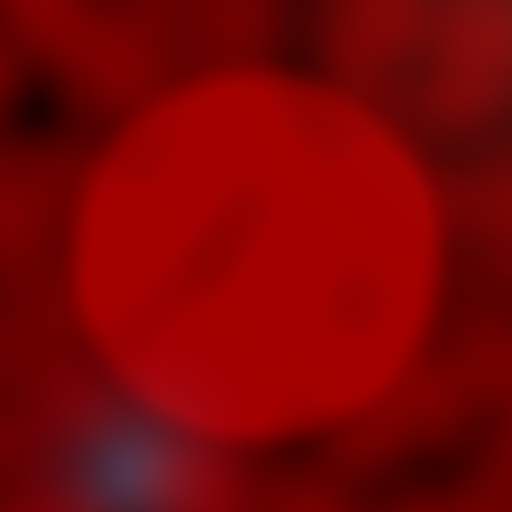}};
                \node[white] at (0, 0) {\textbf{\large (l)}};
            \end{tikzpicture}
        \end{subfigure}
    \end{minipage}
    \caption{%
        Comparison between simulation results (a-d) and realistic observations (e-h) for four storm types: single cell, multicell, squall line, and supercell. This layout highlights the structural similarities between our simulated storms and their real-world counterparts. Additionally, the last row (i-l) illustrates the ground properties (including the temperature field and vapor field initialized with a consistent pattern based on a flat heightfield) associated with each thunderstorm type.
    }
    \label{fig:thundercloud_comparison}
\end{figure}

These simulated variations allow us to explore the diverse behaviors and impacts of thunderstorms within a MCS. Figure~\ref{fig:thundercloud_comparison} presents a visual comparison along with the corresponding parameter sets, showcasing the structural differences and distinctive features of these thunderstorm types.

\subsection{Severe Weather Phenomena}
\label{sec:severe}

To explore the impact of MCSs in real-world scenarios, we reference severe weather events from geographically diverse locations using the Storm Events Database provided by the National Oceanic and Atmospheric Administration (NOAA)\footnote{\url{https://www.ncdc.noaa.gov/stormevents}}. These events highlight the variability and intensity of thunderstorms across different environmental conditions and seasons:

% \begin{itemize}
%     \item \textbf{Biscayne National Park, Florida:} A thunderstorm occurred over this region on July 15, 2023, characterized by heavy rainfall and frequent cloud-to-ground lightning strikes.
%     \item \textbf{Chiricahua Mountain, New Mexico:} On August 8, 2023, a thunderstorm swept through this mountainous area, producing intense hail and localized flash flooding.
%     \item \textbf{Fuji-Hakone-Izu National Park, Japan:} On September 12, 2023, a thunderstorm formed in this region, featuring intra-cloud lightning and strong wind gusts associated with a passing typhoon.
%     \item \textbf{Monterey Bay,California:} On October 5, 2023, an atmospheric river brought heavy rainfall and thunderstorm activity along the California coast, causing significant disruptions.
% \end{itemize}

\begin{itemize}
    \item \textbf{Biscayne National Park, Florida (25.3692°N, -80.243°W):} On March 13, 1993, a tropical storm occurred in this region during the spring, bringing heavy rainfall and strong winds. 
    \item \textbf{Chiricahua Mountains, New Mexico (35.7943°N, -106.443°W):} On July 14, 1989, a severe thunderstorm swept through this mountainous area during the summer, producing heavy rain, hail, and strong winds. 
    \item \textbf{Fuji-Hakone-Izu National Park, Japan (37.745°N, -119.54°W):} On January 10, 2011, a sudden winter thunderstorm formed in this region, resulting in brief but intense snowfall.
    \item \textbf{Monterey Bay, California (36.6197°N, -121.906°W):} On December 2, 2012, an atmospheric river brought heavy rainfall and thunderstorms along the California coast, accompanied by strong gusty winds.
\end{itemize}

\begin{figure}[htbp]
    \centering
    % 第一行
    \begin{minipage}{\textwidth}
        \centering
        \begin{subfigure}[t]{0.24\textwidth}
            \centering
            \begin{tikzpicture}
                \node[anchor=north west, inner sep=0] (image) at (-0.5,0.5)  
                {\includegraphics[width=3.5cm,height=2cm]{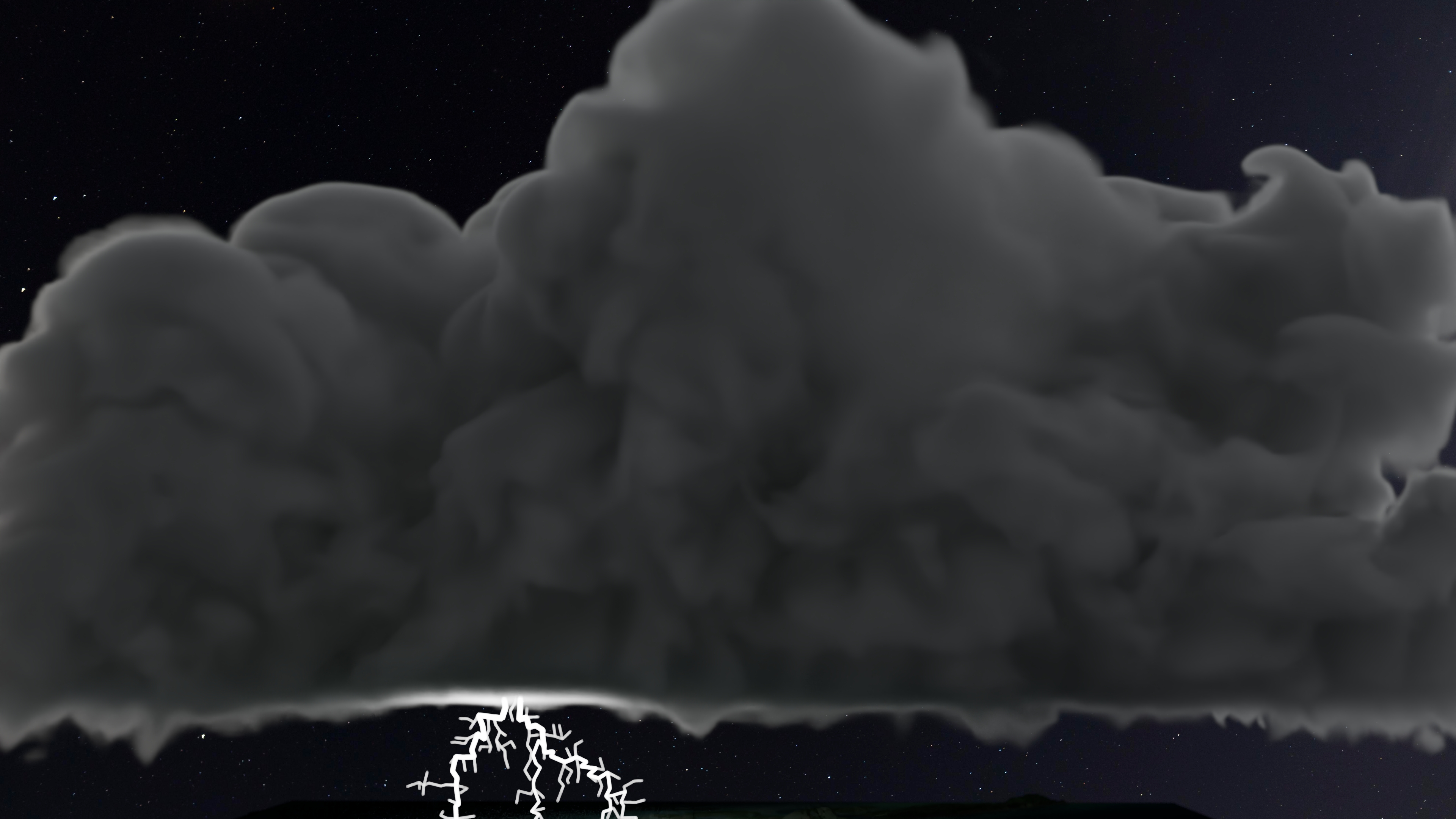}};
                \node[white] at (0, 0) {\textbf{\large (a)}};
            \end{tikzpicture}
        \end{subfigure}
        \hfill
        \begin{subfigure}[t]{0.24\textwidth}
            \centering
            \begin{tikzpicture}
                \node[anchor=north west, inner sep=0] (image) at (-0.5,0.5)  
                {\includegraphics[width=3.5cm,height=2cm]{illstrate/Teaser/NewMexico_Thunder_T.jpg}};
                \node[white] at (0, 0) {\textbf{\large (b)}};
            \end{tikzpicture}
        \end{subfigure}
        \hfill
        \begin{subfigure}[t]{0.24\textwidth}
            \centering
            \begin{tikzpicture}
                \node[anchor=north west, inner sep=0] (image) at (-0.5,0.5)  
                {\includegraphics[width=3.5cm,height=2cm]{illstrate/Teaser/japan_thunder_T.jpg}};
                \node[white] at (0, 0) {\textbf{\large (c)}};
            \end{tikzpicture}
        \end{subfigure}
        \hfill
        \begin{subfigure}[t]{0.24\textwidth}
            \centering
            \begin{tikzpicture}
                \node[anchor=north west, inner sep=0] (image) at (-0.5,0.5)  
                {\includegraphics[width=3.5cm,height=2cm]{illstrate/Teaser/california_thunder_T.jpg}};
                \node[white] at (0, 0) {\textbf{\large (d)}};
            \end{tikzpicture}
        \end{subfigure}
    \end{minipage}
    % 第二行
    \begin{minipage}{\textwidth}
        \centering
        \begin{subfigure}[t]{0.24\textwidth}
            \centering
            \begin{tikzpicture}
                \node[anchor=north west, inner sep=0] (image) at (-0.5,0.5) 
                {\includegraphics[width=3.5cm,height=3.5cm]{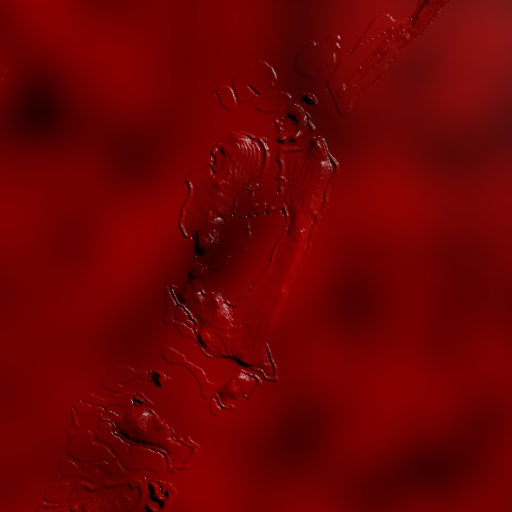}};
                \node[white] at (0, 0) {\textbf{\large (e)}};
            \end{tikzpicture}
        \end{subfigure}
        \hfill
        \begin{subfigure}[t]{0.24\textwidth}
            \centering
            \begin{tikzpicture}
                \node[anchor=north west, inner sep=0] (image) at (-0.5,0.5) 
                {\includegraphics[width=3.5cm,height=3.5cm]{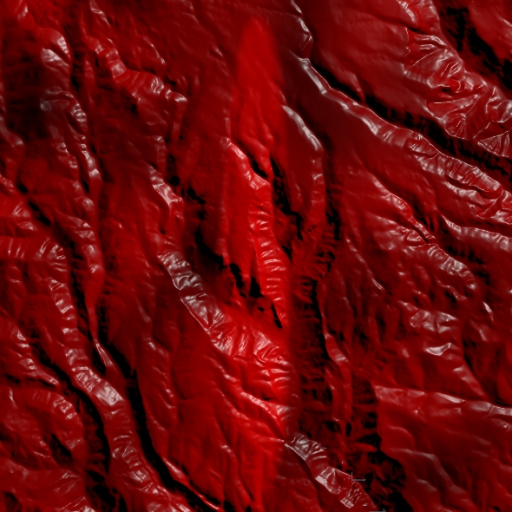}};
                \node[white] at (0, 0) {\textbf{\large (f)}};
            \end{tikzpicture}
        \end{subfigure}
        \hfill
        \begin{subfigure}[t]{0.24\textwidth}
            \centering
            \begin{tikzpicture}
                \node[anchor=north west, inner sep=0] (image) at (-0.5,0.5) 
                {\includegraphics[width=3.5cm,height=3.5cm]{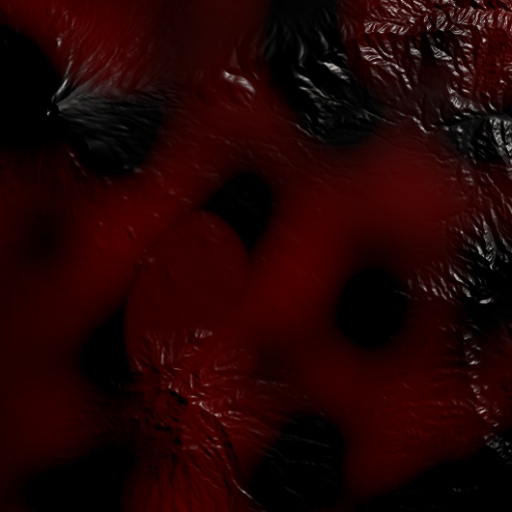}};
                \node[white] at (0, 0) {\textbf{\large (g)}};
            \end{tikzpicture}
        \end{subfigure}
        \hfill
        \begin{subfigure}[t]{0.24\textwidth}
            \centering
            \begin{tikzpicture}
                \node[anchor=north west, inner sep=0] (image) at (-0.5,0.5)  
                {\includegraphics[width=3.5cm,height=3.5cm]{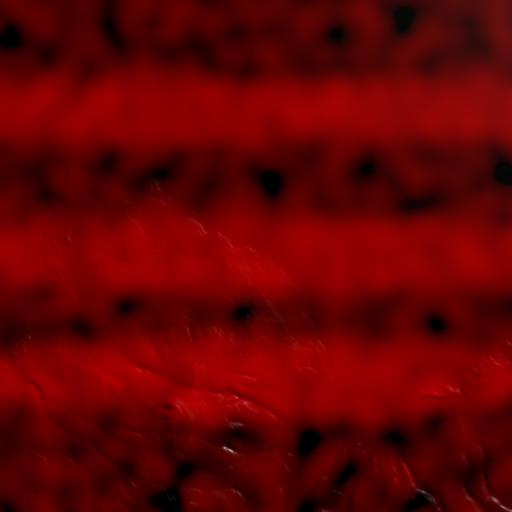}};
                \node[white] at (0, 0) {\textbf{\large (h)}};
            \end{tikzpicture}
        \end{subfigure}
    \end{minipage}
    \\[0.5em] % 行间距
    \caption{%
        Simulation inspired by real-world weather events:(a) Florida’s Biscayne National Park, (b) New Mexico’s Chiricahua Mountains, (c) Japan’s  Fuji-Hakone-Izu National Park, (d) California’s Monterey Bay. These visualizations highlight the geographical diversity and meteorological phenomena captured in our simulation framework.Additionally, the last row (e-h) illustrates the ground properties (including the temperature field and vapor field, both initialized with a consistent pattern derived from a specific regional heightfield) associated with each region.
    }
    \label{fig:weather_simulation}
\end{figure}

These events serve as the basis for our simulations, showcasing the capability of our model to replicate the diverse and complex phenomena associated with severe weather. The corresponding visualizations and parameter sets are presented in Figure~\ref{fig:weather_simulation}.

\section{VALIDATION}
The validation of our model involves presenting spatially simulated thunderstorm structures alongside meteorological characteristics and comparing the temporally simulated results with real-world weather data. This includes analyzing cloud fraction profiles to assess cloud formation and structure, tracking the evolution of cloud coverage against real-life data from national weather services\footnote{\url{https://www.visualcrossing.com/weather/weather-data-services/}}, and evaluating the temporal variation in lightning flash rates.

\begin{figure}[htbp]
    \centering
    \resizebox{\textwidth}{!}{%
        \begin{minipage}{\textwidth}
            \centering
            % Row 1: Cloud Fraction Model
            \begin{subfigure}[t]{0.24\textwidth}
                \centering
                \includegraphics[width=\textwidth]{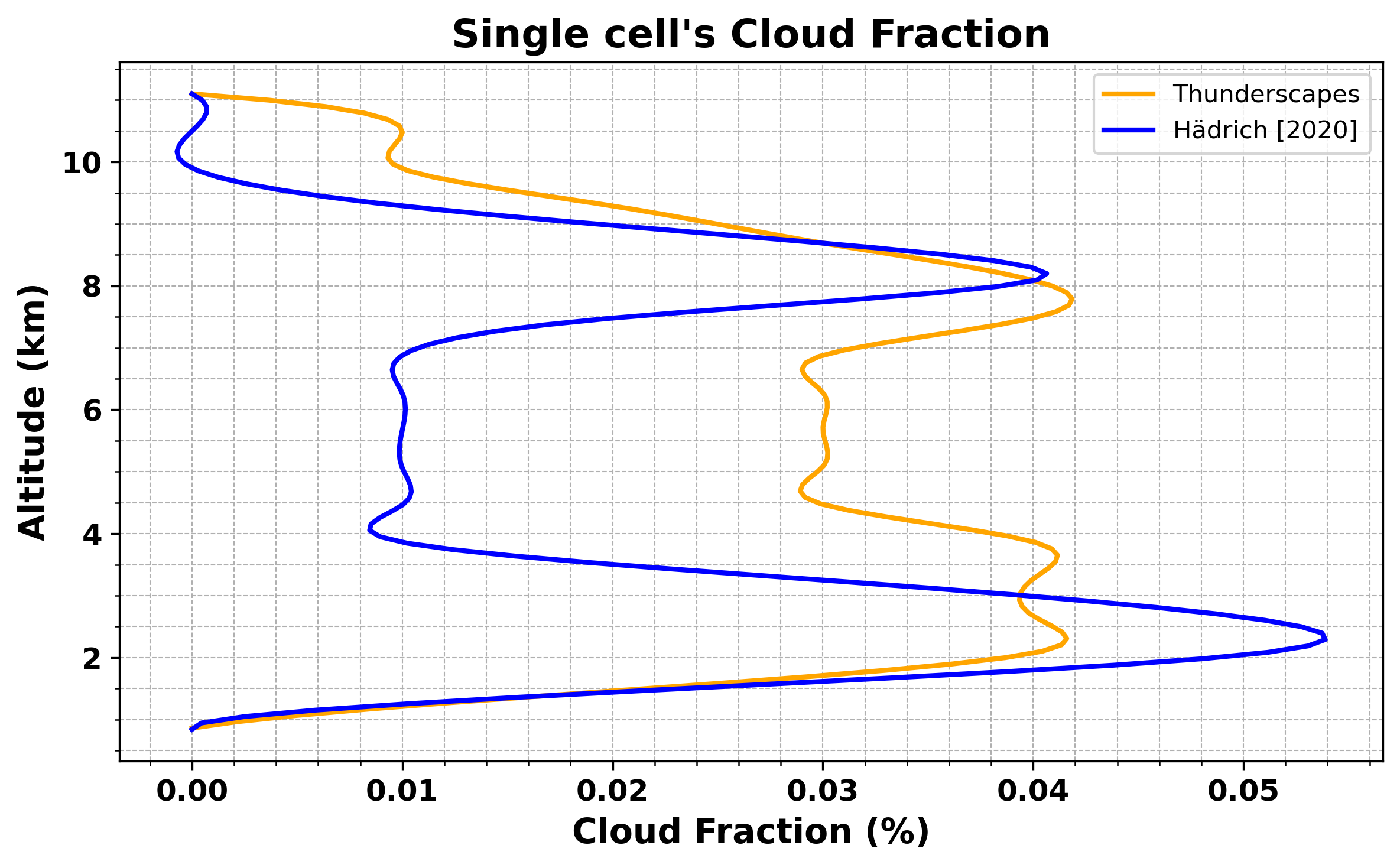}
            \end{subfigure}
            \hfill
            \begin{subfigure}[t]{0.24\textwidth}
                \centering
                \includegraphics[width=\textwidth]{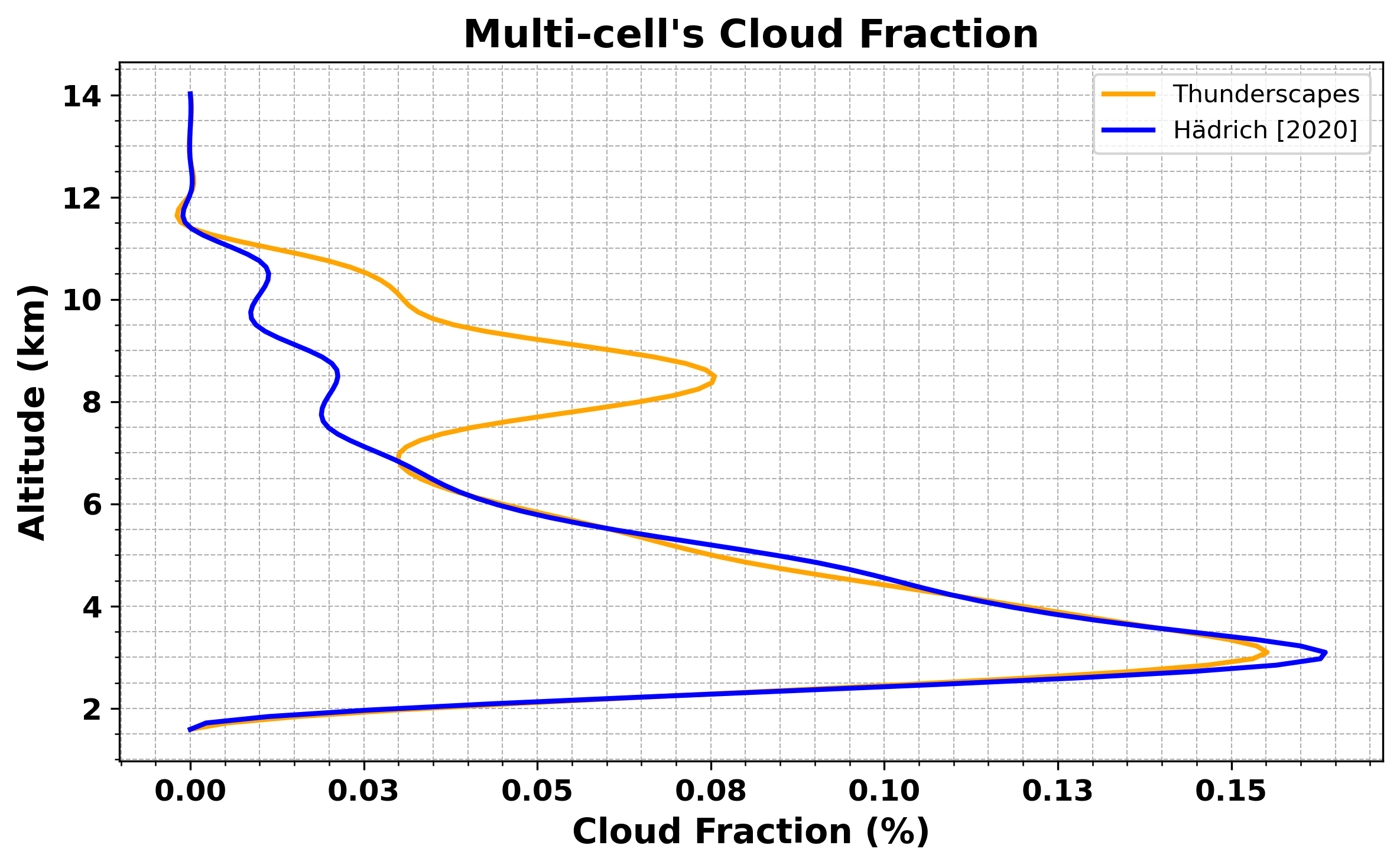}
            \end{subfigure}
            \hfill
            \begin{subfigure}[t]{0.24\textwidth}
                \centering
                \includegraphics[width=\textwidth]{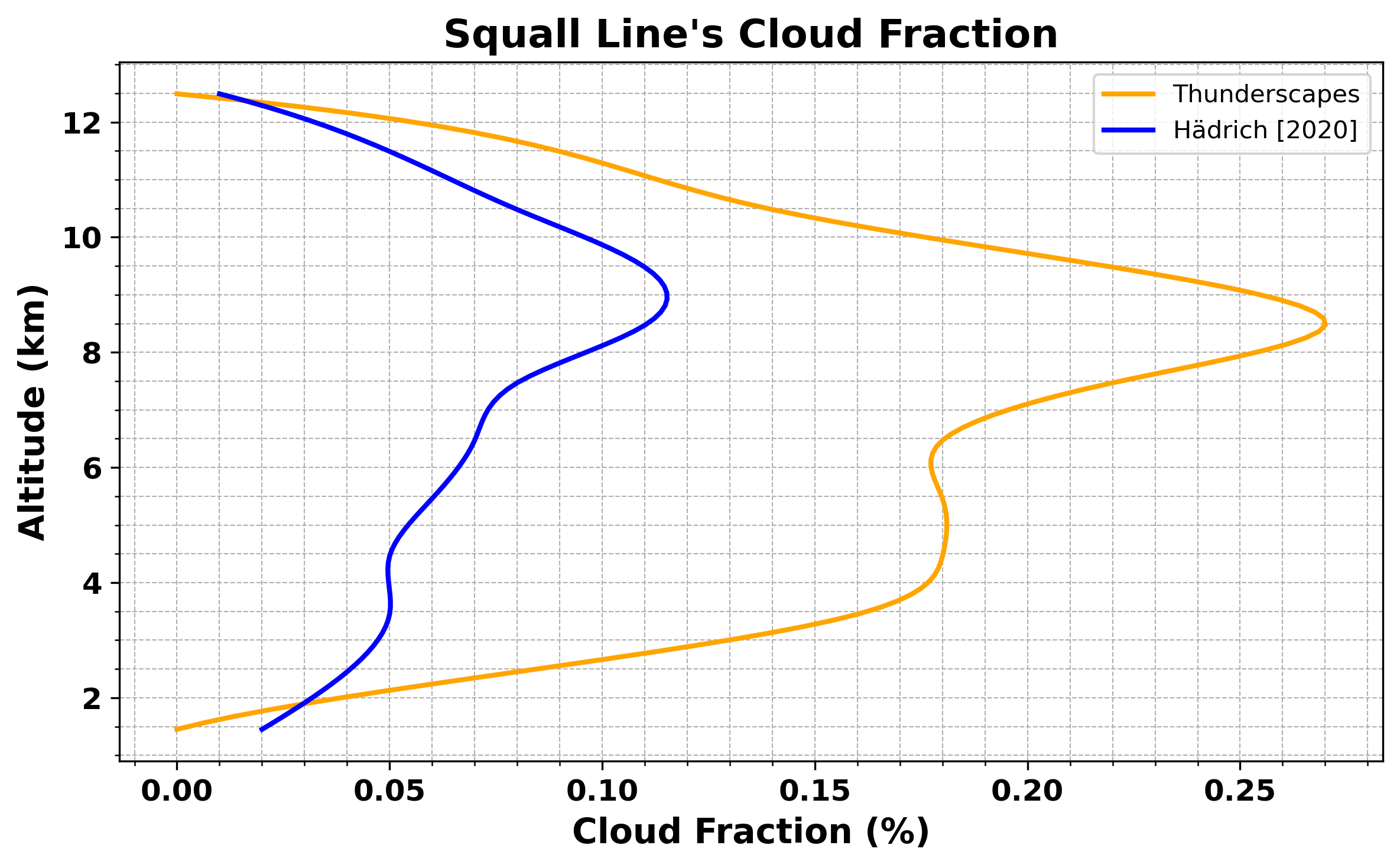}
            \end{subfigure}
            \hfill
            \begin{subfigure}[t]{0.24\textwidth}
                \centering
                \includegraphics[width=\textwidth]{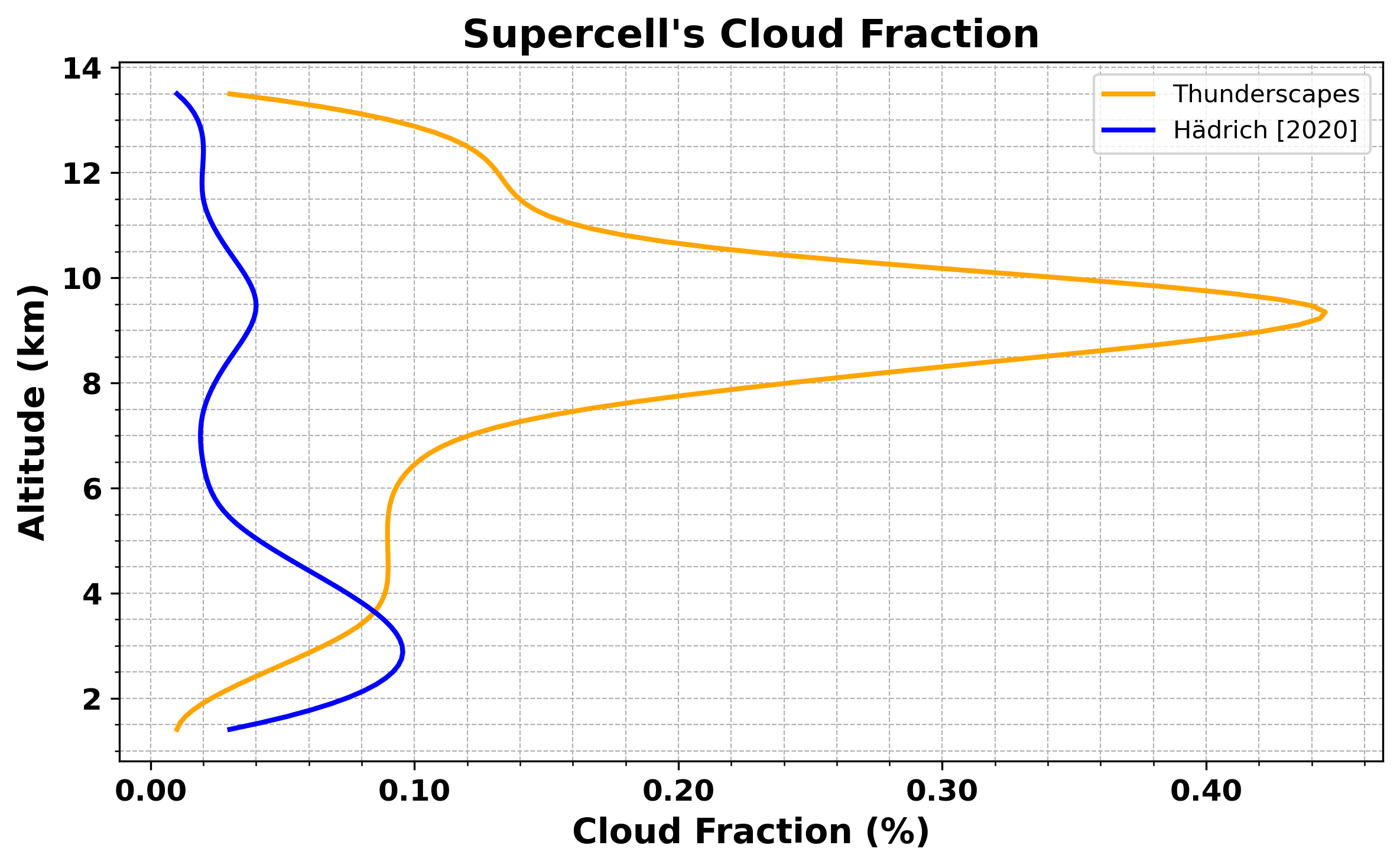}
            \end{subfigure}
            
            % Row 2: Hädrich et al. [2020] Results
            % Add vertical space between rows
            \vspace{1em}
            \begin{subfigure}[t]{0.24\textwidth}
                \centering
                \includegraphics[width=\textwidth]{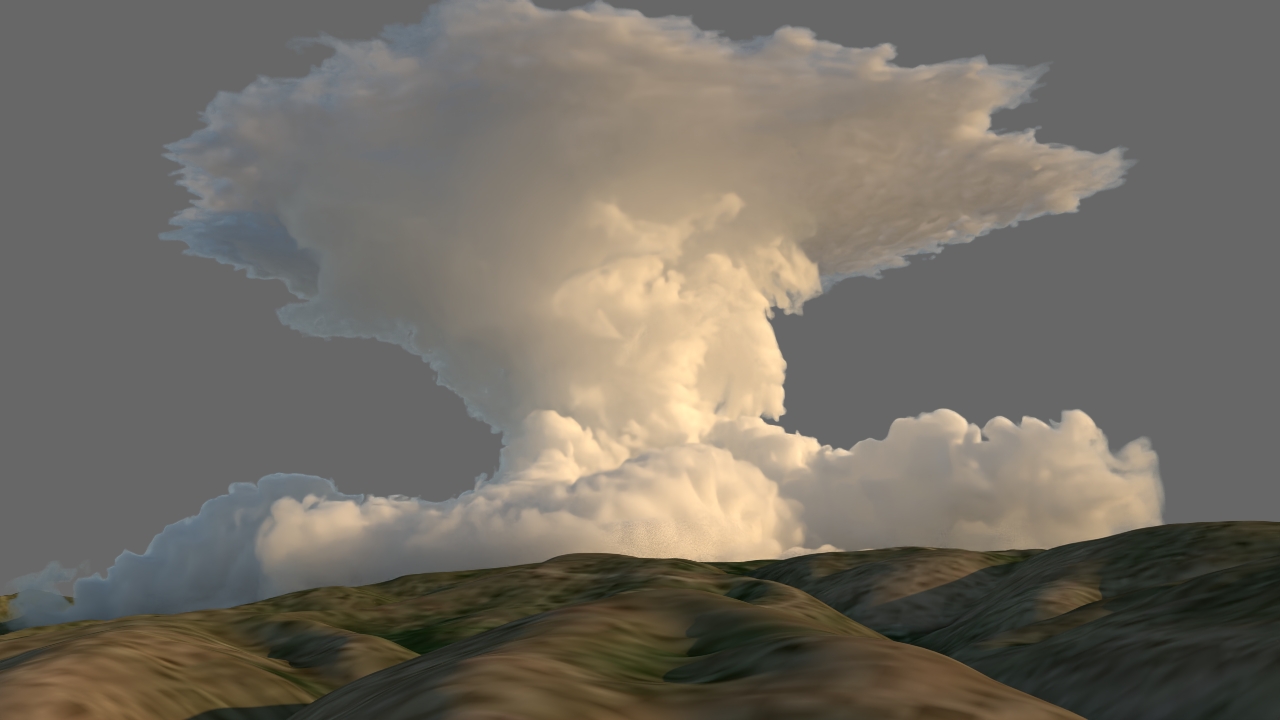}
            \end{subfigure}
            \hfill
            \begin{subfigure}[t]{0.24\textwidth}
                \centering
                \includegraphics[width=\textwidth]{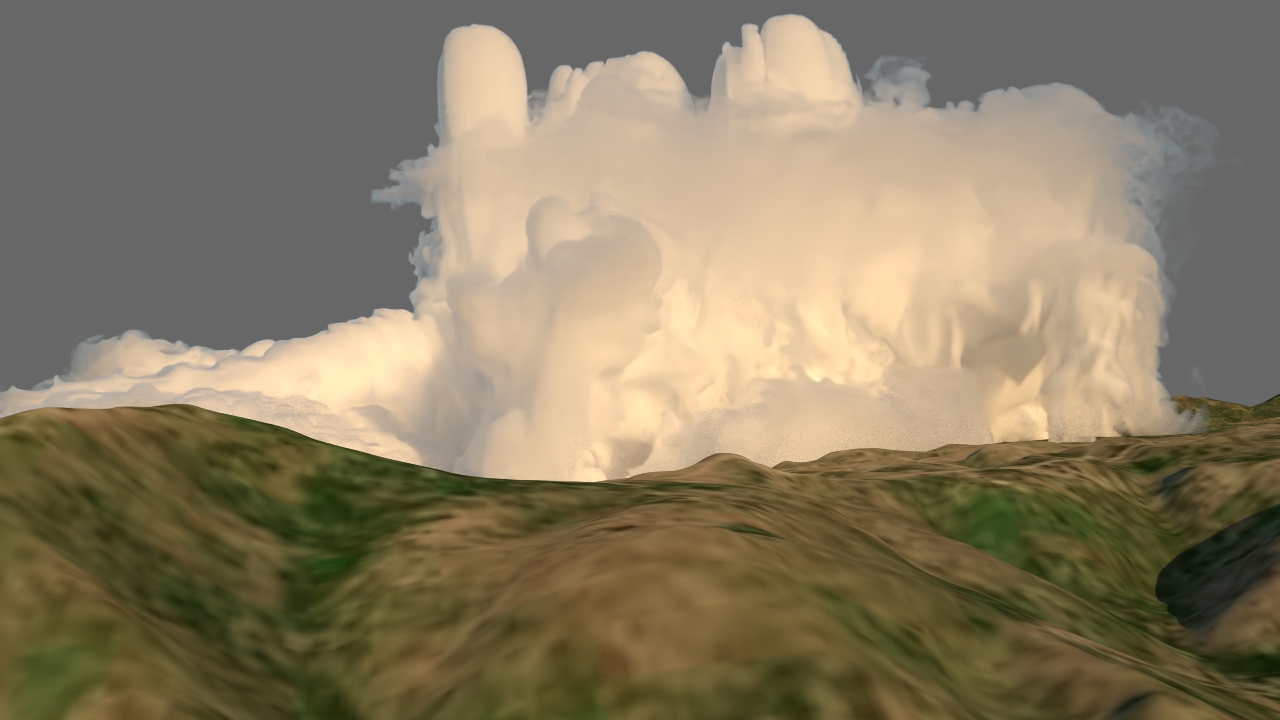}
            \end{subfigure}
            \hfill
            \begin{subfigure}[t]{0.24\textwidth}
                \centering
                \includegraphics[width=\textwidth]{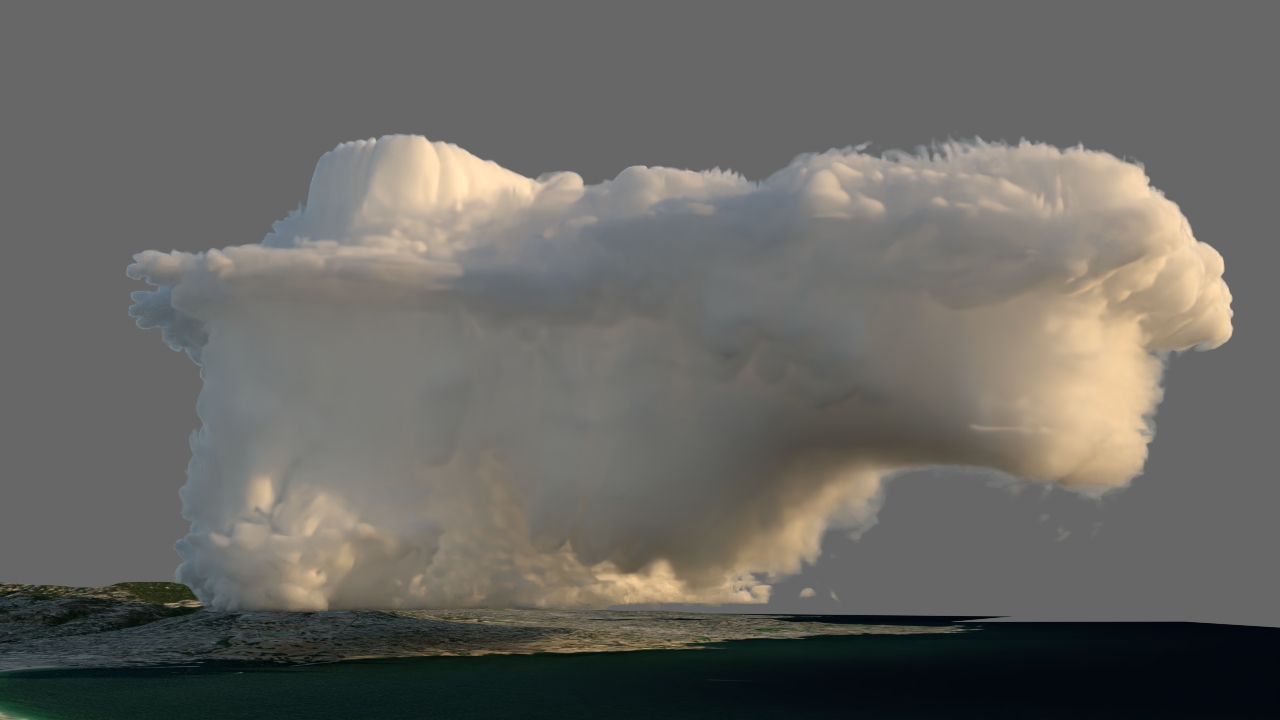}
            \end{subfigure}
            \hfill
            \begin{subfigure}[t]{0.24\textwidth}
                \centering
                \includegraphics[width=\textwidth]{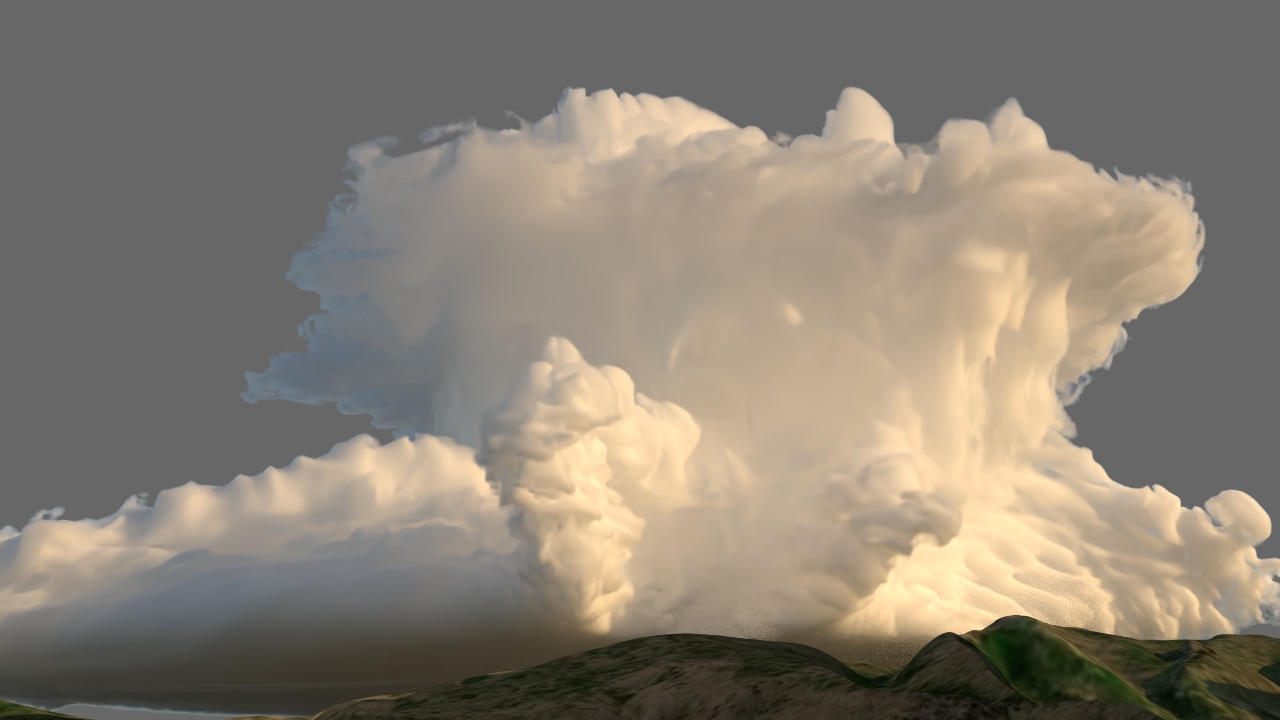}
            \end{subfigure}

            % Row 3: Our Results
           % Add vertical space between rows
            \vspace{1em}
            \begin{subfigure}[t]{0.24\textwidth}
                \centering
                \includegraphics[width=\textwidth]{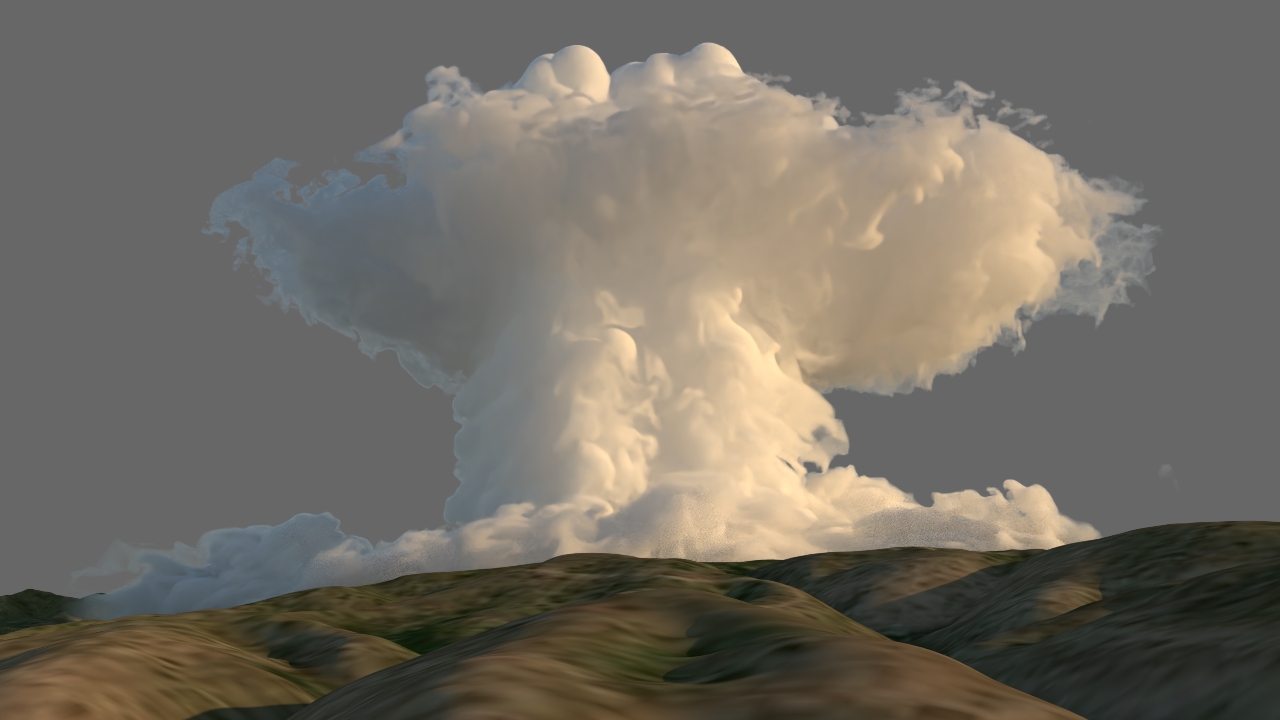}
            \end{subfigure}
            \hfill
            \begin{subfigure}[t]{0.24\textwidth}
                \centering
                \includegraphics[width=\textwidth]{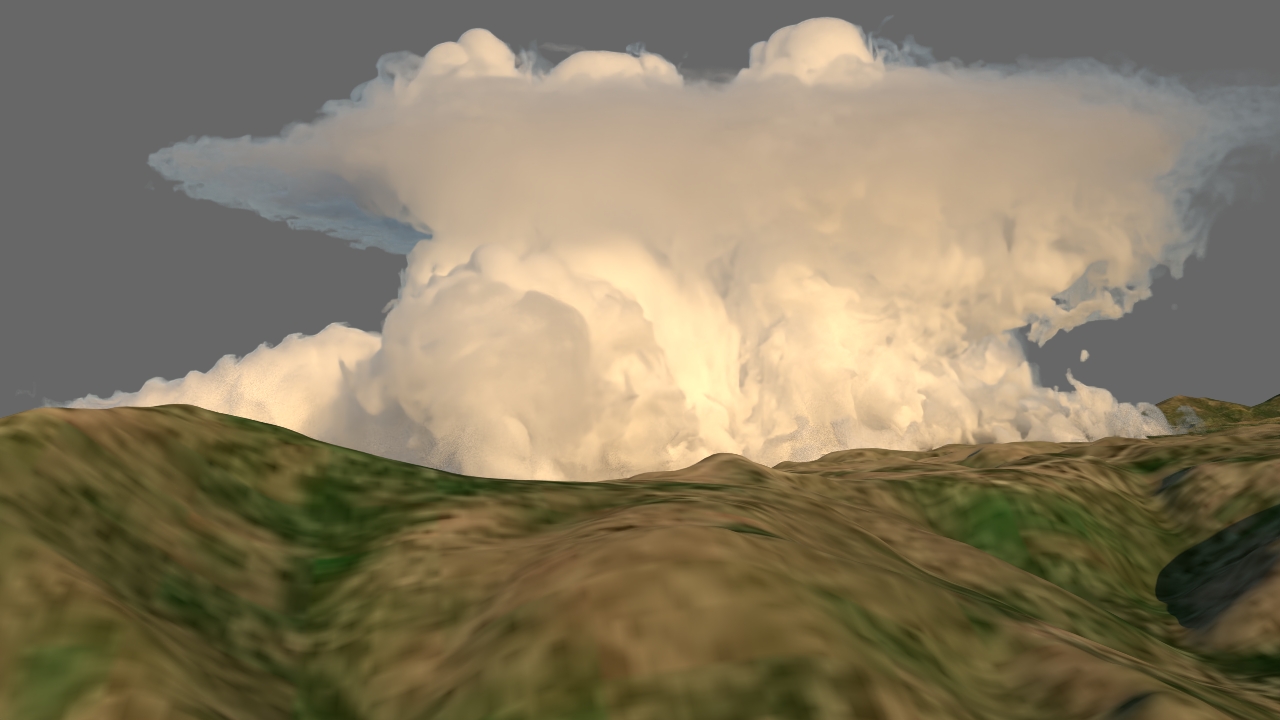}
            \end{subfigure}
            \hfill
            \begin{subfigure}[t]{0.24\textwidth}
                \centering
                \includegraphics[width=\textwidth]{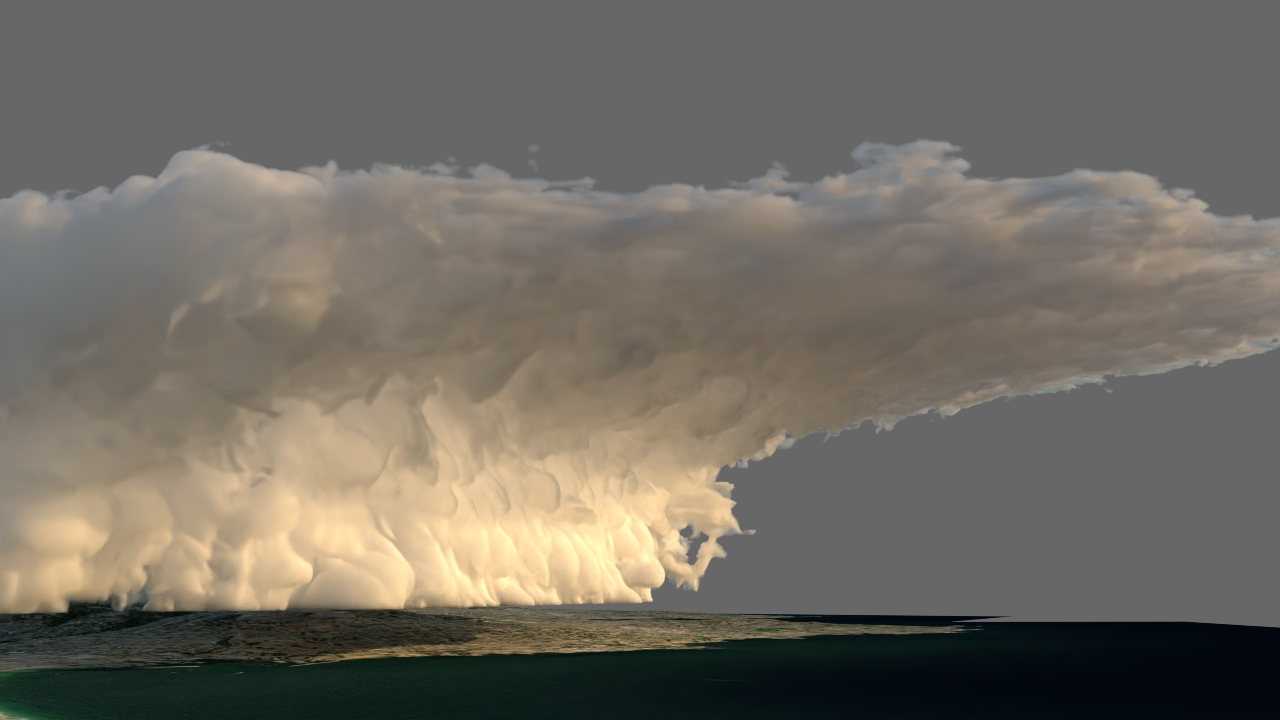}
            \end{subfigure}
            \hfill
            \begin{subfigure}[t]{0.24\textwidth}
                \centering
                \includegraphics[width=\textwidth]{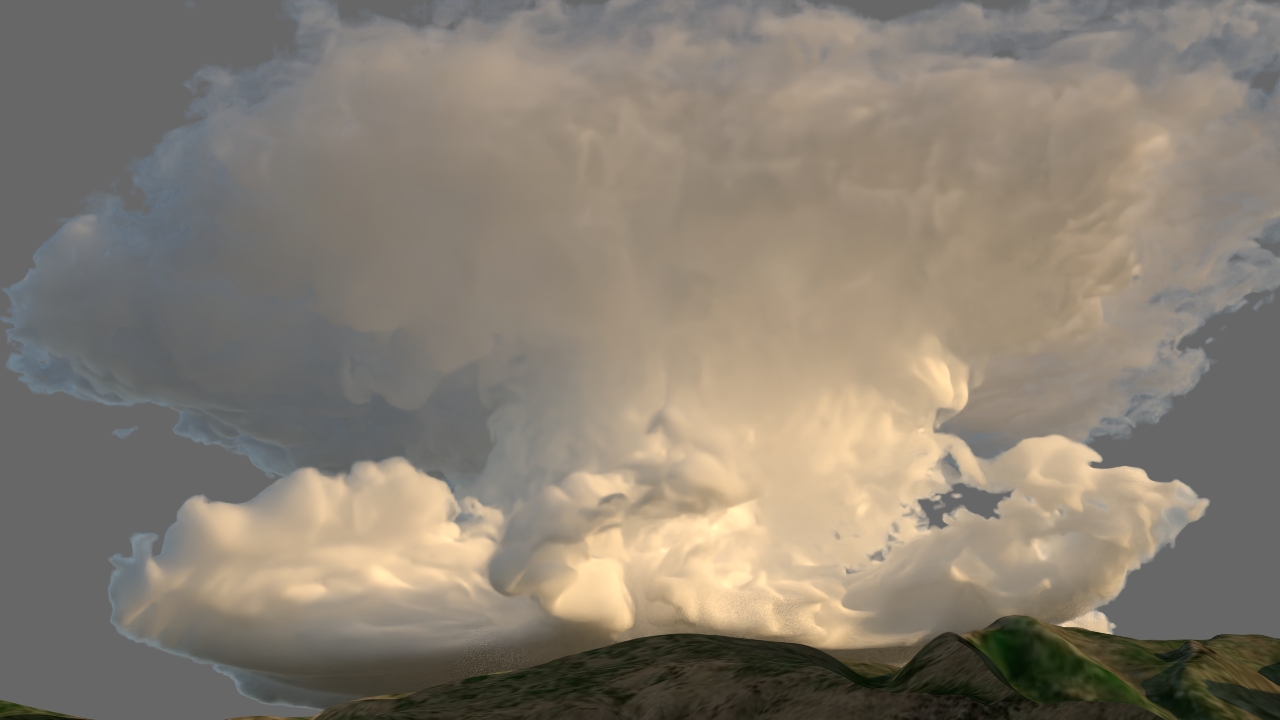}
            \end{subfigure}
        \end{minipage}
    }
    \caption{Visualization comparing cloud fraction models and simulation results for four storm types: single cell, multi-cell, squall line, and supercell. \textbf{Top row}: Cloud fraction models, where orange represents our method, and blue represents the method of Hädrich et al. \cite{hadrich2020stormscapes}. \textbf{Middle row}: Results generated using Hädrich et al. \cite{hadrich2020stormscapes}. \textbf{Bottom row}: Results generated by our method. The visualization highlights differences in cloud fraction structures and thunderstorm characteristics captured by the two approaches.}
    \label{fig:cloud_fraction_comparison}
\end{figure}
%for 6.1 thunderstorm variations:

%TODO:说我们方法为何优秀于stormscape:对于cloud fraction以及模拟的结果看，对应single cell，我们两个的方法都表现不错，但对于更复杂的雷暴种类，即multi-cell,squall line,supercell,在较高海拔下他们的曲线没有得到很好的延展，并不符合气象观测这些雷暴种类的预期,即他们的模型fail to capture the detail of anvil cloud of the thunderstorm，anvil cloud是The anvil is the elongated cloud at the top of the storm that spreads downwind with upper level steering winds. The anvil will appear solid, not wispy, and will have sharp, well defined edges.，因为他们使用的是经典Kessler-style降雨模型，其只考虑了vapor，rain和cloud water的相变过程，而我们参考的Grabowski-style将相变过程可以细分到vapor,rain,snow,cloud water和ice的级别，对大气浮力的建模更加准确，所以证明我们的模型更适合在MCS尺度下进行真实感模拟

% We adopt the evaluation methods outlined in \cite{shen2020statistically} and \cite{cesana2019cumulus}, utilizing altitude-based quantitative data to analyze cloud behavior across different atmospheric layers. The results are presented in Figure~\ref{fig:cloud_fraction_comparison}, which illustrates the relationship between cloud fraction and storm height. Notably, the maximum cloud fraction corresponds to the storm's peak altitude, consistent with the classical vertical distribution of cumulonimbus clouds in meteorology.

%加上我们和积雨云模型\textit{Stormscapes}\cite{hadrich2020stormscapes}进行比较

% The results, presented in Figure~\ref{fig:cloud_fraction_comparison}, illustrate the relationship between cloud fraction and storm height. Notably, the maximum cloud fraction aligns with the storm's peak altitude, reflecting the classical vertical distribution of cumulonimbus clouds in meteorology.
\subsection{Thunderstorm structure}

We adopt the evaluation methods outlined in \cite{shen2020statistically} and \cite{cesana2019cumulus}, utilizing altitude-based quantitative data to analyze cloud behavior across various atmospheric layers. To ensure a fair comparison, we evaluate our framework alongside the cumulonimbus simulation model \textit{Stormscapes} \cite{hadrich2020stormscapes} using identical input fields and consistent initial conditions. 

The results, shown in Figure~\ref{fig:cloud_fraction_comparison}, illustrate the relationship between cloud fraction and the structure of thunderstorms at their maximum development height as modeled by the two approaches. While both models perform well for simpler thunderstorm types, such as single cell thunderstorms, notable differences arise for more complex systems, including multi-cell thunderstorms, squall lines, and supercells. At higher altitudes, the \textit{Stormscapes} struggles to capture the expected extension of cloud coverage, particularly the anvil cloud, which is a critical feature of complex thunderstorms. The anvil, characterized by its solid appearance, sharp, well-defined edges, and downwind spread influenced by upper-level steering winds, is not accurately represented in \textit{Stormscapes}.

This limitation arises because the \textit{Stormscapes} relies on the classical Kessler-style precipitation parameterization, which only accounts for phase transitions among vapor, rain, and cloud water. In contrast, our framework employs a Grabowski-style microphysics scheme, which incorporates a finer classification of phase transitions, including vapor, rain, snow, cloud water, and ice. This enhanced granularity enables more accurate modeling of atmospheric buoyancy and better aligns with meteorological observations, making our model particularly well-suited for realistic MCS simulations.Moreover, the \textit{Stormscapes} model does not account for hydrometeor electrification processes, limiting its ability to simulate lightning dynamics during thunderstorm development and to reproduce consistent atmospheric phenomena.

\begin{figure}[htbp]
    \centering
    \resizebox{\textwidth}{!}{%
        \begin{minipage}{\textwidth}
            % 第一行：Cloud Coverage Evolution
            \begin{subfigure}[t]{0.24\textwidth}
                \centering
                \begin{tikzpicture}
                    \node[anchor=north west, inner sep=0] (image) at (-0.5,0.5)  
                    {\includegraphics[width=\textwidth]{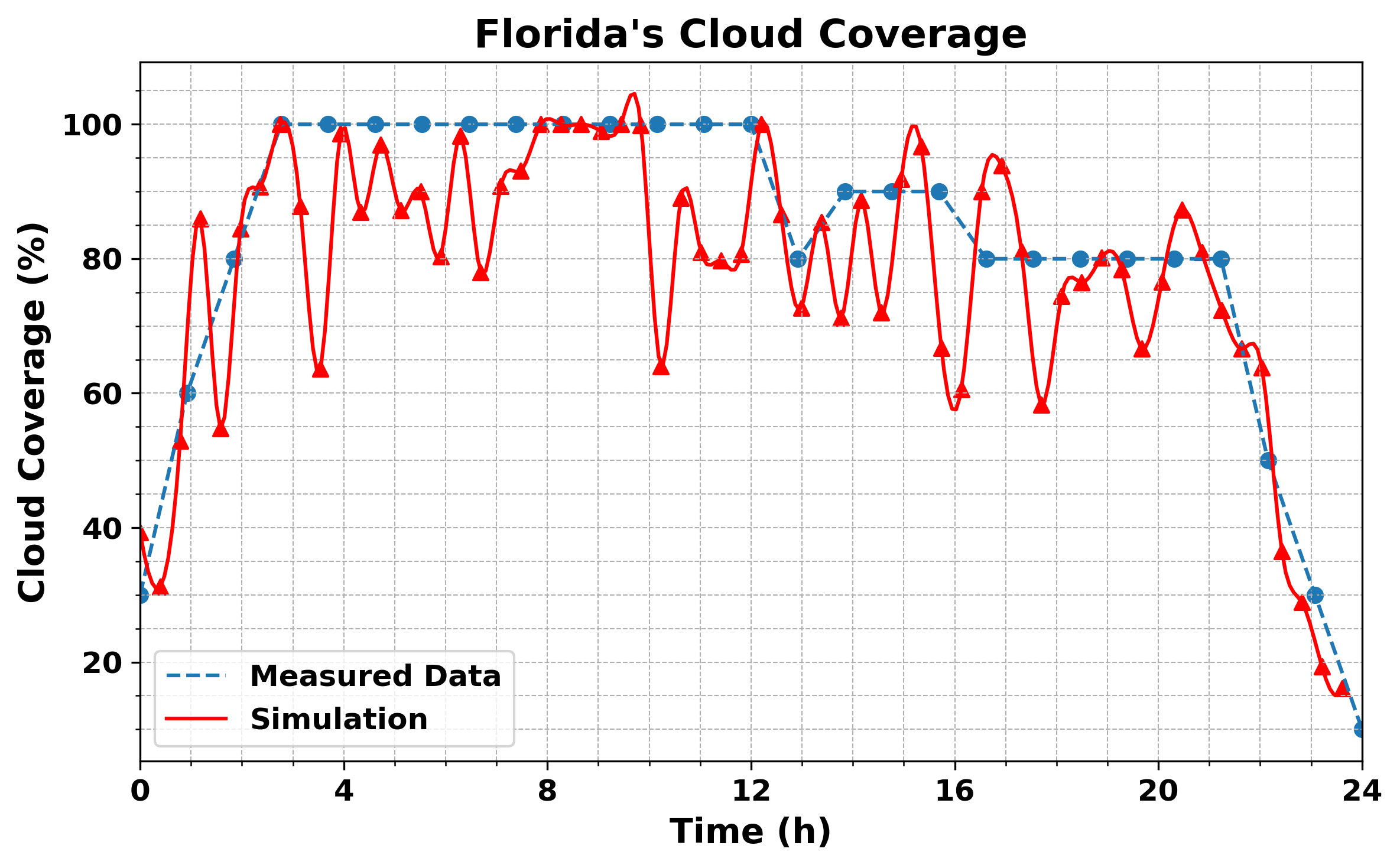}};
                \end{tikzpicture}
            \end{subfigure}
            \hfill
            \begin{subfigure}[t]{0.24\textwidth}
                \centering
                \begin{tikzpicture}
                    \node[anchor=north west, inner sep=0] (image) at (-0.5,0.5)  
                    {\includegraphics[width=\textwidth]{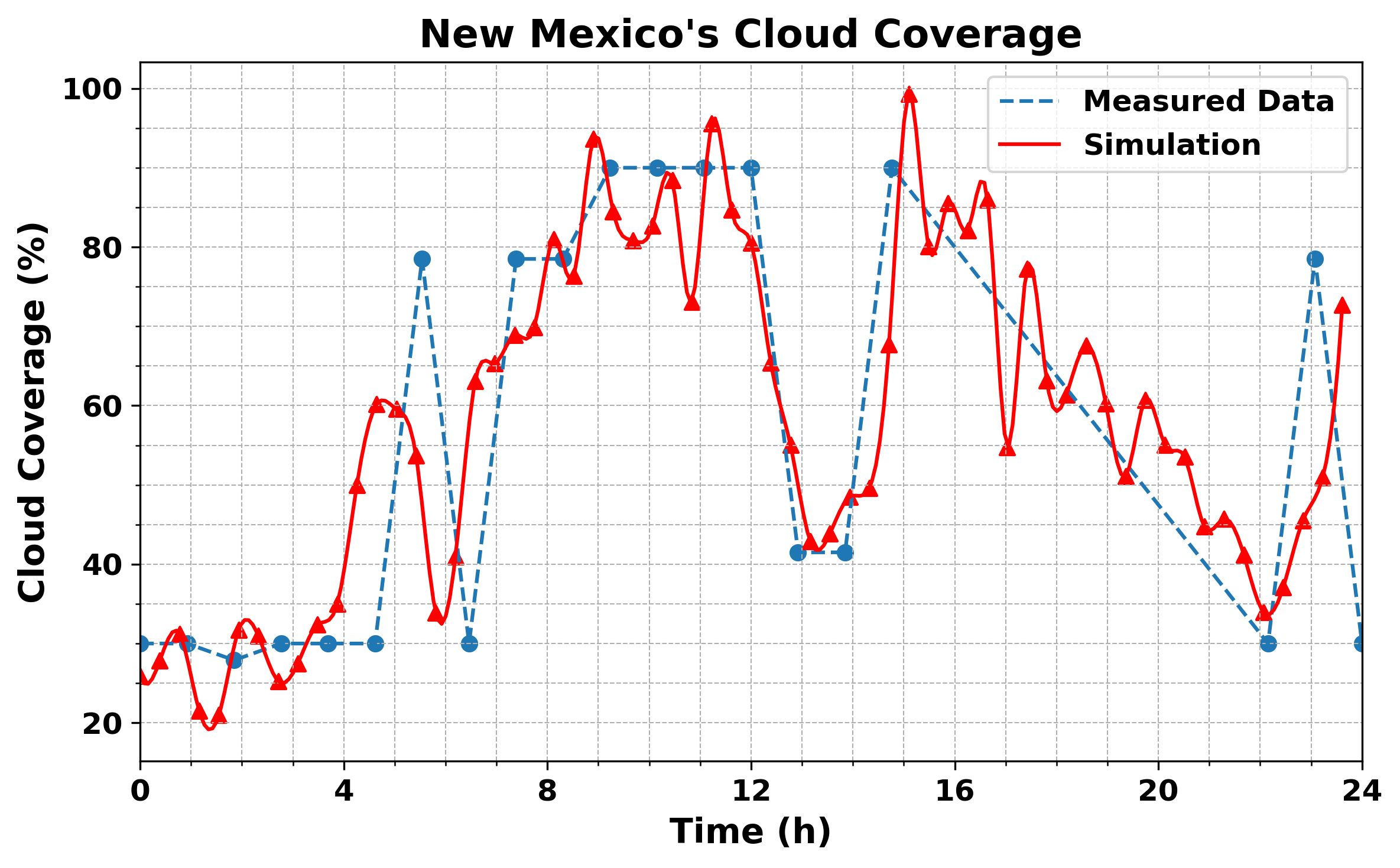}};
                \end{tikzpicture}
            \end{subfigure}
            \hfill
            \begin{subfigure}[t]{0.24\textwidth}
                \centering
                \begin{tikzpicture}
                    \node[anchor=north west, inner sep=0] (image) at (-0.5,0.5)  
                    {\includegraphics[width=\textwidth]{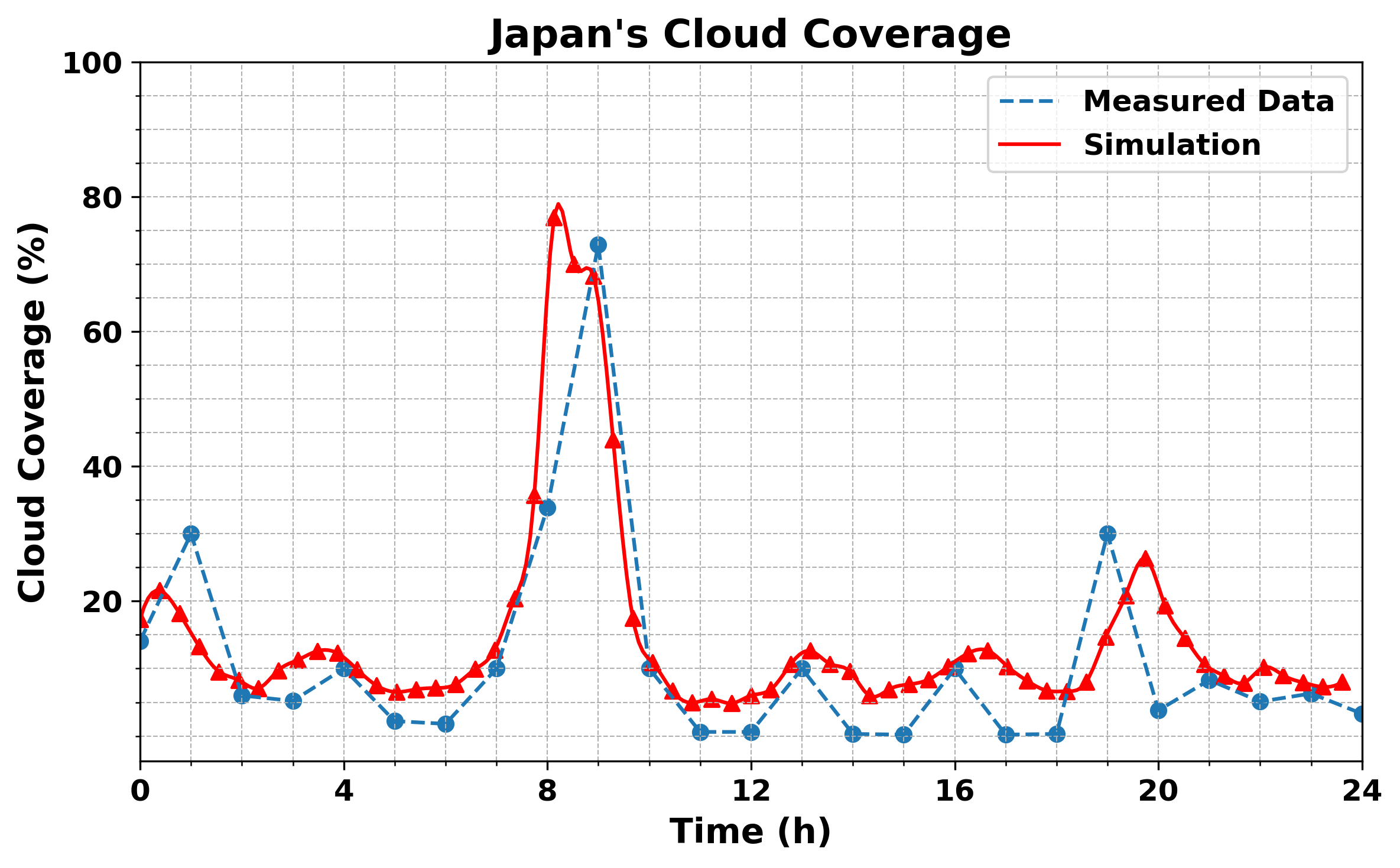}};
                \end{tikzpicture}
            \end{subfigure}
            \hfill
            \begin{subfigure}[t]{0.24\textwidth}
                \centering
                \begin{tikzpicture}
                    \node[anchor=north west, inner sep=0] (image) at (-0.5,0.5)  
                    {\includegraphics[width=\textwidth]{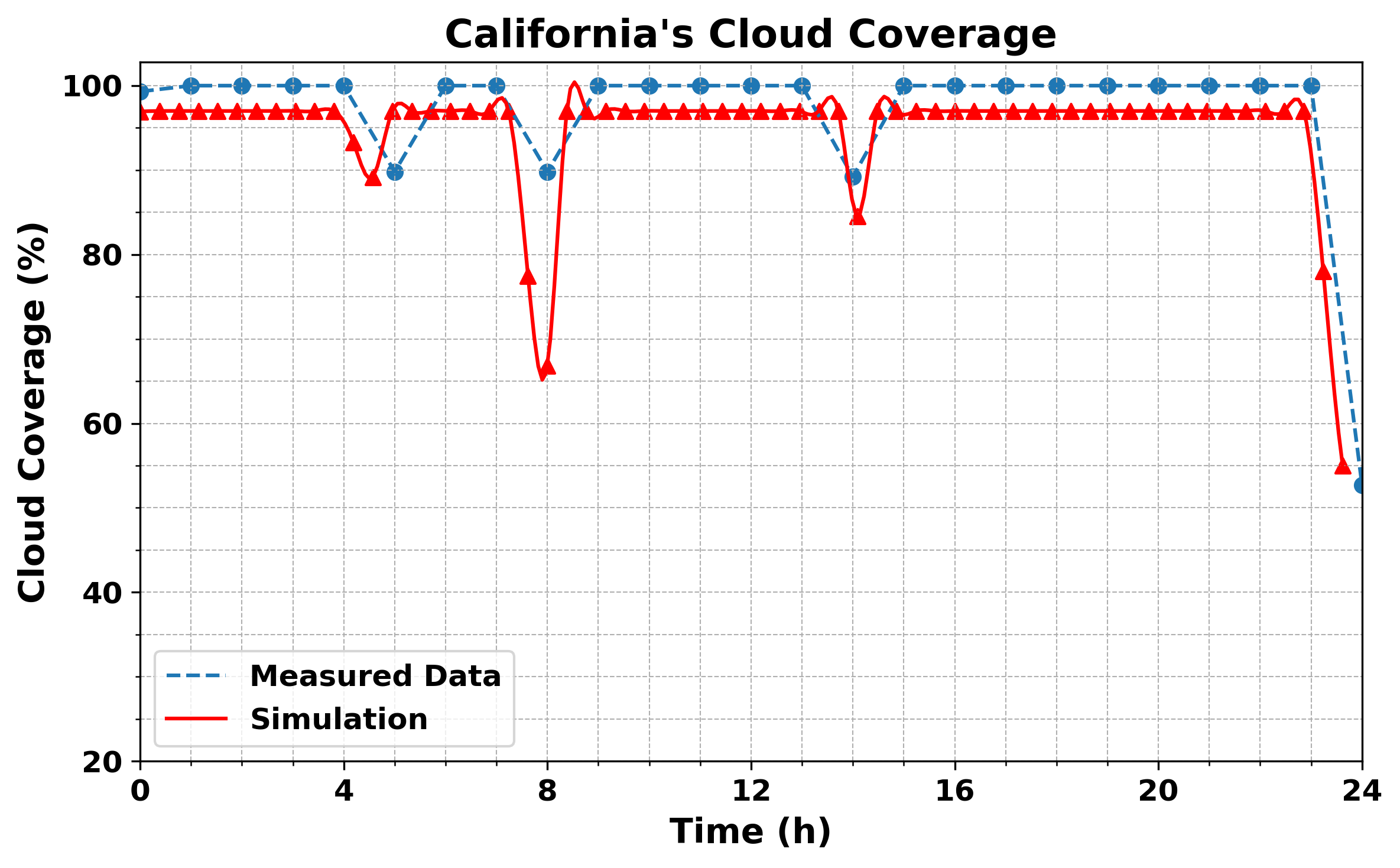}};
                \end{tikzpicture}
            \end{subfigure}
            \\
            % 第二行：Cloud Fraction
            \begin{subfigure}[t]{0.24\textwidth}
                \centering
                \begin{tikzpicture}
                    \node[anchor=north west, inner sep=0] (image) at (-0.5,0.5)  
                    {\includegraphics[width=\textwidth]{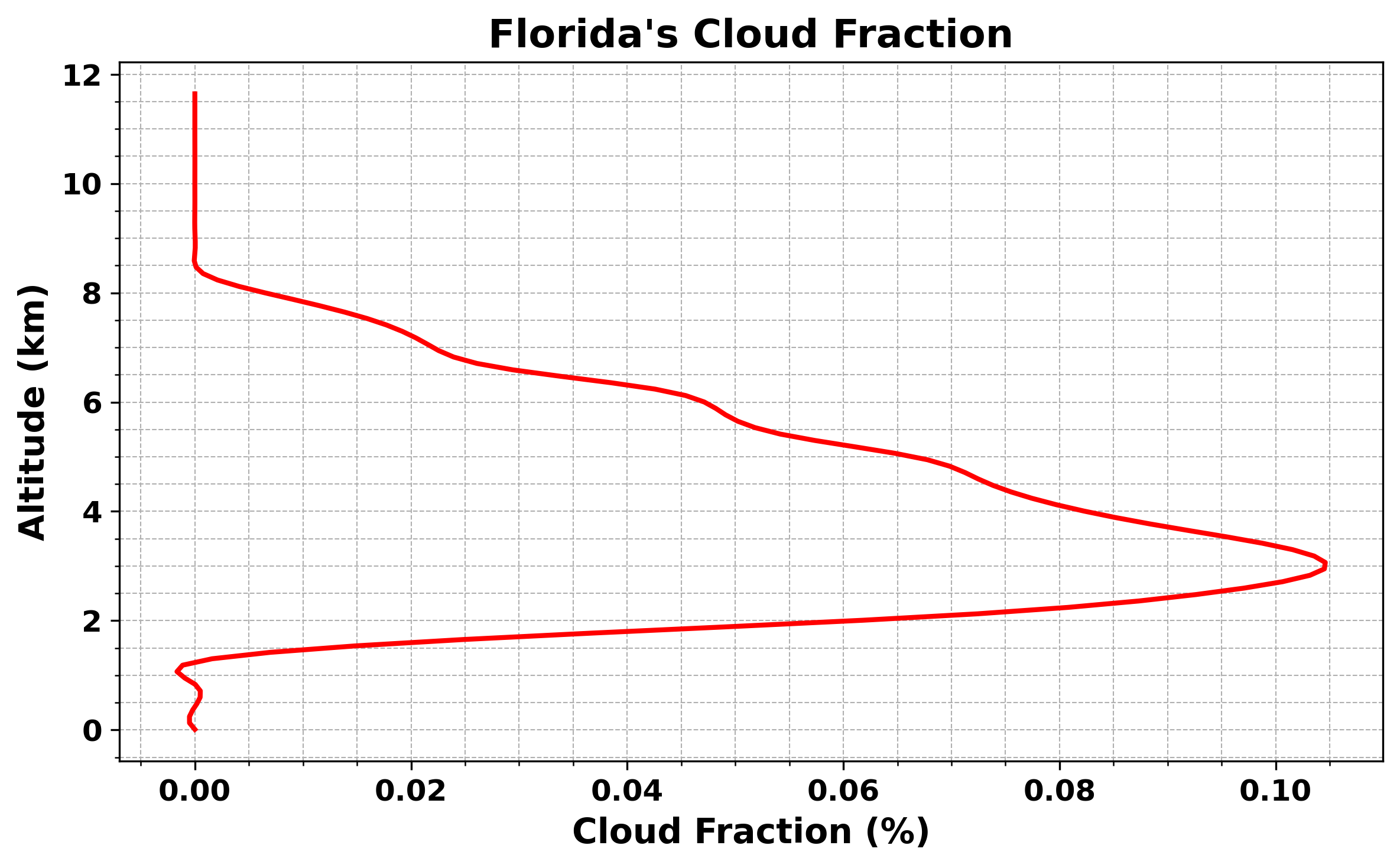}};
                \end{tikzpicture}
            \end{subfigure}
            \hfill
            \begin{subfigure}[t]{0.24\textwidth}
                \centering
                \begin{tikzpicture}
                    \node[anchor=north west, inner sep=0] (image) at (-0.5,0.5)  
                    {\includegraphics[width=\textwidth]{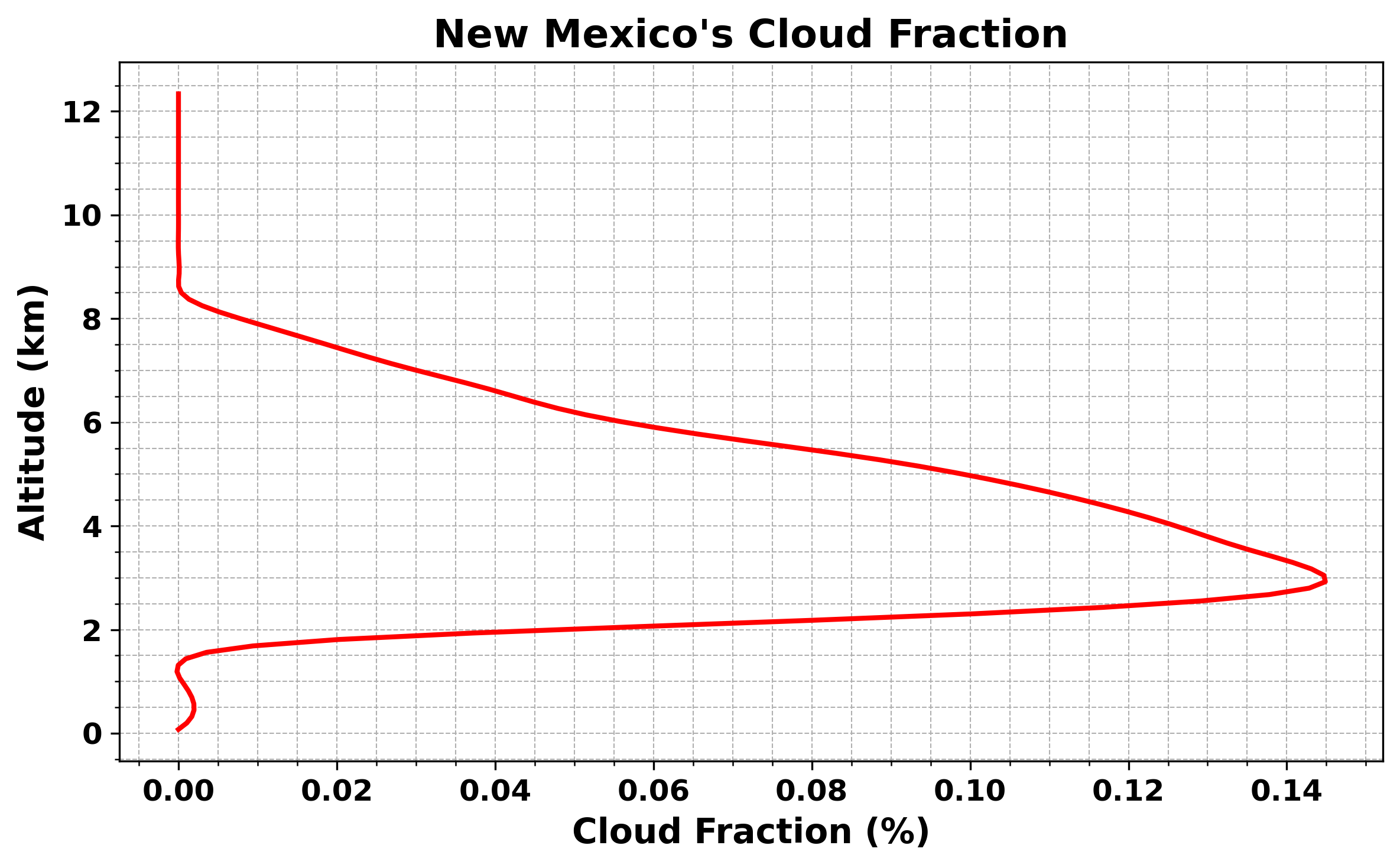}};
                \end{tikzpicture}
            \end{subfigure}
            \hfill
            \begin{subfigure}[t]{0.24\textwidth}
                \centering
                \begin{tikzpicture}
                    \node[anchor=north west, inner sep=0] (image) at (-0.5,0.5)  
                    {\includegraphics[width=\textwidth]{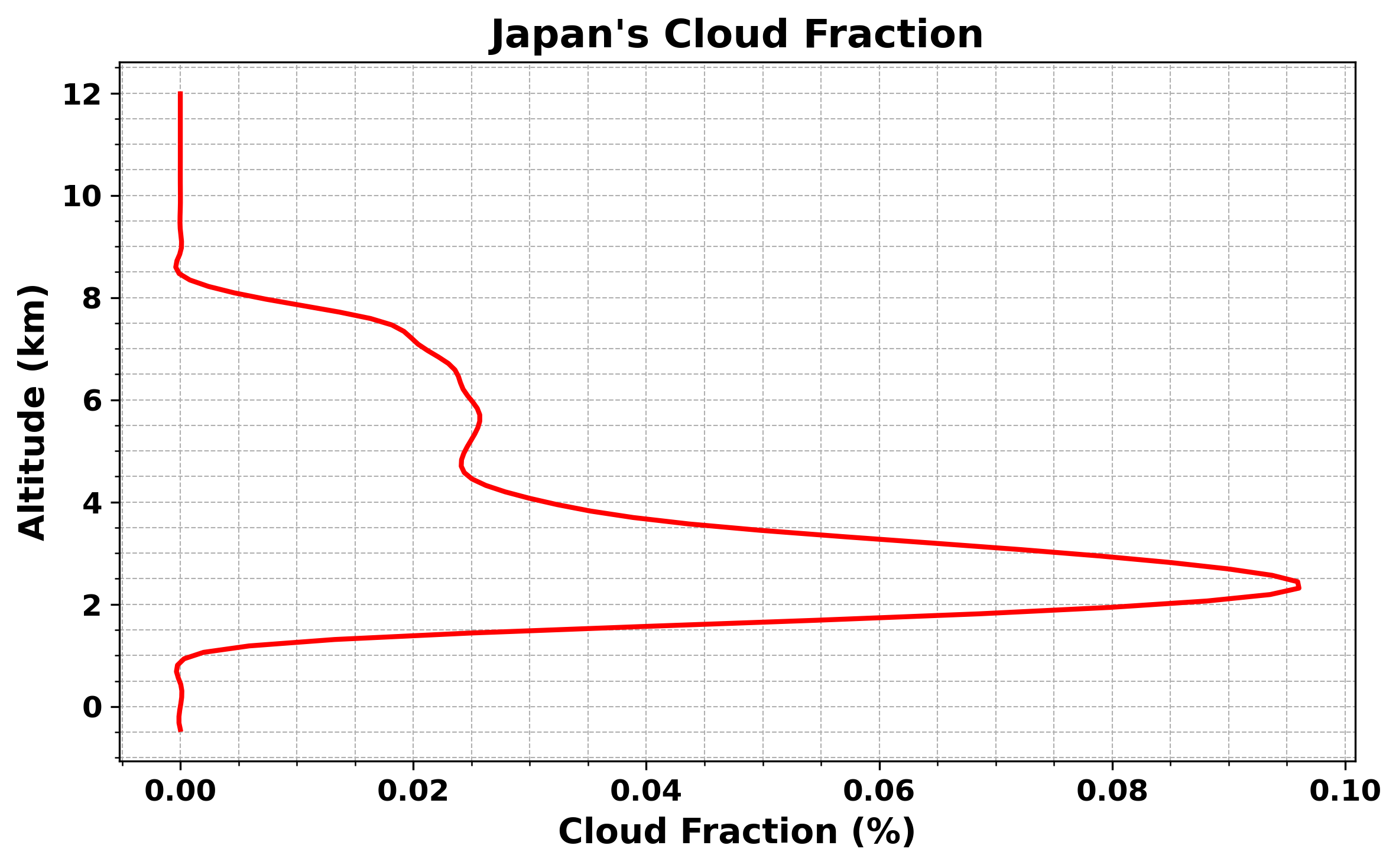}};
                \end{tikzpicture}
            \end{subfigure}
            \hfill
            \begin{subfigure}[t]{0.24\textwidth}
                \centering
                \begin{tikzpicture}
                    \node[anchor=north west, inner sep=0] (image) at (-0.5,0.5)  
                    {\includegraphics[width=\textwidth]{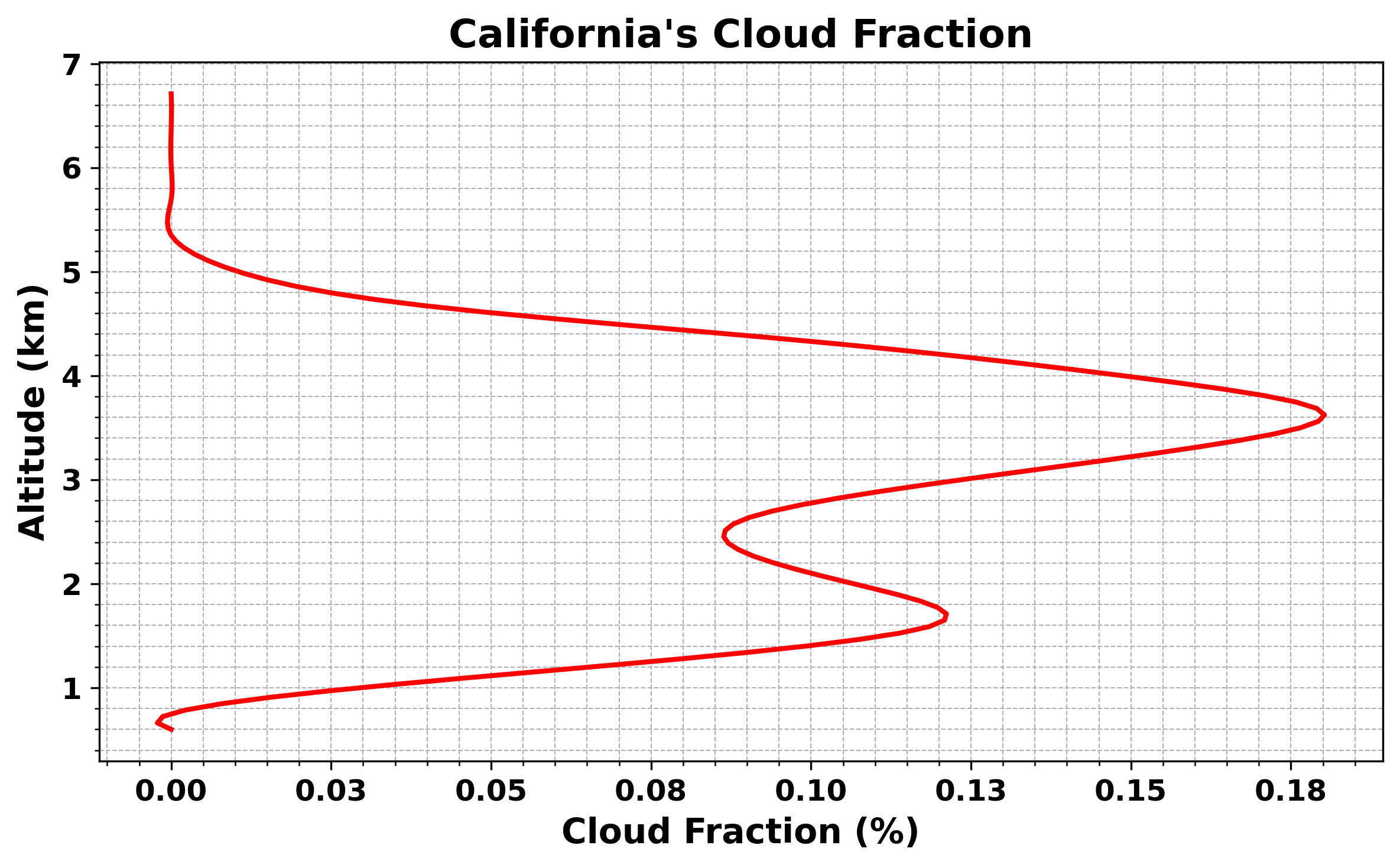}};
                \end{tikzpicture}
            \end{subfigure}
            % 第三行：Lightning Evolution
            \begin{subfigure}[t]{0.24\textwidth}
                \centering
                \begin{tikzpicture}
                    \node[anchor=north west, inner sep=0] (image) at (-0.5,0.5)  
                    {\includegraphics[width=\textwidth]{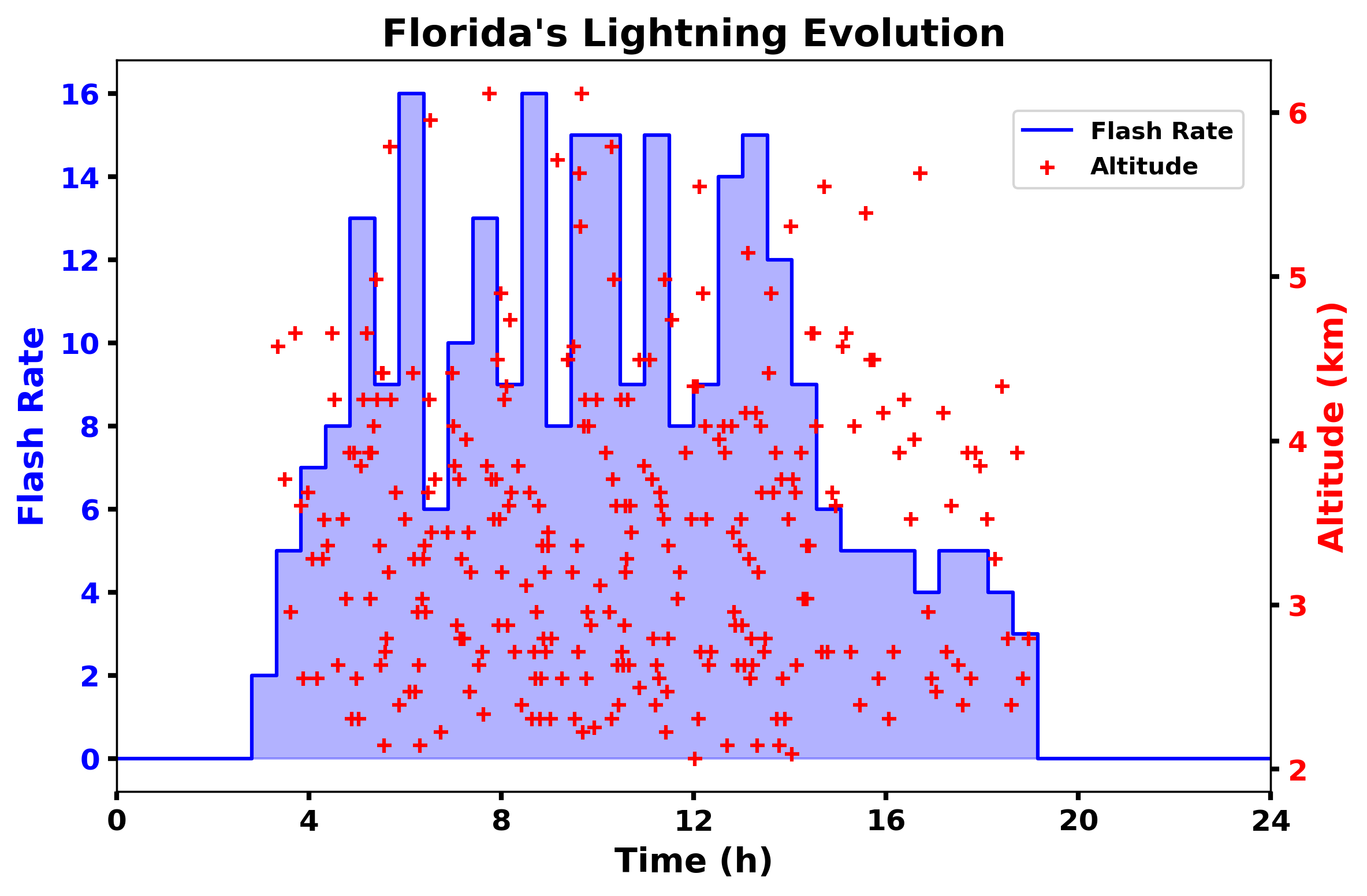}};
                \end{tikzpicture}
            \end{subfigure}
            \hfill
            \begin{subfigure}[t]{0.24\textwidth}
                \centering
                \begin{tikzpicture}
                    \node[anchor=north west, inner sep=0] (image) at (-0.5,0.5)  
                    {\includegraphics[width=\textwidth]{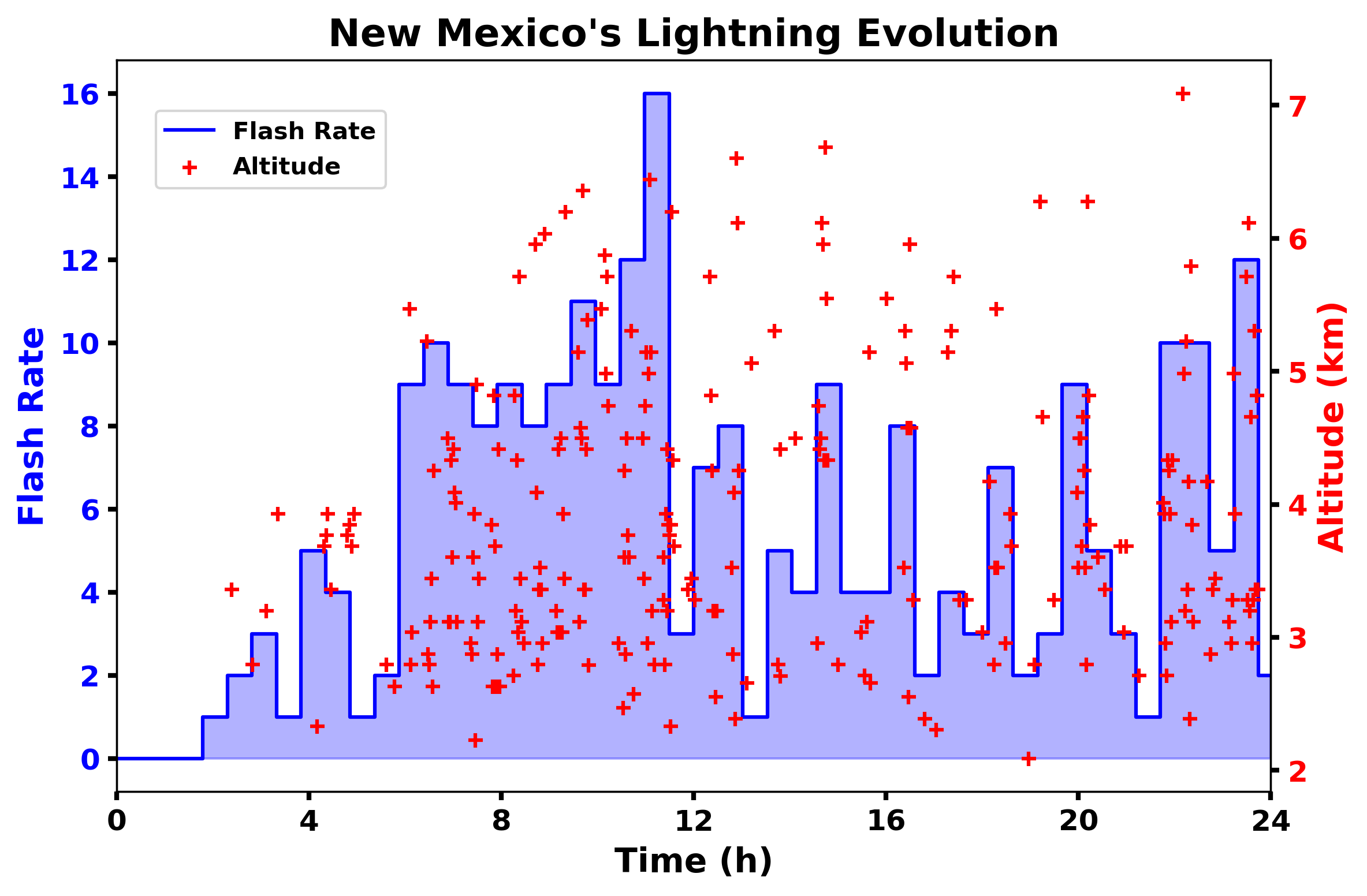}};
                \end{tikzpicture}
            \end{subfigure}
            \hfill
            \begin{subfigure}[t]{0.24\textwidth}
                \centering
                \begin{tikzpicture}
                    \node[anchor=north west, inner sep=0] (image) at (-0.5,0.5)  
                    {\includegraphics[width=\textwidth]{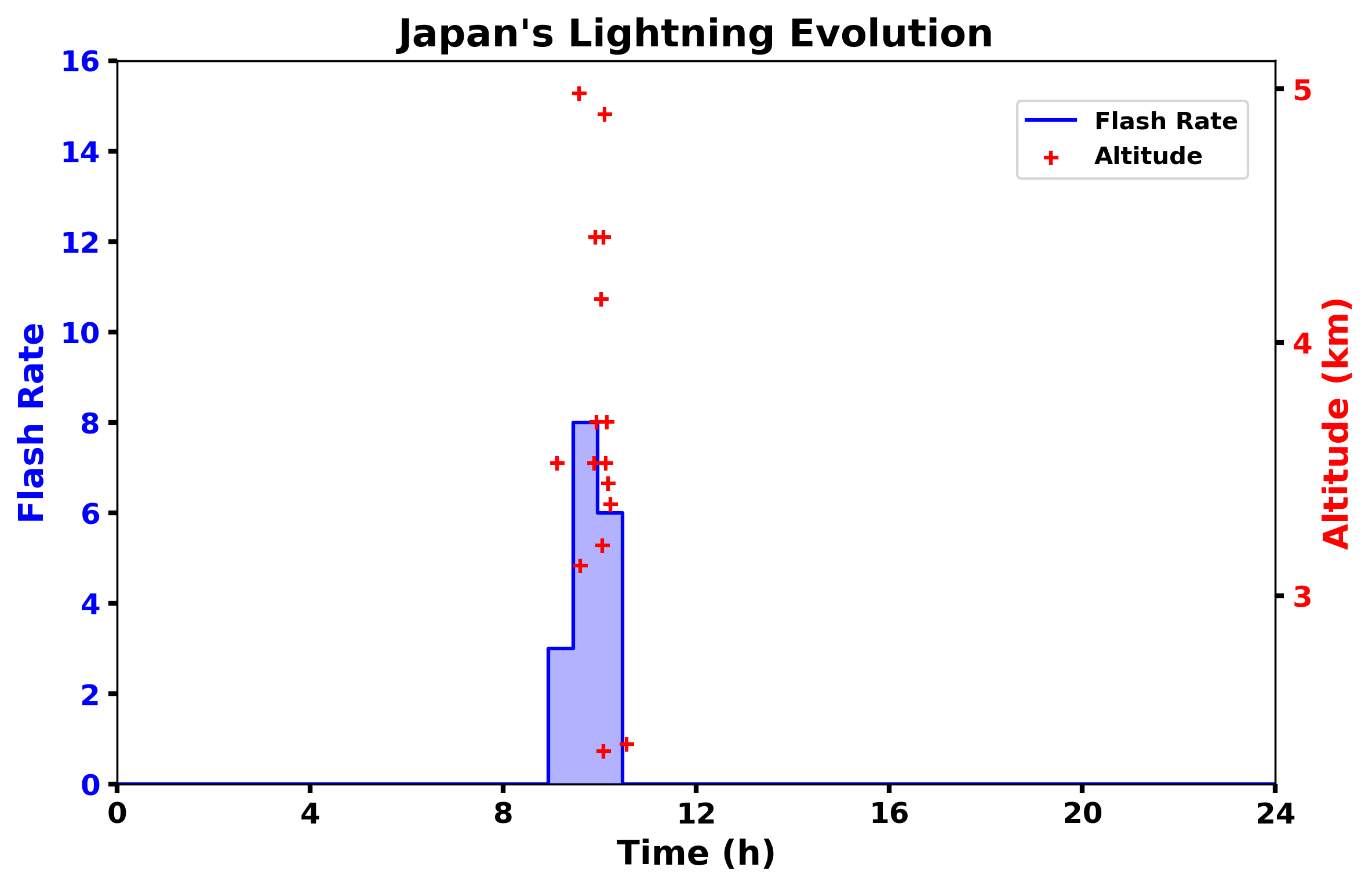}};
                \end{tikzpicture}
            \end{subfigure}
            \hfill
            \begin{subfigure}[t]{0.24\textwidth}
                \centering
                \begin{tikzpicture}
                    \node[anchor=north west, inner sep=0] (image) at (-0.5,0.5)  
                    {\includegraphics[width=\textwidth]{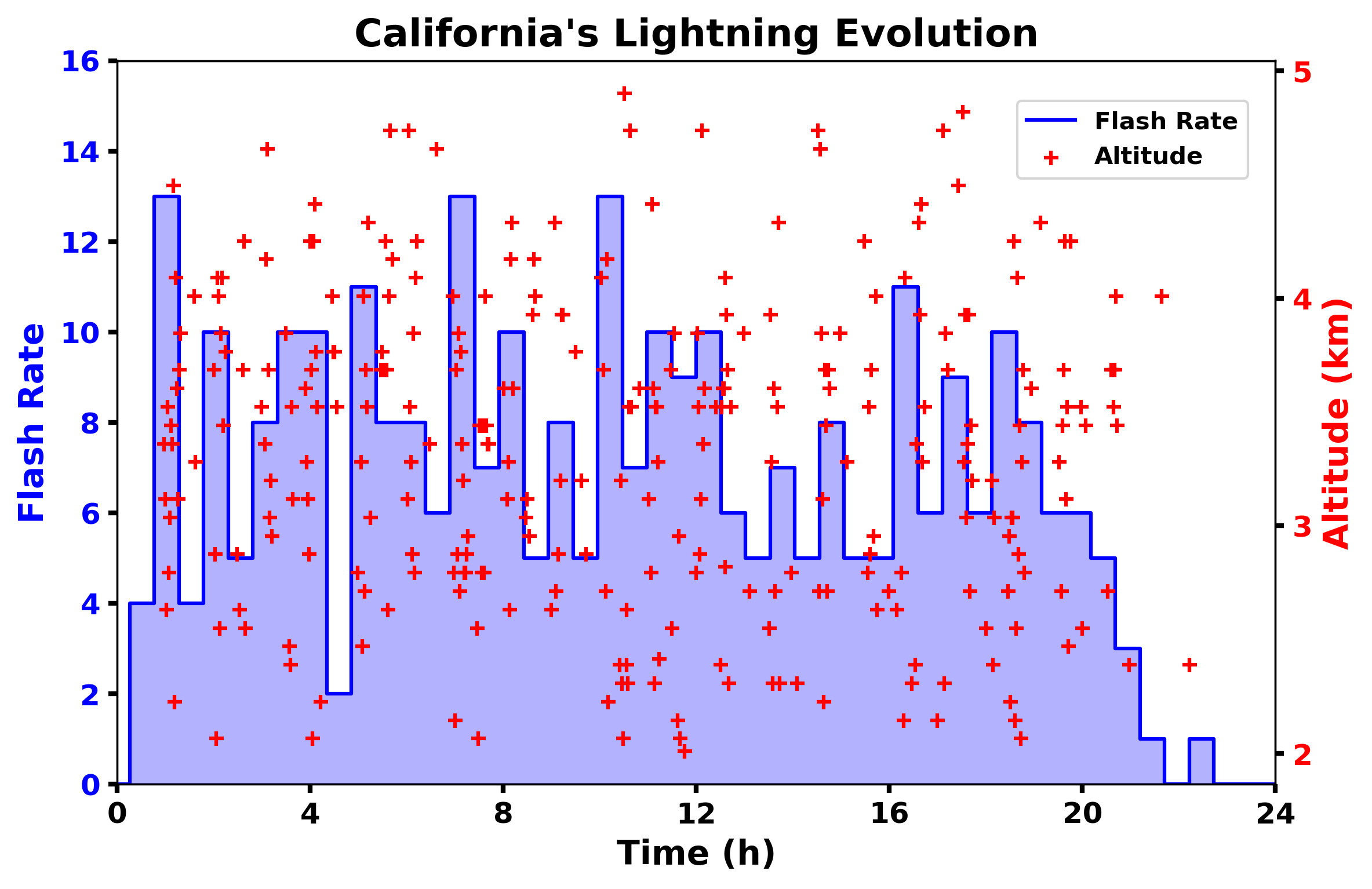}};
                \end{tikzpicture}
            \end{subfigure}
        \end{minipage}
    }
    \caption{Comparison of MCS across four distinct regions: Florida, New Mexico, Japan, and California. \textbf{Top row}: Cloud coverage evolution over a 24-hour period during thunderstorm events. \textbf{Middle row}: Cloud fraction analysis, illustrating the structure and distribution of clouds at the maximum development height of thunderstorms. \textbf{Bottom row}: Lightning activity evolution, capturing the variations in lightning activity as thunderstorms develop. The figures highlight the unique patterns shaped by regional environmental conditions.}
    \label{fig:cloud_real}
    
\end{figure}

\subsection{Consistent Thunderstorm System}

\begin{figure}[htbp]
    \centering
    % 第一行：Florida
    \begin{minipage}{\textwidth}
        \centering
        \begin{subfigure}[t]{0.24\textwidth}
            \centering
            \includegraphics[width=3.5cm,height=2cm]{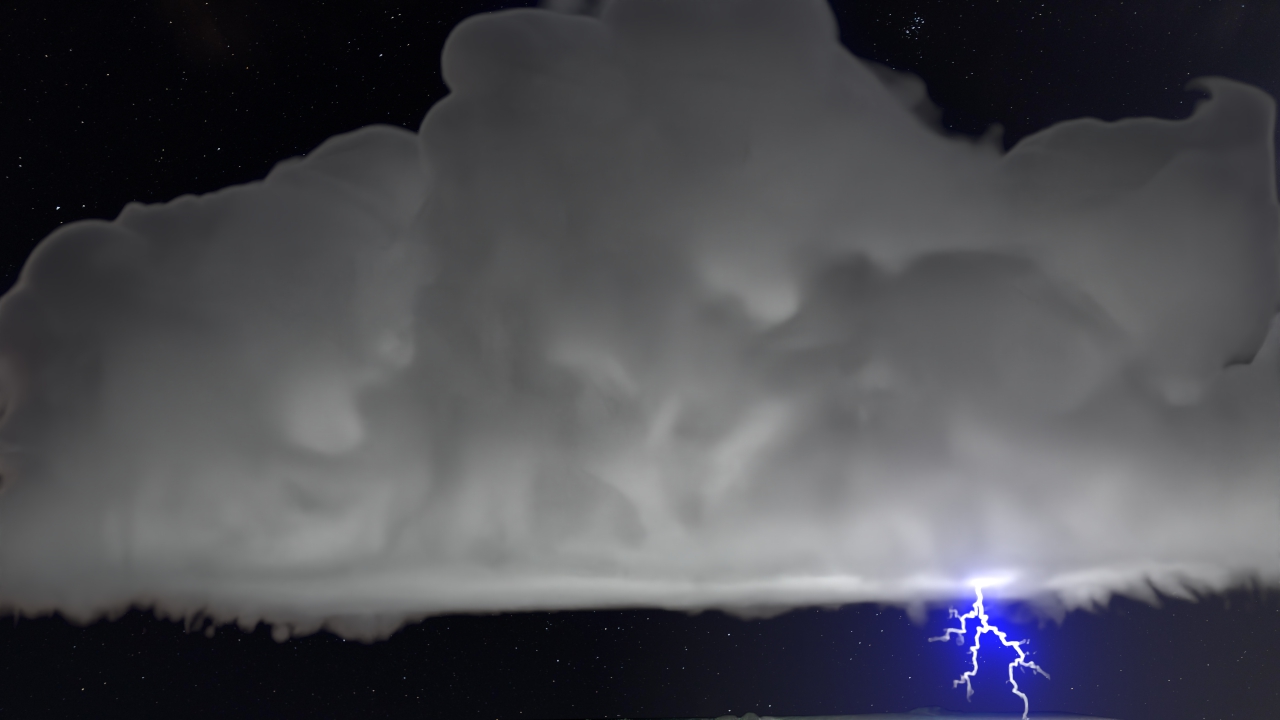}
        \end{subfigure}
        \hfill
        \begin{subfigure}[t]{0.24\textwidth}
            \centering
            \includegraphics[width=3.5cm,height=2cm]{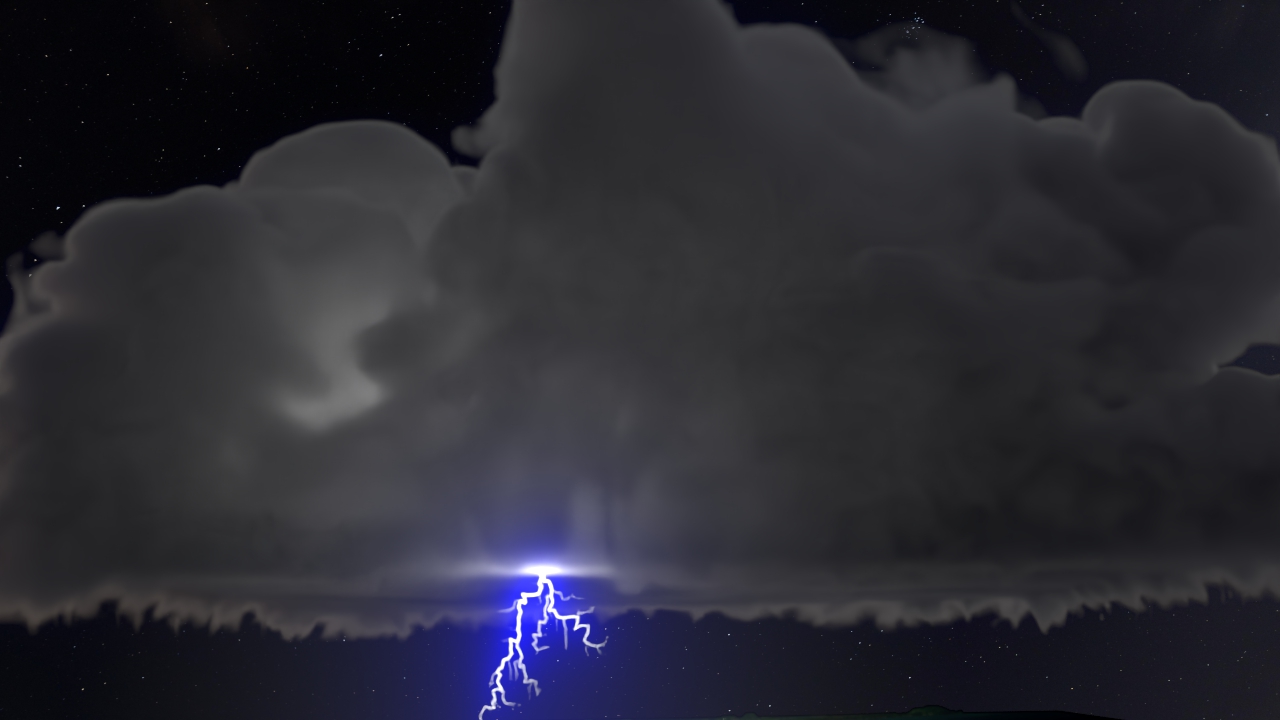}
        \end{subfigure}
        \hfill
        \begin{subfigure}[t]{0.24\textwidth}
            \centering
            \includegraphics[width=3.5cm,height=2cm]{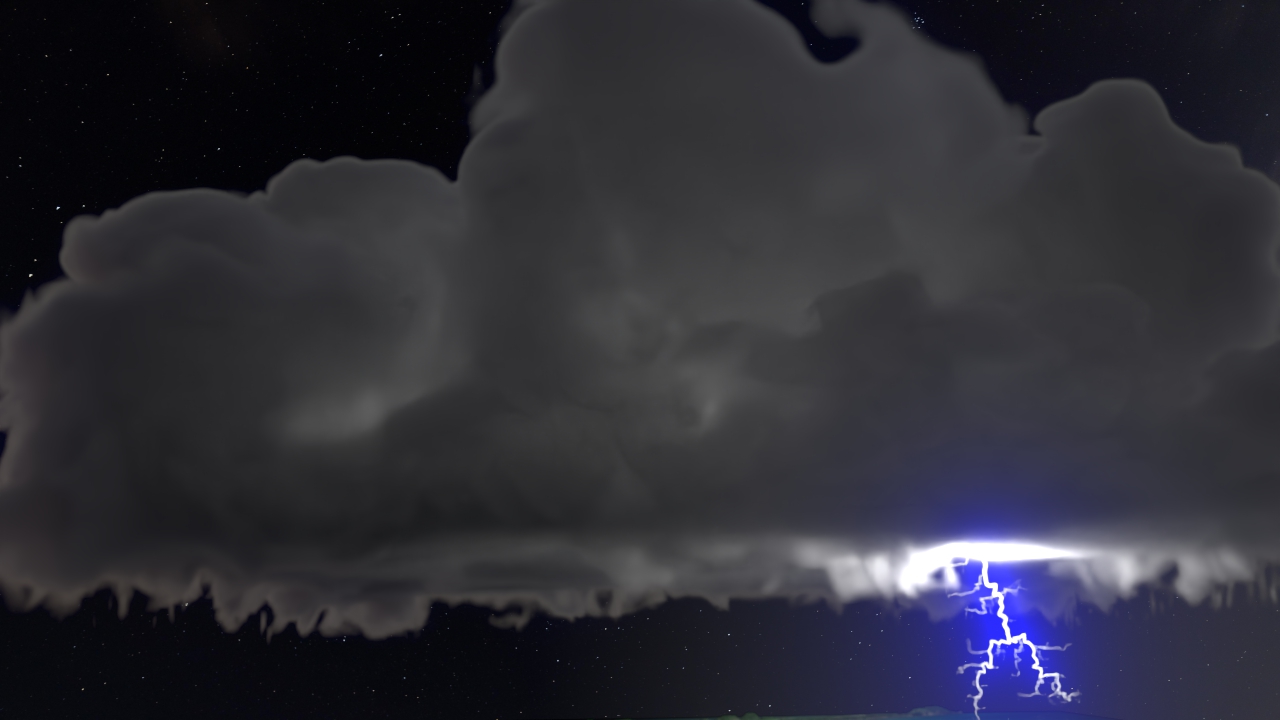}
        \end{subfigure}
        \hfill
        \begin{subfigure}[t]{0.24\textwidth}
            \centering
            \includegraphics[width=3.5cm,height=2cm]{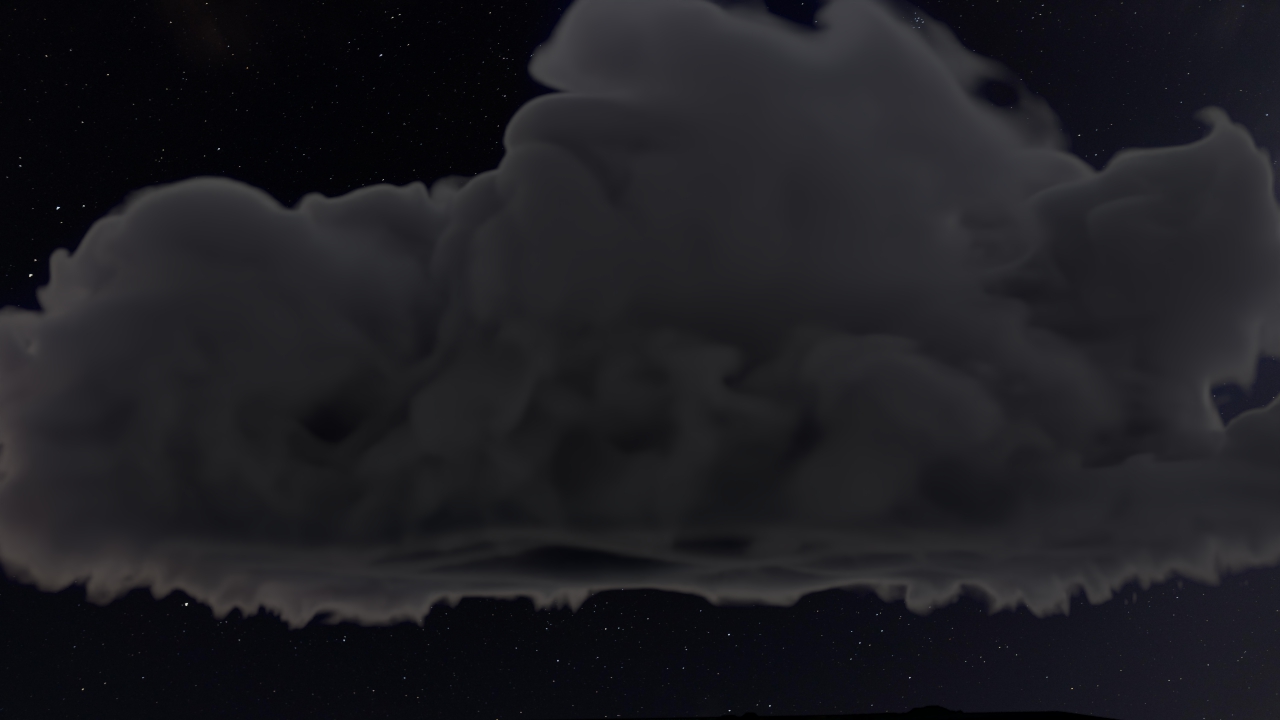}
        \end{subfigure}
    \end{minipage}
    \\[0.5em] % 行间距
    % 第二行：New Mexico
    \begin{minipage}{\textwidth}
        \centering
        \begin{subfigure}[t]{0.24\textwidth}
            \centering
            \includegraphics[width=3.5cm,height=2cm]{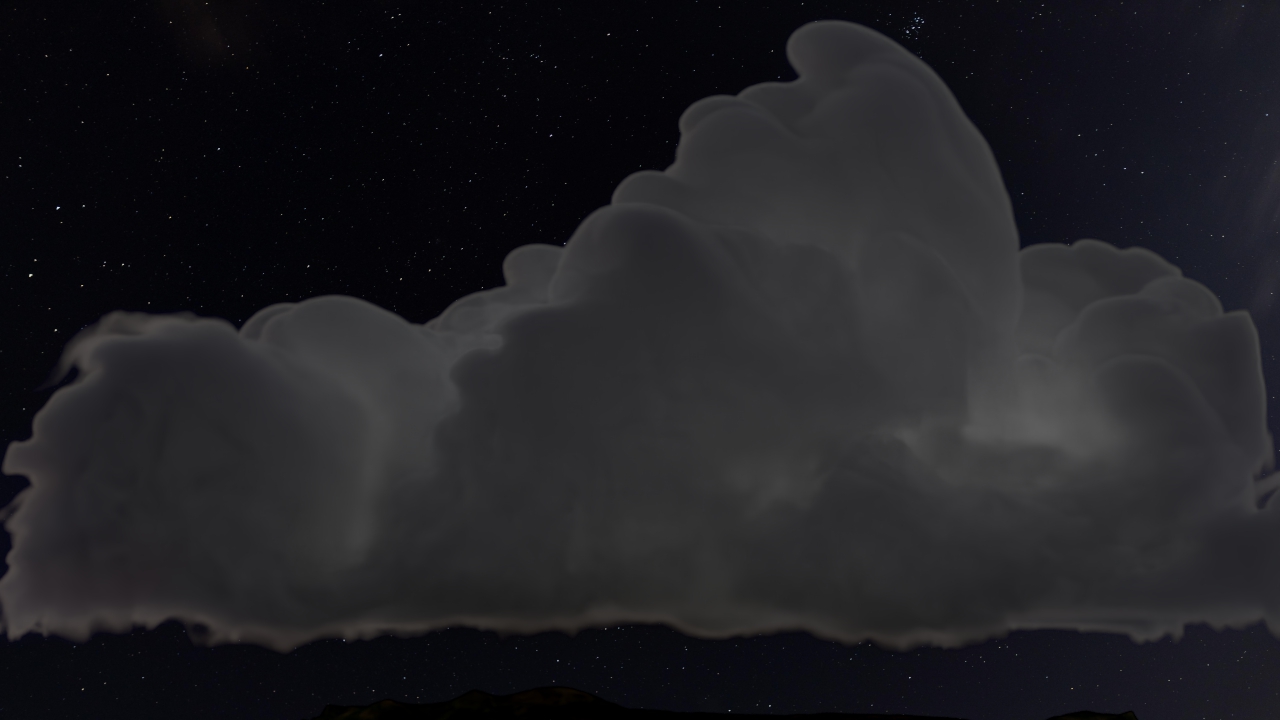}
        \end{subfigure}
        \hfill
        \begin{subfigure}[t]{0.24\textwidth}
            \centering
            \includegraphics[width=3.5cm,height=2cm]{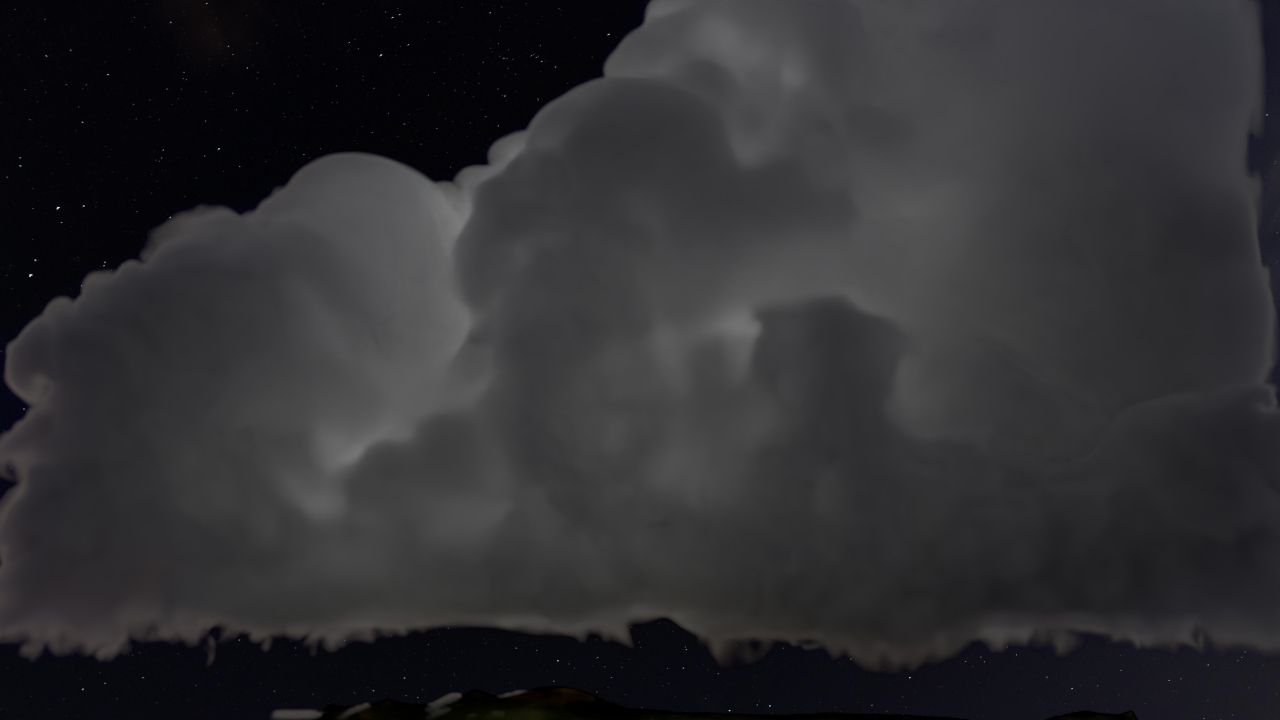}
        \end{subfigure}
        \hfill
        \begin{subfigure}[t]{0.24\textwidth}
            \centering
            \includegraphics[width=3.5cm,height=2cm]{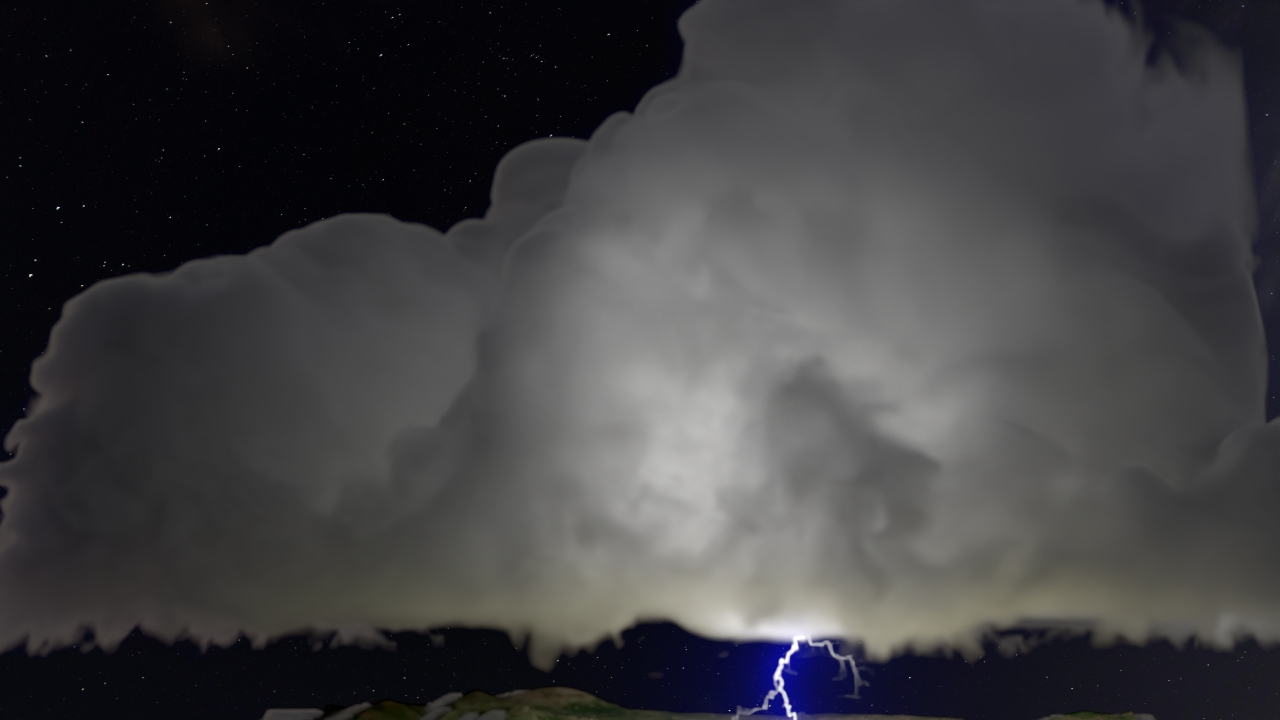}
        \end{subfigure}
        \hfill
        \begin{subfigure}[t]{0.24\textwidth}
            \centering
            \includegraphics[width=3.5cm,height=2cm]{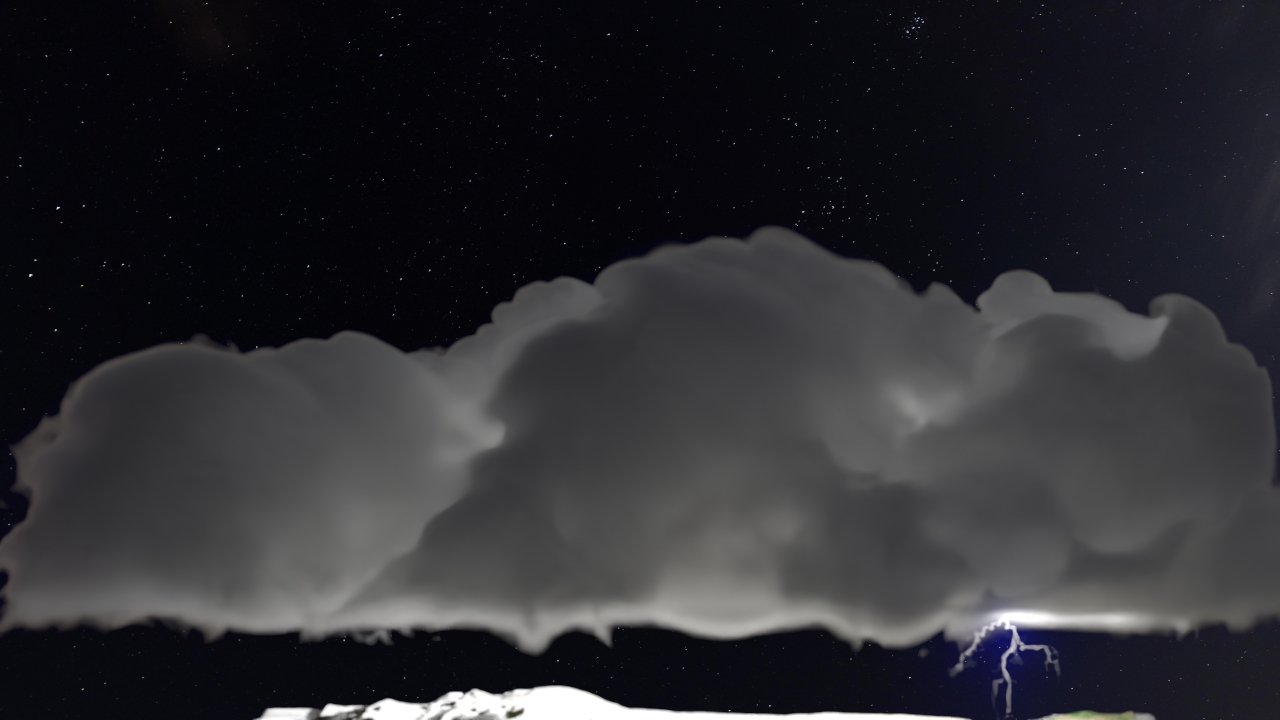}
        \end{subfigure}
    \end{minipage}
    \\[0.5em] % 行间距
    % 第三行：Japan
    \begin{minipage}{\textwidth}
        \centering
        \begin{subfigure}[t]{0.24\textwidth}
            \centering
            \includegraphics[width=3.5cm,height=2cm]{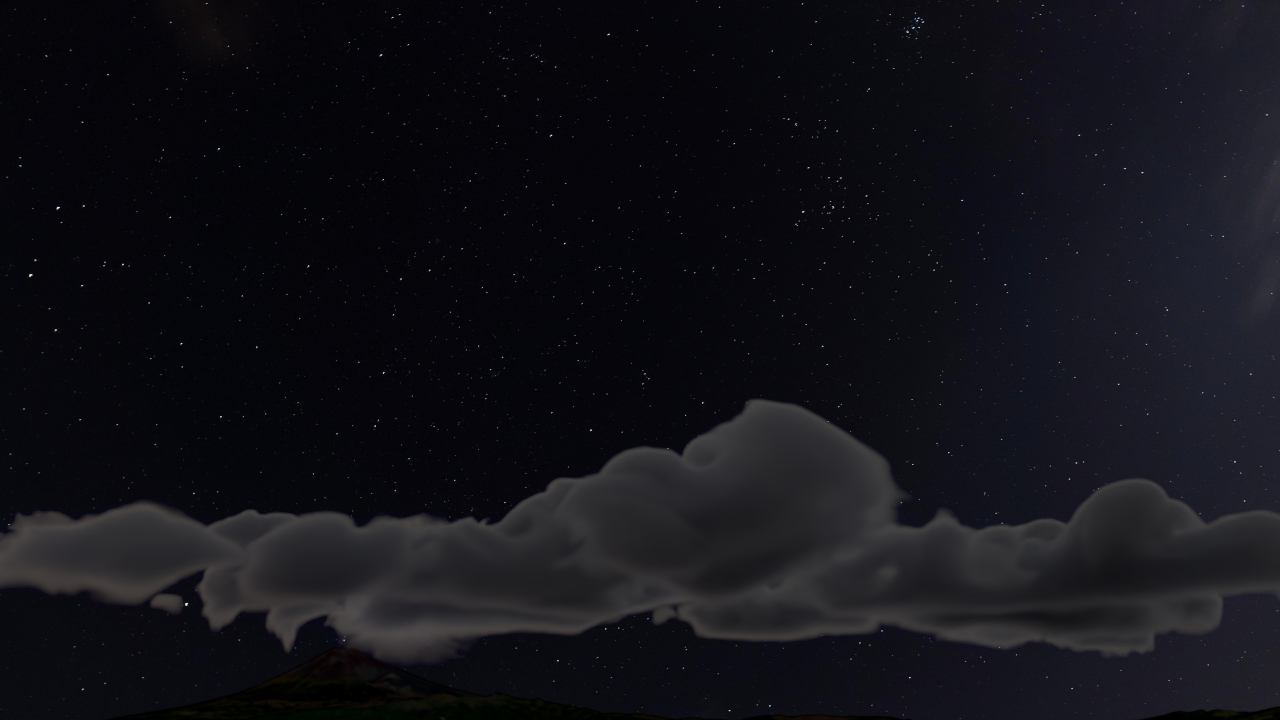}
        \end{subfigure}
        \hfill
        \begin{subfigure}[t]{0.24\textwidth}
            \centering
            \includegraphics[width=3.5cm,height=2cm]{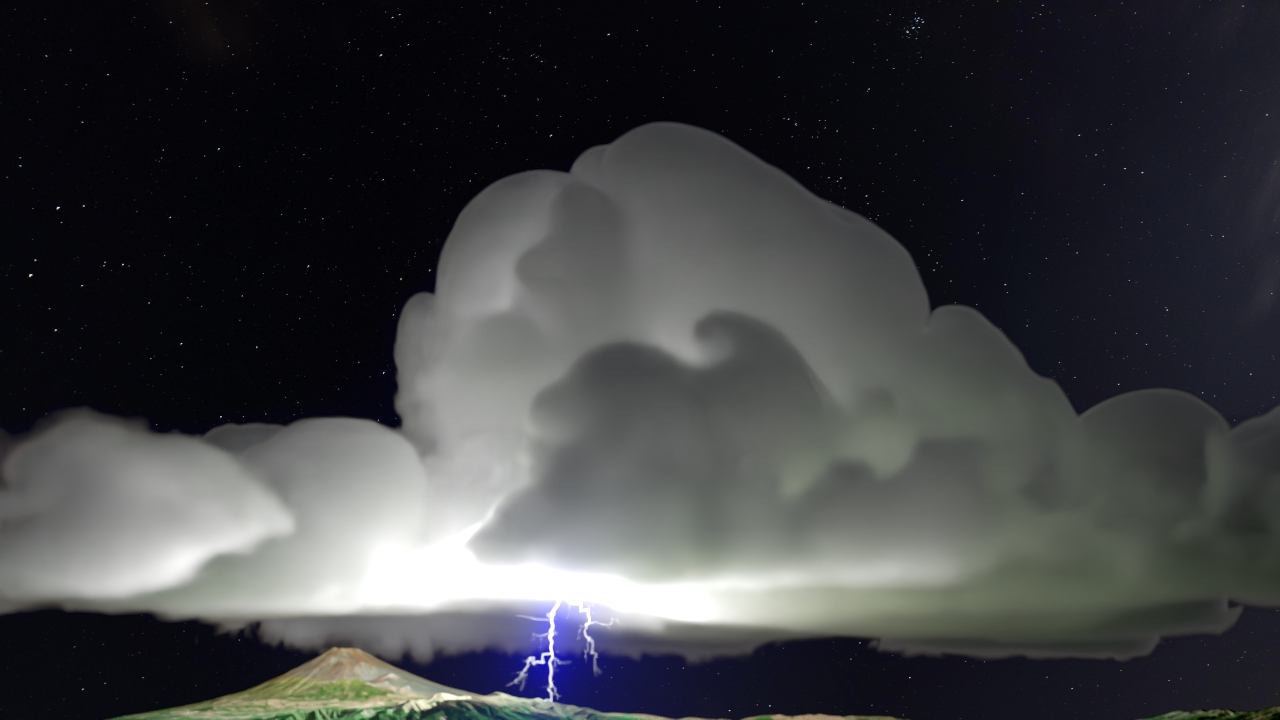}
        \end{subfigure}
        \hfill
        \begin{subfigure}[t]{0.24\textwidth}
            \centering
            \includegraphics[width=3.5cm,height=2cm]{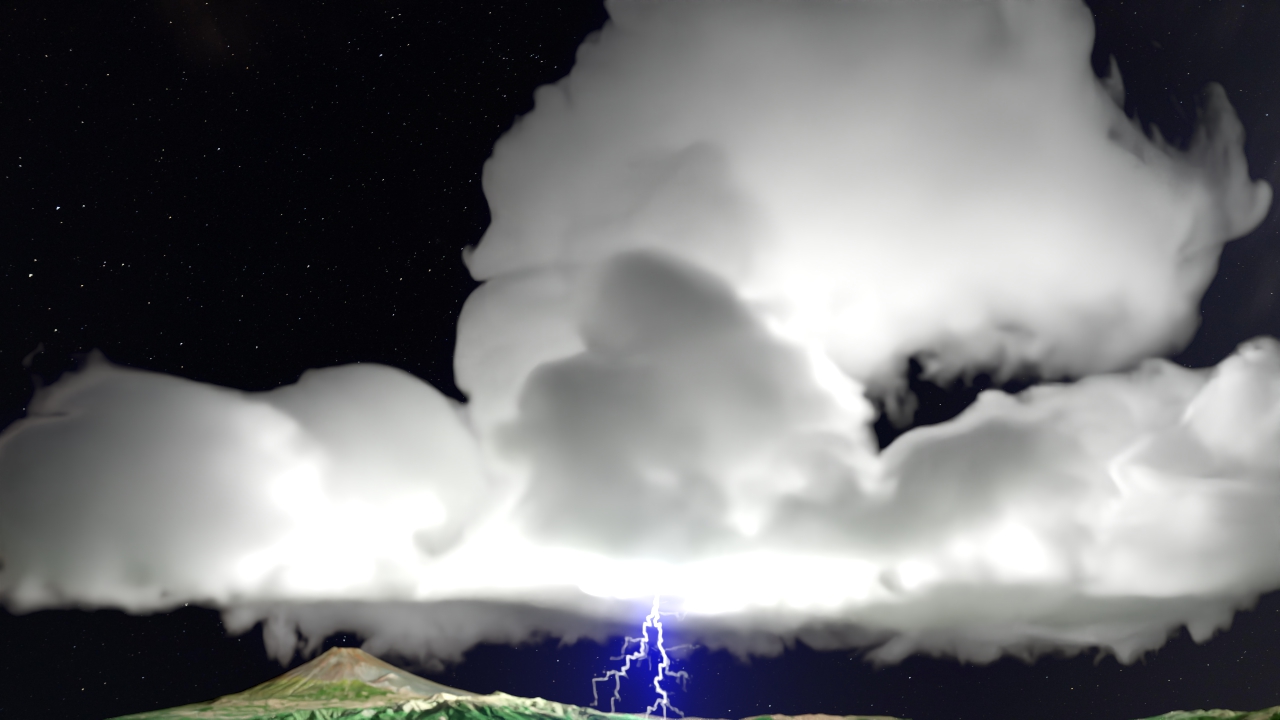}
        \end{subfigure}
        \hfill
        \begin{subfigure}[t]{0.24\textwidth}
            \centering
            \includegraphics[width=3.5cm,height=2cm]{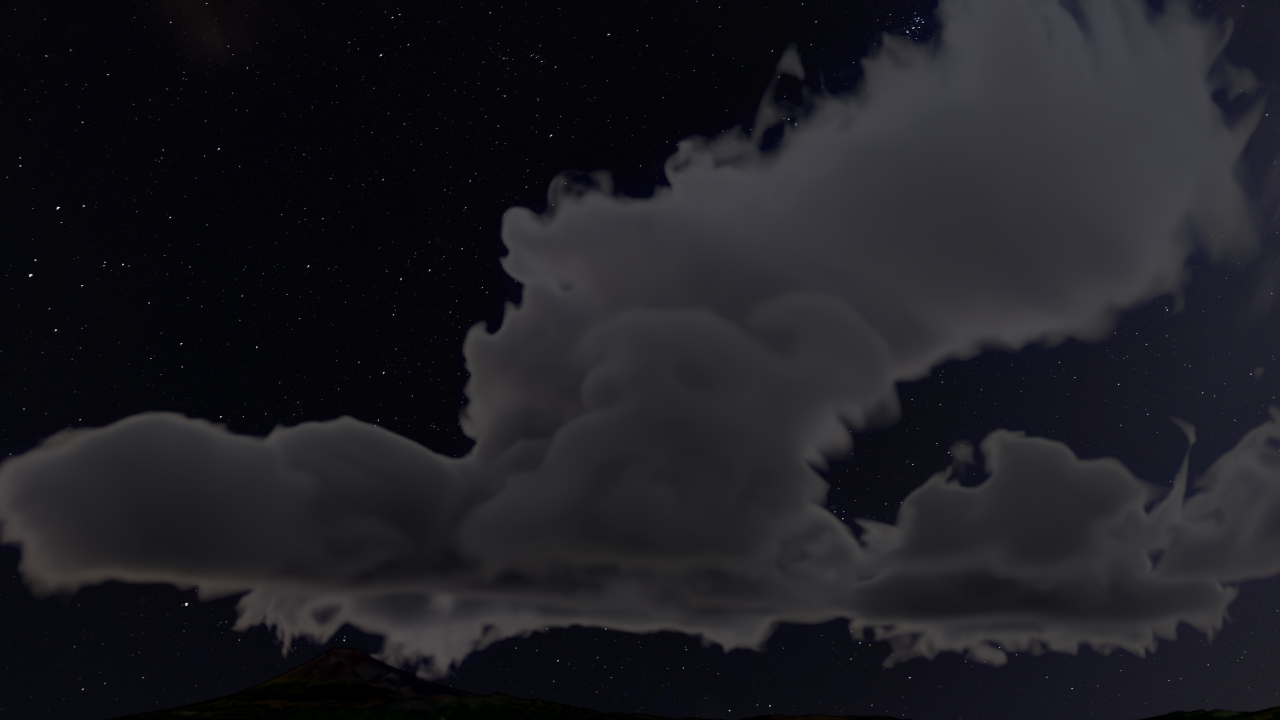}
        \end{subfigure}
    \end{minipage}
    \\[0.5em] % 行间距
    % 第四行：California
    \begin{minipage}{\textwidth}
        \centering
        \begin{subfigure}[t]{0.24\textwidth}
            \centering
            \includegraphics[width=3.5cm,height=2cm]{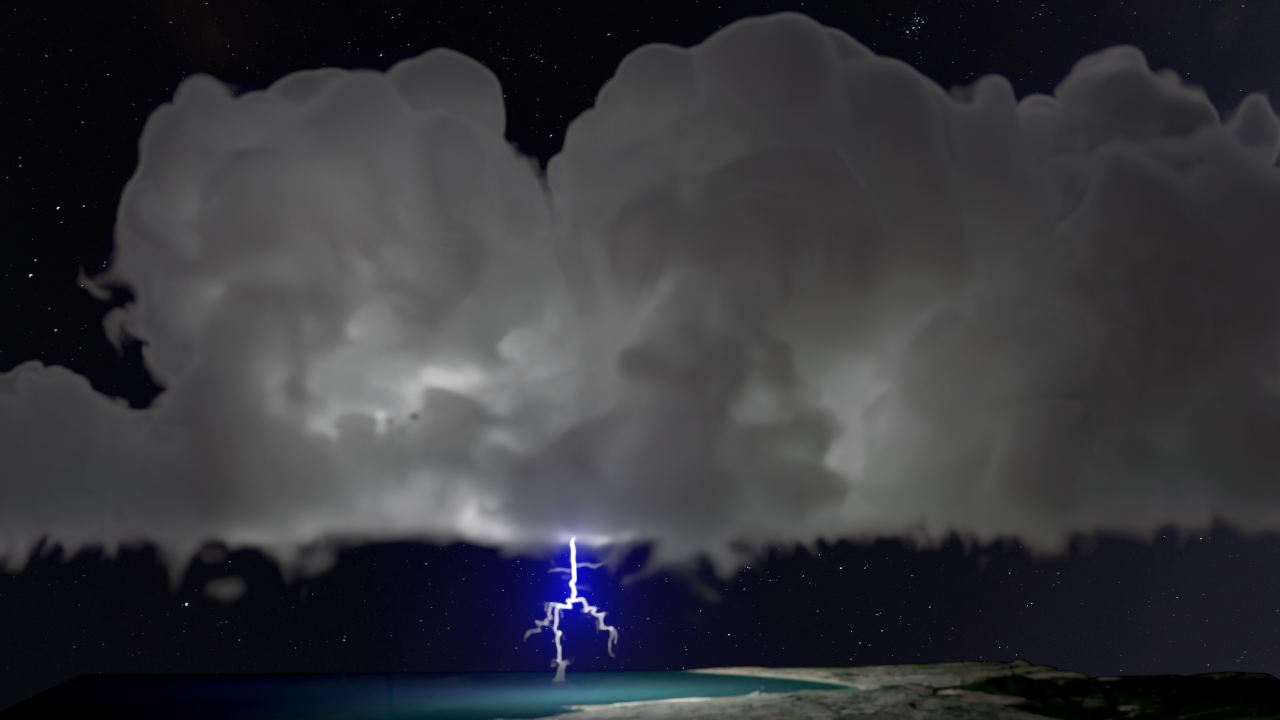}
        \end{subfigure}
        \hfill
        \begin{subfigure}[t]{0.24\textwidth}
            \centering
            \includegraphics[width=3.5cm,height=2cm]{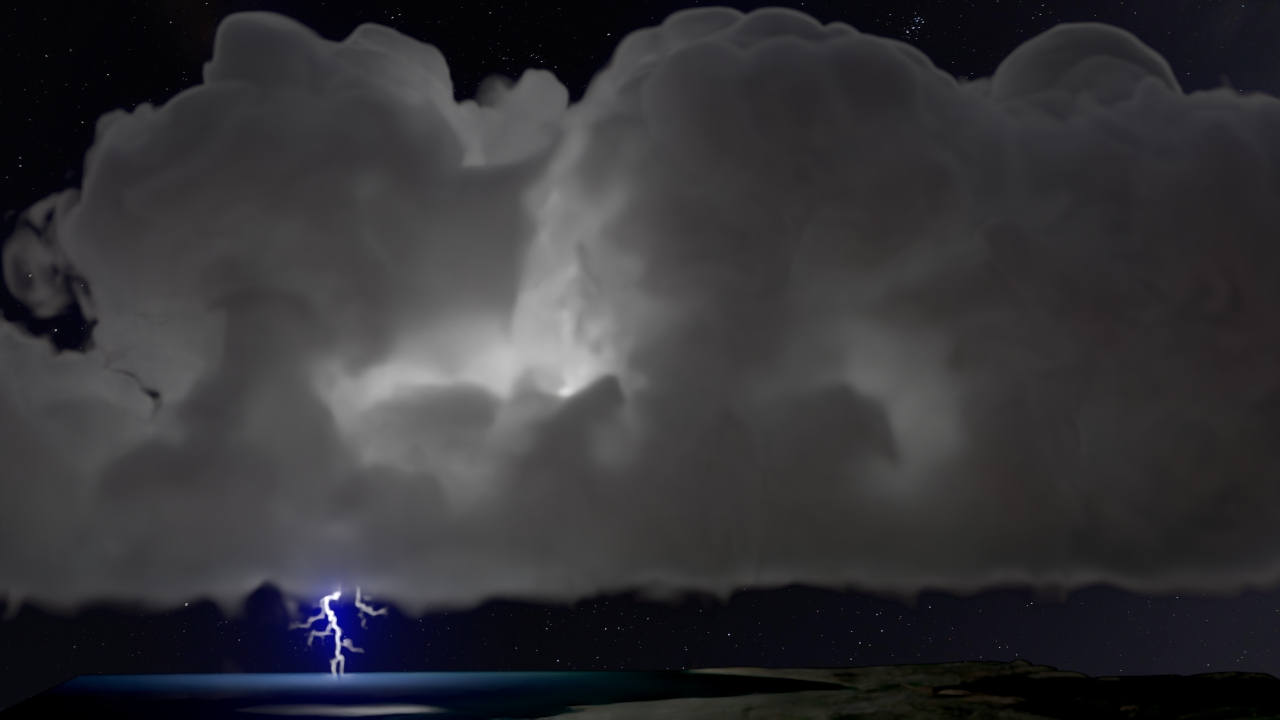}
        \end{subfigure}
        \hfill
        \begin{subfigure}[t]{0.24\textwidth}
            \centering
            \includegraphics[width=3.5cm,height=2cm]{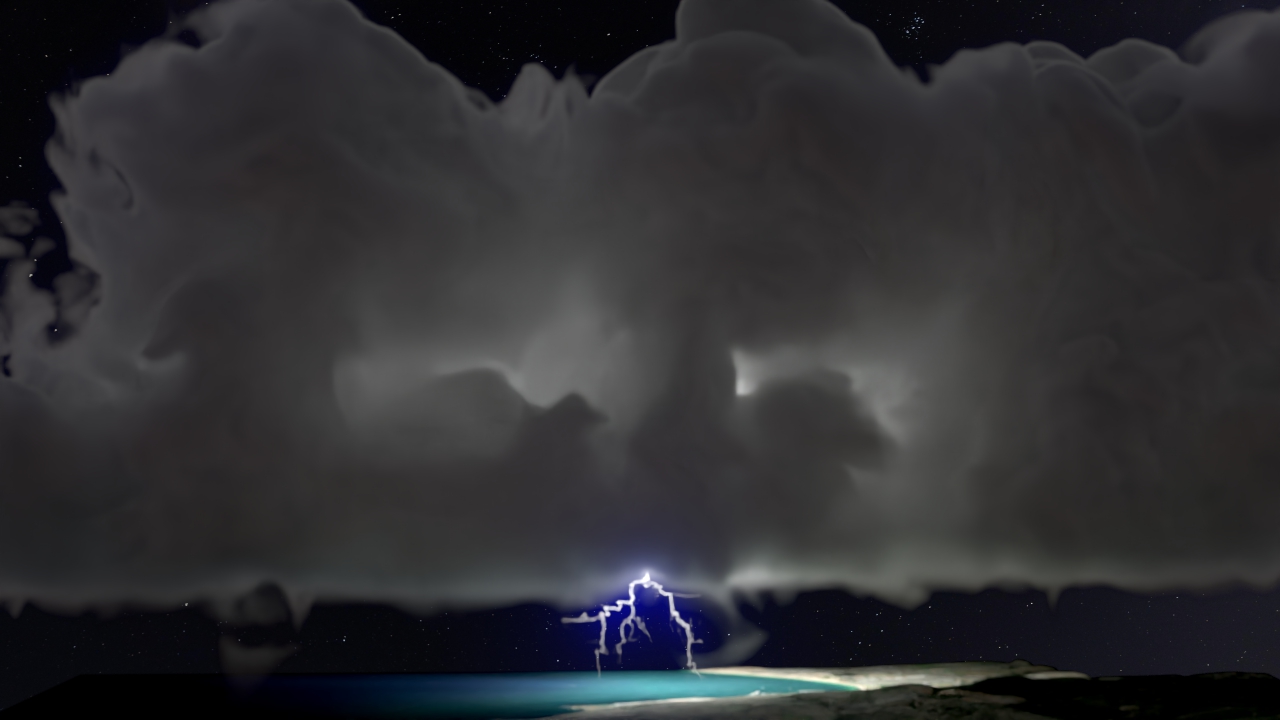}
        \end{subfigure}
        \hfill
        \begin{subfigure}[t]{0.24\textwidth}
            \centering
            \includegraphics[width=3.5cm,height=2cm]{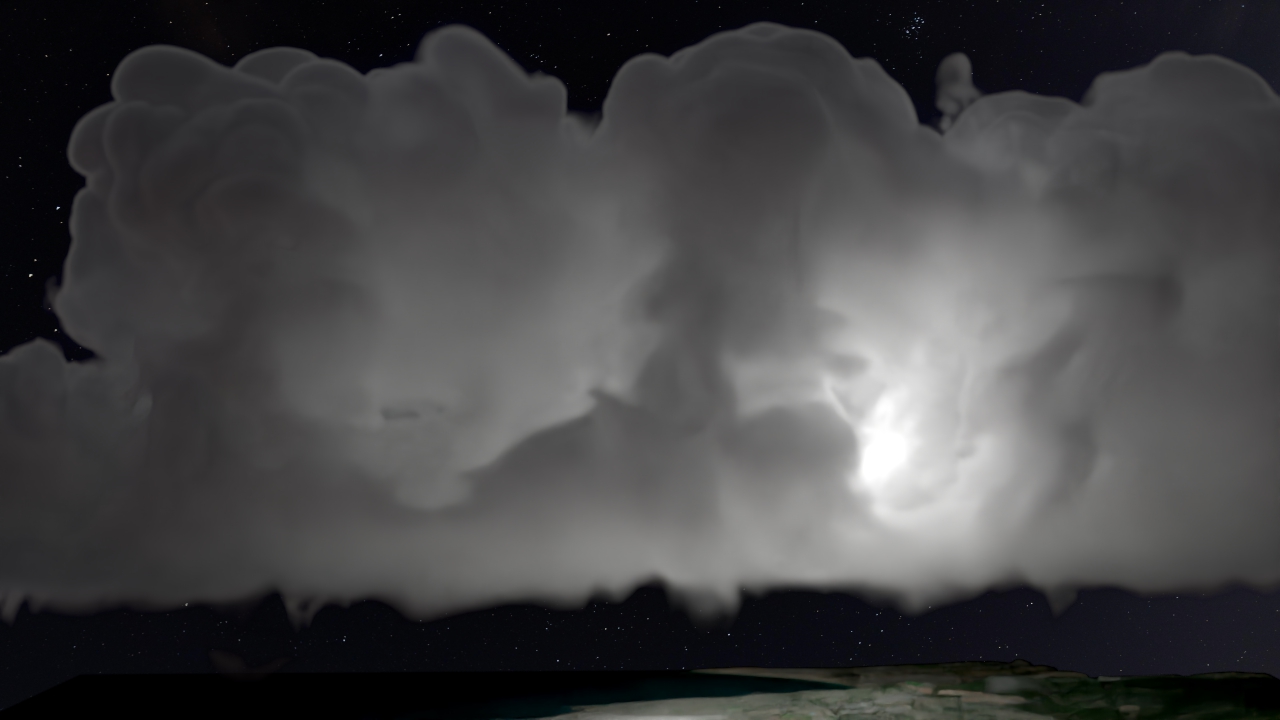}
        \end{subfigure}
    \end{minipage}
    \caption{
    Different regions' MCS evolution over time. First row: Florida at times $t_1 = 6.56\text{h}, t_2 = 9.58\text{h}, t_3 = 13.44\text{h}, t_4 = 20.51\text{h}$. 
    Second row: New Mexico at times $t_1 = 6.12\text{h}, t_2 = 11.45\text{h}, t_3 = 17.70\text{h}, t_4 = 22.32\text{h}$. 
    Third row: Japan at times $t_1 = 5.36\text{h}, t_2 = 9.34\text{h}, t_3 = 10.36\text{h}, t_4 = 10.38\text{h}$. 
    Fourth row: California at times $t_1 = 9.89\text{h}, t_2 = 11.04\text{h}, t_3 = 11.80\text{h}, t_4 = 16.26\text{h}$.
    }

    \label{fig:storm_evolution}
\end{figure}

We proceed to analyze and quantify the simulation data discussed in Section \hyperref[sec:severe]{6.2}.Figure~\ref{fig:cloud_real} provides a comparative overview of MCS characteristics across four regions: Florida, New Mexico, Japan, and California. Additionally, Figure~\ref{fig:storm_evolution} illustrates the development of MCS over specific evolution times, offering insights into temporal dynamics.

Our analysis begins by comparing the simulated cloud coverage attributes with 24-hour observational meteorological data. Next, we assess the cloud fraction at the maximum development height of thunderstorms in each region, providing critical insights into the vertical structure and dynamics of the thunderstorm systems.

Using the evaluation methodology described in \cite{formenton2013using}, we further analyze the evolution of lightning flash rates to capture the electrical activity within the thunderstorms. The flash rate follows a characteristic trend: increasing during the thunderstorm's intensification phase, peaking at maturity, and gradually declining as the system dissipates. Additionally, we examine the altitude of lightning generation, revealing the vertical distribution of electrical activity. The alignment of lightning flash rates with the spatial and temporal evolution of the thunderstorm systems demonstrates the model's consistency in realistically simulating the life cycle of MCS.

\section{CONCLUSION AND FUTURE WORK}
%TODO:每个地区加入三个时刻渲染图（三列四行）
We have developed a physically based simulation framework for visually realistic MCS. This framework integrates a Grabowski-style extended warm cloud microphysics scheme with hydrometeor electrification processes to ensure the consistent coupling of thunderstorms formation and lightning dynamics. Our framework reproduces various thunderstorms types within a MCS.Validation against real-world weather data demonstrates the model’s consistency in simulating cloud structure, cloud coverage, and lightning flash rates, with results benchmarked against national weather services.

%adapt more professional atmospheric concept about thunderstorm microphysics and electrification process
%support more thunderstorm varation:Mesoscale convective complex (MCC),Mesoscale convective vortex (MCV),derecho 
%support more lightning type:Anvil crawlers，Bolt from the blue，Sheet lightning
Future work will focus on expanding the framework to incorporate advanced atmospheric concepts and support a broader range of thunderstorm and lightning phenomena. Enhancements to thunderstorm microphysics, such as more detailed electrification processes and turbulence modeling, will improve the realism of lightning generation. The framework will also be extended to simulate additional thunderstorm types, including Mesoscale Convective Complex (MCC), Mesoscale Convective Vortex (MCV), and derechos, broadening its coverage of mesoscale systems. Furthermore, new lightning types such as anvil crawlers, bolts from the blue, and sheet lightning will be introduced, requiring refined electric field and branching models. Finally, scalability and performance will be enhanced by leveraging advanced numerical techniques and improving hardware utilization. Currently, our numerical method employs a uniform grid.Transitioning to an adaptive grid approach \cite{raateland2022dcgrid} appears to be a promising direction for achieving higher performance.

\renewcommand{\refname}{REFERENCES}  % 如果是 article 类

\bibliographystyle{ACM-Reference-Format}

%\bibliographystyle{plainnat}  % 选择 plainnat 样式来显示为 [Gagnon et al. 2019]
%\bibliography{sample-base}


\end{document}